\documentclass[letterpaper, english, aps, prl, reprint, twocolumn, superscriptaddress]{revtex4-2}
\usepackage[T1]{fontenc}
\usepackage{color}
\usepackage{amsmath}
\usepackage{amssymb}
\usepackage{graphicx}
\usepackage{multirow}
\usepackage{booktabs}
\usepackage{capt-of}
\usepackage[latin9]{inputenc}
\usepackage{babel}
\usepackage{calc}
\usepackage{float}
\makeatletter

\usepackage[bookmarks=true,colorlinks,linkcolor=black,anchorcolor=green,citecolor=blue,urlcolor=blue]{hyperref}

\renewcommand{\fnum@figure}{FIG. \thefigure}
\renewcommand{\fnum@table}{TABLE. \thetable}

\makeatother

\begin{document}
	\title{Second Order Topological Insulator State in Hexagonal Lattices and its Abundant Material Candidates}
	
	\author{Shifeng Qian}
	\affiliation{Centre for Quantum Physics, Key Laboratory of Advanced Optoelectronic Quantum Architecture and Measurement (MOE), School of Physics, Beijing Institute of Technology, Beijing, 100081, China}
	\affiliation{Beijing Key Lab of Nanophotonics \& Ultrafine Optoelectronic Systems, School of Physics, Beijing Institute of Technology, Beijing, 100081, China}
	
	\author{Cheng-Cheng Liu}
	\email{ccliu@bit.edu.cn}
	\affiliation{Centre for Quantum Physics, Key Laboratory of Advanced Optoelectronic Quantum Architecture and Measurement (MOE), School of Physics, Beijing Institute of Technology, Beijing, 100081, China}
	\affiliation{Beijing Key Lab of Nanophotonics \& Ultrafine Optoelectronic Systems, School of Physics, Beijing Institute of Technology, Beijing, 100081, China}
	
	\author{Yugui Yao}
	\email{ygyao@bit.edu.cn}
	\affiliation{Centre for Quantum Physics, Key Laboratory of Advanced Optoelectronic Quantum Architecture and Measurement (MOE), School of Physics, Beijing Institute of Technology, Beijing, 100081, China}
	\affiliation{Beijing Key Lab of Nanophotonics \& Ultrafine Optoelectronic Systems, School of Physics, Beijing Institute of Technology, Beijing, 100081, China}
	
	
	\begin{abstract}
		We propose two mechanisms to realize the second order topological insulator (SOTI) state in spinless hexagonal lattices, viz., chemical modification and anti-Kekul\'e/Kekul\'e distortion of hexagonal lattice. Correspondingly, we construct two models and demonstrate the nontrivial band topology of the SOTI state characterized by the second Stiefel-Whitney class $w_2$ in the presence of inversion symmetry (\textit{P}) and time-reversal symmetry (\textit{T}). Based on the two mechanisms and using first-principles calculations and symmetry analysis, we predict three categories of real light element material candidates, i.e., hydrogenated and halogenated 2D hexagonal group IV materials XY (X=C, Si, Ge, Sn, Y=H, F, Cl), 2D hexagonal group V materials (blue phosphorene, blue arsenene, and black phosphorene, black arsenene), and the recent experimentally synthesized anti-Kekul\'e/Kekul\'e order graphenes and the counterparts of silicene/germanene/stanene. We explicitly demonstrate the nontrivial topological invariants and existence of the protected corner states with fractional charge for these candidates with giant bulk band gap (up to 3.5 eV), which could facilitate the experimental verification by STM. Our approaches and proposed abundant real material candidates will greatly enrich 2D SOTIs and promote their intriguing physics research.
	\end{abstract}
	\maketitle
	
	\paragraph{\textcolor{blue}{Introduction.}\textemdash{}}
	Topological classification of states and search for the material candidates are among the most activate research fields in condensed matter physics \cite{TIRev1, TIRev2, fenlei1, fenlei2, fenlei3}. Recently, high-order topological insulators (HOTIs) have been proposed and attracted great interest \cite{HOTI2, HOTI3, HOTI4, HOTI1, HOTI5,  HOTIREV,  BeathKogome, HOTIA, HOTIGeneral, HOTIPC, Fengliu, Fengliugraphene}. The high-order of topological insulator (TI) is reflected in special bulk-boundary correspondence. A TI in \emph{d} dimensions has protected gapless states on its (\emph{d}-1)-dimensional boundary, while an \emph{n}th (\emph{n} = 2, 3 ...)-order (high-order) TI in \emph{d} dimensions has protected gapless states on its (\emph{d}-\emph{n})-dimensional boundary. For instance, a three dimensional (3D) second-order topological insulator (SOTI)  displays 1D gapless modes along its hinges, and a 2D SOTI hosts zero-energy states localized at its corners. Until now, SOTIs have only been realized in 3D bismuth single crystal \cite{HOTI_Bi} and some artificial systems \cite{mechanic, acoustic1, acoustic2, photonic1, photonic2, photonic3, elect1, elect2, elect3, micro, HOTISonic1, HOTIElast}. In addition, in the literature there are a few theoretical material proposals for 3D and 2D SOTIs \cite{V_Topo, graphdiyne1, graphdiyne2, graphyne1,graphyne2, TBG1, TBG2, MnBiTe,HOTI, MnBiTe, XiDai1, XiDai2}. However, a simple and feasible program to achieve SOTI with abundant real material candidates is still absent, which greatly impeded the experimental and further theoretical studies on SOTIs, especially for 2D.
	
	2D hexagonal lattice systems provide us a platform to realize intriguing phenomena and have always received widespread attention \cite{Macdonald, XiaoDi, yao1, hexamater, valleyhall, yao2, congjun1,congjun2, yongxu, yao3, Das, cclprb2014}. On one hand, graphene, as a superstar material, takes a hexagonal lattice structure, opening the curtain of the 2D material research boom \cite{geim1}. On the other hand, the important typical topological phases originate in hexagonal lattice, such as quantum anomalous Hall effect \cite{haldane1}, valley Hall effect \cite{semenoff} and quantum spin Hall effect \cite{kanemele}.

	In this Letter, we put forward three ways to realize SOTI states in hexagonal lattice based on two different mechanisms, and propose three categories of real materials. First, we construct two kinds of spinless tight-binding (TB) models and demonstrate that the two mechanisms are feasible to realize the SOTI state. The nontrivial band topology is characterized by the second Stiefel-Whitney class in the presence of inversion symmetry (\textit{P}) and time-reversal symmetry (\textit{T}). Based on combined density function theory (DFT) simulation and theoretical analysis,  we predict three major types of real light elememt material families in which spin-orbit coupling (SOC) is irrelevent, i.e., hydrogenated and halogenated 2D hexagonal group IV materials XY (X=C, Si, Ge, Sn, Y=H, F, Cl) \cite{graphane, halide, halide2, SiH}, 2D hexagonal group V materials (blue phosphorene, blue arsenene, and black phosphorene, black arsenene) \cite{blueph, blackP1, blackP2, blueP1, blackAs1, blackAs2, blueAs1}, and the recent experimentally synthesized anti-Kekul\'e/Kekul\'e order graphenes \cite{Kekule1, kekuexp1, kekuexp2, kekuexp3} and Kekul\'e order silicene/germanene/stanene. These materials have giant bulk band gap (up to 3.5 eV), which will facilitate the experimental detection and exploration of the SOTI phase.

	\paragraph{\textcolor{blue}{Symmetry and topological classifaction.}\textemdash{}}
	Let us first review the symmetry of hexagonal lattices. A planar hexagonal lattice is made of two equivalent sublattices and has six fold rotation symmetry, \textit{P} and \textit{T}. For a buckled hexagonal lattice, $\textit{P}$ and $\textit{T}$ are preserved and the six fold rotation symmetry is degraded to three fold rotation symmetry, as shown in Fig. \ref{fig:1}(a). Here, we focus on the light elememt systems with $\textit{P}$ and $\textit{T}$ symmetry.
	
	In the spinless system with $\textit{P}$$\textit{T}$ symmetry, the corresponding Bloch wave functions is real-valued. $\textit{P}$$\textit{T}$ is an antiunitary symmetry operator that is local in momentum space and satisfies the relation of $(\textit{P}\textit{T})^2=1$. The $\textit{P}$$\textit{T}$ can be represented by $\textit{P}$$\textit{T} = K$, where $K$ is the complex conjugation. The invariance of the Hamiltonian $H(\boldsymbol{k})$ under $\textit{P}$$\textit{T}$ imposes the reality condition to $H(\textbf{k})$ and real eigenstates $|u(\boldsymbol{k})\rangle$ \cite{PT}, such that $(\textit{PT}) H(\boldsymbol{k}) (\textit{PT})^{-1}=H^{*}(\boldsymbol{k})=H(\boldsymbol{k})$, $(\textit{PT})|u(\boldsymbol{k})\rangle=|u(\boldsymbol{k})\rangle^{*}=|u(\boldsymbol{k})$. Since the real occupied states are orientable on a sphere, transition functions can be restricted to special orthogonal group $\mathrm{SO}\left(N_{\mathrm{occ}}\right)$ with the number of occupied bands $N_{\mathrm{occ}}$. From the mathematical results of homotopy groups, $\pi_{1}[\mathrm{SO}(2)]=\mathbb{Z}$, $\pi_{1}\left[\mathrm{SO}\left(N_{\mathrm{occ}}>2\right)\right]=\mathbb{Z}_{2}$. There exist $\textit{P}$$\textit{T}$-invariant topological insulators in 2D, characterized by topological invariant corresponding to the above homotopy class. In the case of $N_{\mathrm{occ}}=2$, the topological invariant is thus $\mathbb{Z}$-valued, which corresponds to the Euler characteristic class. The Euler class is not stable in $K$-theory but adding a trivial valence band can collapse the $\mathbb{Z}$ into $\mathbb{Z}_2$ topological invariants, i.e., the so-called second Stiefel-Whitney number $w_{2}$. The $w_{2}$ can measure the higher-order band topology of $\textit{P}$-symmetric spinless fermion systems \cite{PT1, PT2, PT3}. 
	
	One can use three methods to calculate the second Stiefel-Whitney number $w_{2}$, namely, the parity criterion, Wilson loop method, and nested Wilson loop method \cite{parity, PT3, axion1}. The value of $w_{2}$ can be calculated by parity eigenvalues in the presence of $\textit{P}$ symmetry,
	\begin{equation}
		(-1)^{w_{2}}=\prod_{i=1}^{4}(-1)^{\left \lfloor N_{\text {occ }}^{-}\left(\Gamma_{i}\right) / 2\right \rfloor},
	\end{equation}
	where $N_{\text {occ }}^{-}\left(\Gamma_{i}\right)$ is the number of occupied bands with odd parity at time-reversal invariant momentum (TRIM) $\Gamma_{i}$ and $\left \lfloor \right \rfloor$ is the floor function. We also use Wilson loop method and nested Wilson loop method to check the second Stiefel-Whitney number $w_{2}$ \cite{SuppMater}.
	
	\paragraph{\textcolor{blue}{Two mechanisms, three ways to achieve the second order topological insulator state in hexagonal lattices.}\textemdash{}} The parity formula provides us an intuitive knob to engineer the second Stiefel-Whitney number $w_2$. For pristine hexagonal lattice (e.g. graphene), $w_2=1$ for the $\sigma$ bands ($sp^2$) alone, while $w_2=0$ for the $\pi$ bands ($p_z$) alone. Fortunately, the $w_2$ of the $\pi$ bands can be engineered to nontrivial value through a simple $\sqrt{3}\times\sqrt{3}$ distortion. As a result, we propose two physical mechanisms to approach the SOTI state, i.e., chemical modification to remove the $p_z$ orbitals with a gap left and anti-Kekul\'e/Kekul\'e distortion[Fig. \ref{fig:1}]. According to the two mechanisms, we give three ways, i.e., group IV hexagonal lattices with chemical modification, group V hexagonal lattices with appropriate electron filling, and group IV hexagonal lattices anti-Kekul\'e/Kekul\'e distortion, to realize the SOTI state in planar, buckled or puckered hexagonal lattices.
	
	\begin{figure}
		\begin{center}
			\includegraphics[width=1\linewidth]{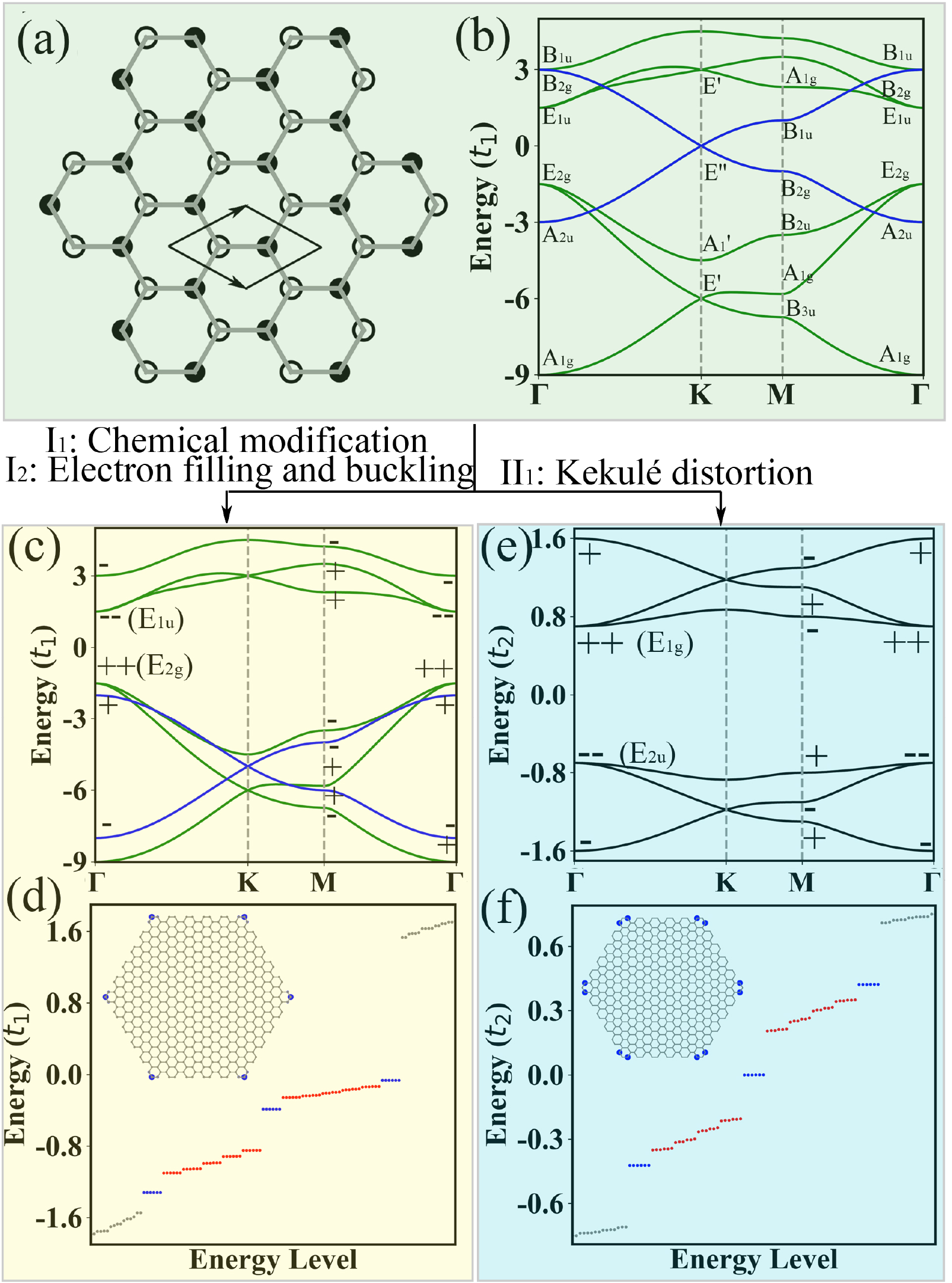}
		\end{center} 
		\caption{Schematic illustration of two mechanisms and three ways to achieve the second order topological insulator state. (a) Hexagonal lattices with solid and hollow circles labeling two sublattices. (b) Band structure of the eight-band model. For a planar hexagonal structure, the $p_z$ orbitals (blue) is decoupled with the $sp^2$ ($s$, $p_x$, $p_y$) orbitals (green). (c) Same as (b) but with $p_z$ orbitals (blue) shifted under the Fermi level. (d) The energy spectrum of the finite-size hexagonal flake plotted in the inset. The charge spatial distribution of the six zero-energy modes, which are localized at corners, is shown in the inset. Blue, red and grey dots represent corner, edge and bulk states, respectively. (e) (f) Same as (c) (d) but for anti-Kekul\'e distortion hexagonal lattices. The parameters for (b)-(f) are given in Suppl. Mater. \cite{SuppMater}.}  \label{fig:1}
	\end{figure}
	
	For Mechanism I, we build an eight-band TB model (Model I) for both planar or buckled hexagonal lattices \cite{SuppMater}. Figure \ref{fig:1}(b) shows the band structure and irreducible representations (IR) at high symmetric points for Model I, similar to graphene. The $p_z$ orbitals (blue lines) are decoupled with $sp^2$ orbitals (green lines). There is a Dirac cone at the Fermi level located at the $K$ point. One can not open a gap at $K$ point between two $p_z$ orbitals with the $\textit{P}\textit{T}$ preserved in the spinless case. However, one can shift the two $p_z$ orbitals entirely up or down the Fermi level to obtain a gap by chemical modification. The corresponding band structure and parity eigenvalues at TRIMs are exhibited in Fig. \ref{fig:1}(c), where the onsite energy of $p_z$ orbitals is set to $-5t_1$ to force the $p_z$ orbitals shift down the Fermi level. No matter the two $p_z$ orbitals is shifted up or down, which depends on the relative electronegativity, the second Stiefel-Whitney number $w_{2}$ of the bands under the Fermi level is nontrivial according to the parity eigenvalues or Wilson loop spectrum, indicating a SOTI state. The nontrivial band topology is also preserved in the buckled structure as long as the buckling height does not close the gap. To explore the hallmark corner states of 2D SOTIs, we calculate the discrete energy spectrum of a hexagonal finite-size flake, as shown in Fig. \ref{fig:1}(d). There are six states degenerate at zero energy in the spectrum, whose charge spatial distribution shown in the inset is well located at the corners of the flake, corresponding to the corner states. The red (gray) dots represent the surface (bulk) states and the blue dots represent the corner states which are situated up, middle and down the surface states \cite{SuppMater}.
	
	For Mechanism II, we develop anti-Kekul\'e and Kekul\'e distortion hexagonal lattice model (Model II) \cite{SuppMater}. The $\pi$ bands at original $K$ point are folded onto the new $\Gamma$ point with a gap opened. Figure \ref{fig:1}(e) shows band structure of the anti-Kekul\'e distortion hexagonal lattice with a gap opened at $\Gamma$. From the parity criterion or two Wilson loop methods, the second Stiefel-Whitney number $w_{2}$ is also nontrivial, which means a SOTI state. We build a hexagonal flake and calculate its discrete energy spectrum (Fig. \ref{fig:1}(f)). The charge spatial distribution of the six zero-energy states shown in the inset are well located at the flake's corners, i.e. the corner states.
	
	\paragraph{\textcolor{blue}{Material Realization with abundant real candidates.}\textemdash{}}
	Based on the above two mechanisms and three ways to achieve the SOIT state, we propose the following three major types of real material candidate systems. 
	
	\begin{figure}
		\begin{center}
			\includegraphics[width=1\linewidth]{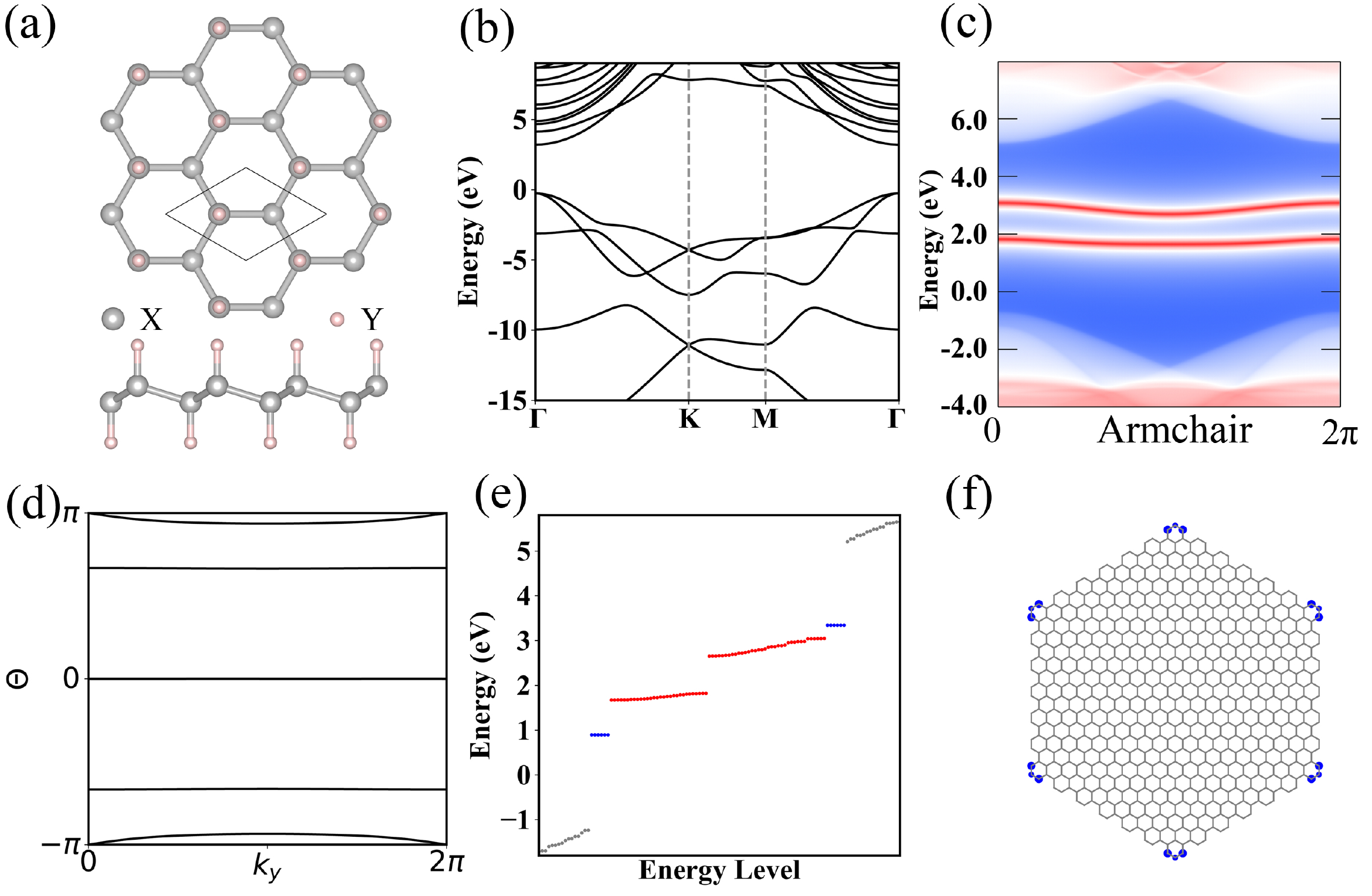}
		\end{center}
		\caption{(a) Top view and side view of geometry structures of hydrogenated or halogenated 2D hexagonal group IV materials XY (X = C, Si, Ge, Sn, Y = H, F, Cl). (b) (c) Bulk bands and edge states of graphane from DFT calculation. (d) Wilson loop of graphane. The number of crossing on $\theta=\pi$ is 1, therefore the $w_2$ is 1. (e) Energy spectrum of hexagonal finite-size flake shown in (f). Blue, red and grey dots represent the corner, edge, and bulk states, respectively. (f)  Charge spatial distribution of blue states in (e).} \label{fig:2}
	\end{figure}
	\paragraph{Hydrogenated and halogenated 2D hexagonal group IV materials XY(X=C, Si, Ge, Sn, Y=H, F, Cl).\textemdash{}} These group IV hydrides and halides take a buckled hexagonal  geometry [Fig. \ref{fig:2}(a)]. We take graphane, i.e., hydride-graphene, as an example. Figure \ref{fig:2}(b) shows the band structure of graphane with a direct band gap (3.5 eV) at $\Gamma$. There are two gapped armchair edge states at the middle of the gap, as shown in Fig. \ref{fig:2}(c). Next, we explore the SOTI state in graphane. The nontrivial second Stiefel-Whitney number is demonstrated by the Wilson loop spectrum as shown in Fig. \ref{fig:2}(d), where the spectrum exhibits a crossing point at $k_y = 0$, $\theta = \pi$,  indicating $w_2 = 1$. The nonzero $w_2$ is checked  by the nested Wilson loop method [Table \ref{table: 1}] and parity criterion [Table. S1]. The $w_2$ of all SOTI candidate materials are verified in the three methods. To explicitly reveal the corner states, we calculate the discrete spectrum [Fig. 2(e)] of the hexagonal finite-size flake based on DFT calculation and Wannier function. Up and below the edge states exit six degenerate states (blue dots in Fig. \ref{fig:2}(d)), whose charge spatial distribution is plotted in Fig. \ref{fig:2}(e). Such six degenerate states are well located at the six corners of the flake [Fig. \ref{fig:2}(f)], i.e., the corner states. The bulk bands, edge states and corner states of other 2D group IV hydrides/halides are given in Suppl. Mater. \cite{SuppMater}.
	
	\begin{figure}
		\begin{center}
			\includegraphics[width=1\linewidth]{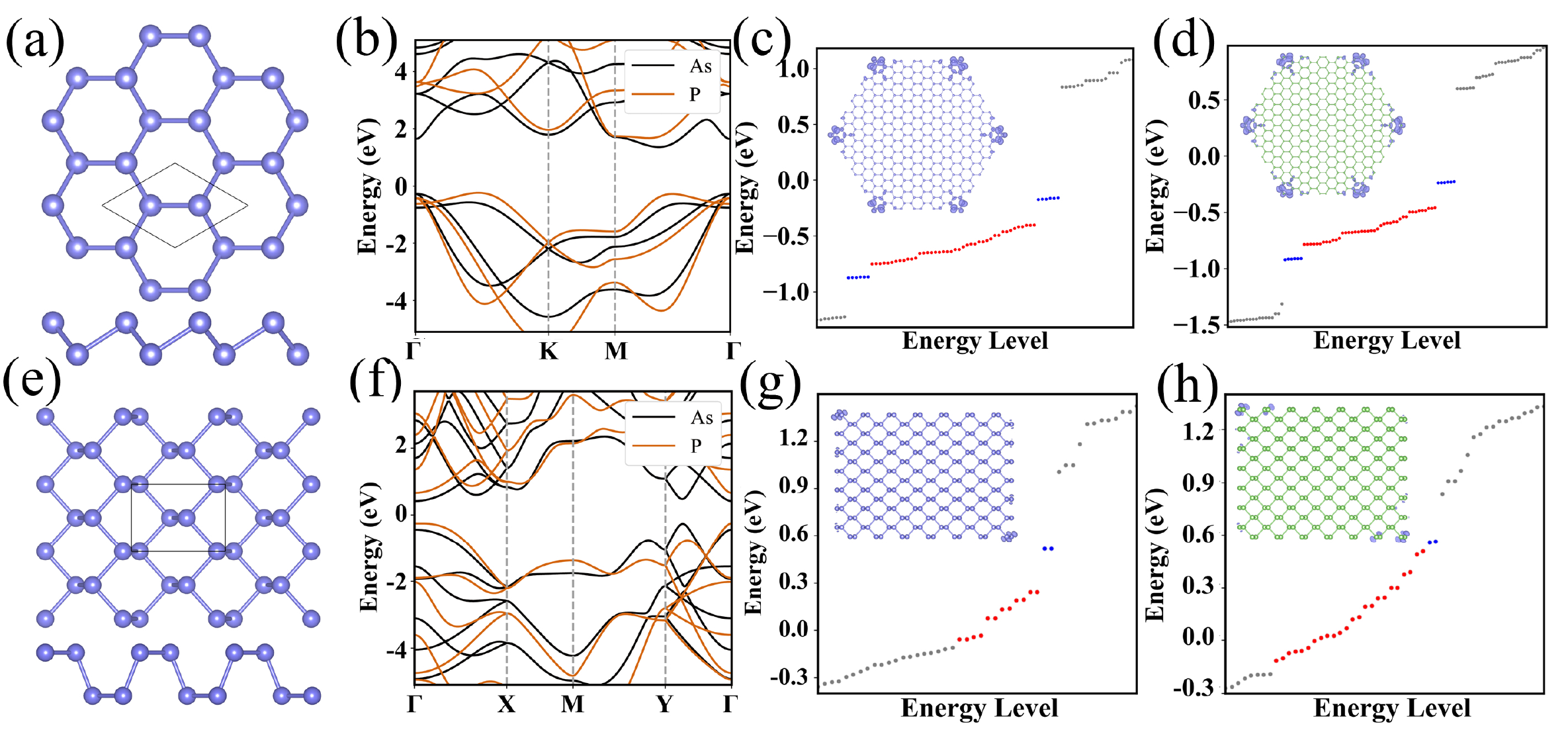}
		\end{center}
		\caption{Top view and side view of geometry structures of blue phosphene or blue arsenene (a) and black phosphene or arsenene (e). Bulk band structures of blue phosphene (orange), blue arsenene (black) (b), and black phosphene (orange) and black arsenene (black) (f). Energy discrete spectra of the finite-size hexagonal/rectangle flakes of blue/black phosphene (c)/(g) and blue/black arsenene (d)/(h) calculated from DFT. The corresponding charge spatial distribution of blue states in (c), (d), (g) and (h) plotted in their insets are well localized at the corners, i.e., the corner states, and the blue, red and grey dots represent the corner, edge, and bulk states, respectively. } \label{fig:3}
	\end{figure}
	
	\paragraph{2D hexagonal group V materials (blue phosphorene, blue arsenene, and black phosphorene,  black  arsenene).\textemdash{}} The blue phosphorene and arsenene [Fig. \ref{fig:3}(a)] and black phosphorene and arsenene [Fig. \ref{fig:3}(e)] are also good SOTI candidates. Although black phosphorene and arsenene do not take a normal hexagonal lattice, they can be seen as puckered honeycomb structures with $\textit{P}$ and $\textit{T}$ symmetry. The band structures of blue phosphene, blue arsenene, black phosphene and black arsenene  are given in Figs. \ref{fig:3}(b, f). The second Stiefel-Whitney number $w_2$ of these four 2D hexagonal group V materials are all nontrivial. We calculate the energy spectra of the finite-size flakes for blue phosphene, blue arsenene, black phosphene and black arsenene, and their hallmark corner states are explicitly displayed from DFT calculation, as shown in Figs. \ref{fig:3}(c, d) and Figs. \ref{fig:3}(g, h). We notice that black phosphorene was also predicted as a SOTI based on TB model recently \cite{blackPmodel1,blackPmodel2}.
	
	\begin{figure}
		\begin{center}
			\includegraphics[width=1\linewidth]{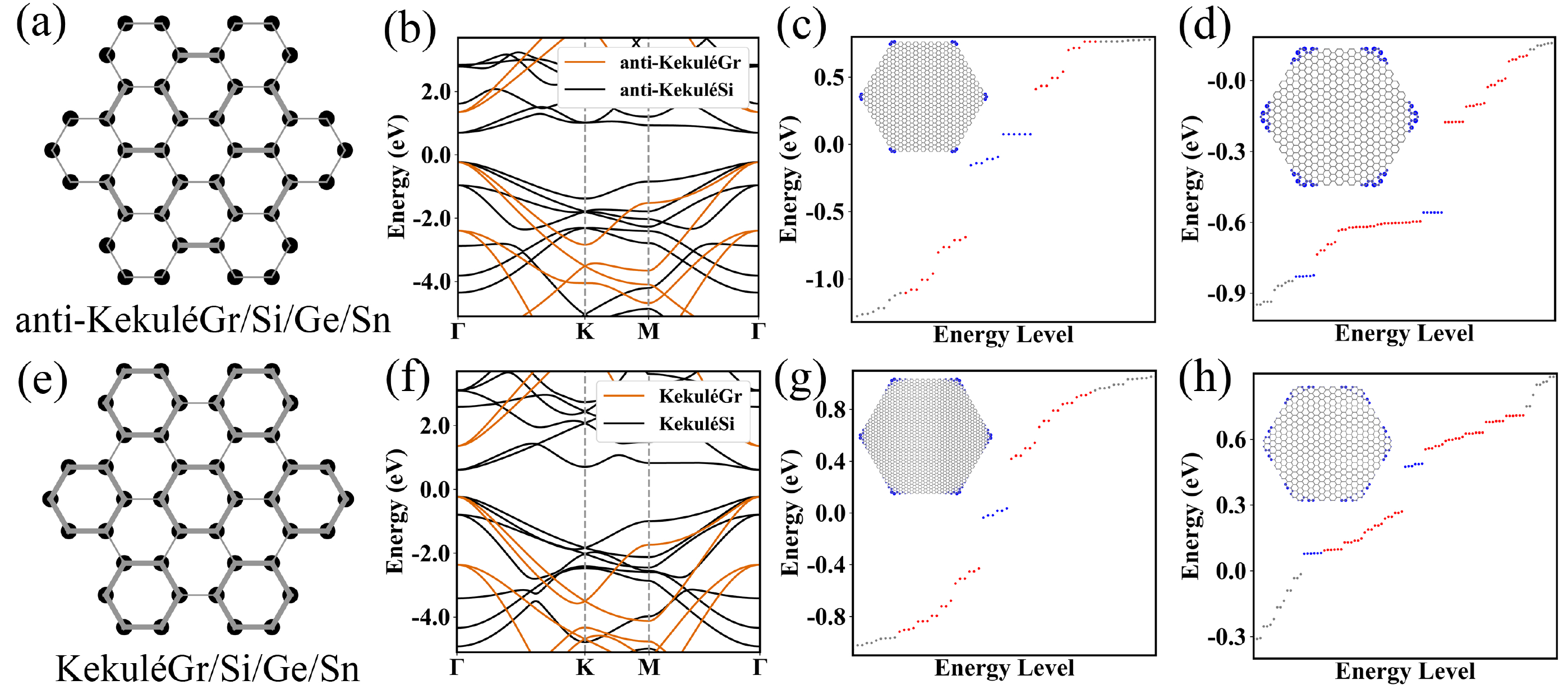}
		\end{center}
		\caption{Structures of anti-Kekul\'e (a) and Kekul\'e (e) distortion lattices of graphene and silicene/germanene/stanene. Black dots label the atoms, and thick and thin lines represent the strong and weak bonds. Bulk bands of anti-Kekul\'e graphene/silicene (orange/black) (b), and the Kekul\'e ones (orange/black) (f). Energy discrete spectra of the hexagonal nanoflakes of anti-Kekul\'e/Kekul\'e graphenes (c)/(g) and silicenes (d)/(h), calculated by DFT and Wanner function. The corresponding charge spatial distribution of blue states in (c), (d), (g) and (h) plotted in their insets are well localized at the corners, i.e., the corner states, and the blue, red and grey dots represent the corner, edge and bulk states, respectively.} \label{fig:4}
	\end{figure}
	
	Bulk band structures of blue phosphene (orange), blue arsenene (black) (b), and black phosphene (orange) and black arsenene (black) (f). Energy discrete spectra of the finite-size hexagonal/rectangle flakes of blue/black phosphene (c)/(g) and blue/black arsenene (d)/(h) calculated from DFT. The corresponding charge spatial distribution of blue states in (c), (d), (g) and (h) plotted in their insets are well localized at the corners, i.e., the corner states, and the blue, red and grey dots represent the corner, edge, and bulk states, respectively.
	
	\paragraph{The experimentally synthesized anti-Kekul\'e/Kekul\'e order graphenes and the counterparts of silicene/germanene/stanene.\textemdash{}} According to Mechanism II, the recent experimentally synthesized anti-Kekul\'e order graphene with a 0.38 eV gap at $\Gamma$ \cite{Kekule1} provides perfect material candidates. Here, we use first-principle calculation to demonstrate that anti-Kekul\'e/Kekul\'e order graphenes and the counterparts of silicene/germanene/stanene are 2D SOTI candidates (See details in Suppl. Mater. \cite{SuppMater}).  We take anti-Kekul\'e/Kekul\'e graphenes (silicenes) as examples, which have $\textit{P}$, $\textit{T}$ and six-fold (three-fold) rotation symmetry [Figs. \ref{fig:4}(a, e)]. The thin and thick lines represent the weak and strong bonds and black dots label the C/Si atoms. The second Stiefel-Whitney number $w_2$ of anti-Kekul\'e and Kekul\'e are 0 and 1 for the occupied bands, indicating that the anti-Kekul\'e (Kekul\'e) order graphene is trivial (nontrivial). However, for anti-Kekul\'e order graphene, the $p_z$ orbitals and $sp^2$ orbitals ($s$, $p_x$, $p_y$) are decoupled and either of the two parts is nontrivial and both parts have corner states \cite{SuppMater}. Therefore, the anti-Kekul\'e graphene is topological nontrivial. To further demonstrate the SOTI state in anti-Kekul\'e/Kekul\'e graphenes and silicenes, we calculate the discrete spectra [Figs. \ref{fig:4}(c, d, g, h)] of their finite-size hexagonal flakes from DFT calculation and Wannier function. The conner states are clearly shown in the insets.

\begin{figure}
		\begin{center}
			\includegraphics[width=1\linewidth]{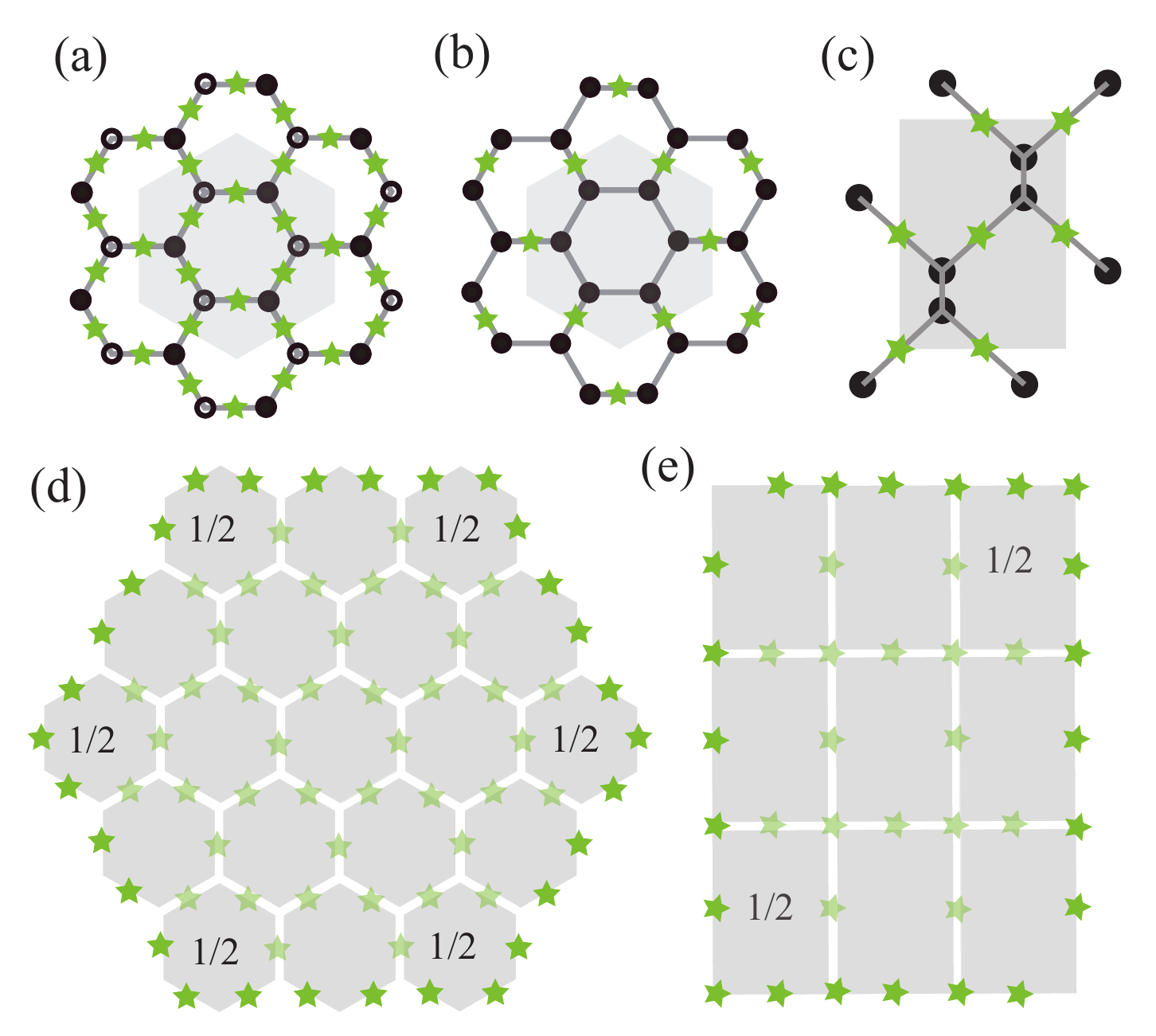}
		\end{center}
		\caption{Atom structure and Wannier centers at maximal Wyckoff position for the SOTI candidates materials based on (a) Mechanism I, (b) Mechanism II, and  for (c) black phosphorene and arsenene. (d) Fractional Corner charge for system (a) and (b). (e) Fractional corner charge for (c). Corner charges are in units of the electron charge $e$ and mod 1. The black solid and hollow circles represent the atoms and green stars denote the Wannier centers. } \label{fig:5}
	\end{figure}

	\paragraph{\textcolor{blue}{The corner states with fractional charge.}\textemdash{}}The topological origin of these corner states can arises from filling anomaly, which keeps track of mismatch between the number of electrons required to simultaneously satisfy charge netrality and the crystal symmetry \cite{kekulemodel3}. From the rotation symmetry eigenvalues and elementary band representations, these systems can be taken as obstructed atomic insulators \cite{ebrs}. The Wannier centers and the atoms do not overlap as shown in Figs. \ref{fig:5}(a-c). For the candidate materials based on Mechanism I, the SOTI state is derived from $sp^2$ orbitals. The Wannier centers of the $sp^2$ orbitals of every unit cell (shadow area) are located at the bond centers [Fig. \ref{fig:5}(a)]. In the finite-size hexagonal flake composed of the unit cells, as shown in Fig. \ref{fig:5}(d), each corner has a charge of $e$/2. The Wanner centers of Model II and anti-Kekul\'e graphene with $p_z$ orbitals are shown in Fig. \ref{fig:5}(b) and Fig. \ref{fig:5}(d), where each corner also has $e$/2. We can use the similar procedure to demonstrate the $e$/2 corner charge for the candidate materials with $p_z$ orbitals based on Mechanism II [Figs. \ref{fig:5}(b, d)], and black phosphorene and arsenene [Figs. \ref{fig:5}(c, e)]. In addition, the Wannier centers are also distributed at the flake's edges, which is the origin of the edge states in the two Models and all the candidate materials.
	
	The fractional corner charge is related with the $C_n$-symmetry representations of the occupied energy bands \cite{kekulemodel3}. The eigenvalues of $n$-fold rotation symmetry at high symmetry points $\mathbf{\Pi}^{(n)}$ is $\Pi_{p}^{(n)}=e^{2 \pi i(p-1) / n}, \quad$ for $p=1,2, \ldots n$. The integer topological invariants are defined by using the rotation eigenvalues at $\mathbf{\Pi}^{(n)}$ compared to a reference point $\boldsymbol{\Gamma}=(0,0)$, which is $\left[\Pi_{p}^{(n)}\right] \equiv \# \Pi_{p}^{(n)}-\# \Gamma_{p}^{(n)}$, where $\# \Pi_{p}^{(n)}$ is the number of energy bands below the Fermi level with eigenvalue $\Pi_{p}^{(n)}$. Our proposed three kinds of SOTI materials have $C_6$, $C_3+I$ and $C_2$ symmetry. Therefore, the fractional corner charge can be given by $Q_{\text {corner }}^{(6)}=\frac{e}{4}\left[M_{1}^{(2)}\right]+\frac{e}{6}\left[K_{1}^{(3)}\right] \bmod e$, $Q_{\text {corner }}^{(3+I)}=\frac{e}{4}\left[M_{1}^{(I)}\right]+\frac{e}{6}\left[K_{1}^{(3)}\right] \bmod e$, and $Q_{\text {corner }}^{(2)}=\frac{e}{4}\left(-\left[X_{1}^{(2)}\right]-\left[Y_{1}^{(2)}\right]+\left[M_{1}^{(2)}\right]\right) \bmod e$, where the superscript $n$ of $Q_{\text {corner }}^{(n)}$ labels the $C_n$ symmetry. 
	We calculated the fractional corner charge of all candidate materials and the corner charge of these materials are all 1/2 [Table. \ref{table: 1}], which agrees with the above filling anomaly analysis and the direct numerical calculation of corner charges \cite{SuppMater}.
	
	\begin{table}[htbp]
		\centering
		{\tabcolsep0.025in
			\caption{Corner charge, second Stiefel-Whitney number $w_2$, nest Wilson loop (det($W_2$)) and bulk gaps of graphane (CH), graphene fluoride (CF), graphene chloride (CCl), silicene hydride (SiH), silicene chloride (SiCl), germanene hydride(GeH), stanene hydride (SnH), blue phosphorene/arsenene (blue P/As), black phosphorene/arsenene (black P/As), anti-Kekul\'e/Kekul\'e graphenes, silicenes, germanenes and stanenes(anti-Kekul\'eGr/Kekul\'eGr, anti-Kekul\'eSi/Kekul\'eSi, anti-Kekul\'eGe/Kekul\'eGe, anti-Kekul\'eSn/Kekul\'eSn). The gap of anti-Kekul\'e distortion graphene is taken as the experimental value of 0.38 eV.} \label{table: 1}
			\begin{tabular}{ccccccc}
				\hline \hline
				& \multicolumn{2}{c}{} & $Q_c$    & $w_2$    & det($W_2$) & Gap (eV) \\ \hline
				\multirow{3}[1]{*}{$C_6$} & \multirow{2}[0]{*}{Anti-KekuleGr} & $sp^2$   & 1/2   & 1     & $\pi$ & \multirow{2}[0]{*}{0.38} \\
				&       & $p_z$    & 1/2   & 1     & $\pi$ &  \\
				& \multicolumn{2}{c}{KekuleGr} & 1/2   & 1     & $\pi$ &  \\
				\cmidrule{2-7}    \multirow{9}[2]{*}{$C_3+I$} & \multicolumn{2}{c}{CH} & 1/2   & 1     & $\pi$ & 3.5 \\
				& \multicolumn{2}{c}{CF} & 1/2   & 1     & $\pi$ & 3.1 \\
				& \multicolumn{2}{c}{CCl} & 1/2   & 1     & $\pi$ & 1.6 \\
				& \multicolumn{2}{c}{SiH} & 1/2   & 1     & $\pi$ & 2.2 \\
				& \multicolumn{2}{c}{SiCl} & 1/2   & 1     & $\pi$ & 1.3 \\
				& \multicolumn{2}{c}{GeH} & 1/2   & 1     & $\pi$ & 1.0 \\
				& \multicolumn{2}{c}{SnH} & 1/2   & 1     & $\pi$ & 0.5 \\
				& \multicolumn{2}{c}{Blue P} & 1/2   & 1     & $\pi$ & 1.9 \\
				& \multicolumn{2}{c}{Blue As} & 1/2   & 1     & $\pi$ & 1.6 \\
				& \multicolumn{2}{c}{Anti-Kekul\'eSi/Ge/Sn}\cite{SuppMater, note1} & 0   & 0     & 0 &   \\
				& \multicolumn{2}{c}{Kekul\'eSi/Ge/Sn}\cite{SuppMater, note2} & 1/2   & 1     & $\pi$ &   \\
				\cmidrule{2-7}    \multirow{2}[1]{*}{$C_2$} & \multicolumn{2}{c}{Black P} & 1/2   & 1     & $\pi$ & 0.9 \\
				& \multicolumn{2}{c}{Black As} & 1/2   & 1     & $\pi$ & 0.7 \\ \hline
			\end{tabular}%
			\label{tab:addlabel}%
		}
	\end{table}%

	We will develop a $k\cdot p$ effective model and edge theory to capture the low-energy physics and corner states. The symmetry group at $\Gamma$ of our systems are $D_{6h}$ or $D_{3d}$ besides $T$ symmetry. The generators for $D_{6h}$ or $D_{3d}$ can be chosen as $C_{3z}$, $P$ and $M_y$. We consider the four low-energy bands with two doublets $(E_{1g}, E_{2u})^T$ or $(E_{2g}, E_{1u})^T$ [Figs. \ref{fig:1} (c, e)], and the symmetry operators can be represented by $C_{3 z}=\tau_{0}e^{-i(\pi / 3) \sigma_{y}} ,\quad \mathcal{P} (C_{2 z})=\tau_{z}\sigma_{0} , \quad \mathcal{M}_{y}=\tau_{0} \sigma_{z}$ with Pauli matrices $\tau$ acting on the two doublets and $\sigma$ acting on the two degenerate states within each doublet. Under these symmetry operations, momentum and pseudospin are transformed as $C_{3}: k_{\pm} \rightarrow e^{\pm i 2 \pi / 3} k_{\pm}, \sigma_{\pm} \rightarrow e^{\pm i 2 \pi / 3} \sigma_{\pm}, \sigma_{y} \rightarrow \sigma_{y}; P (C_2): k_{\pm} \rightarrow -k_{\pm}, \sigma_{\pm} \rightarrow \sigma_{\pm}, \sigma_{y} \rightarrow \sigma_{y}; M_y: k_{\pm} \rightarrow k_{\mp}, \sigma_{\pm} \rightarrow\sigma_{\mp}, \sigma_{y} \rightarrow -\sigma_{y}$. $T = K$ in spinless systems, with $K$ as the complex conjugation. Constrained by these symmetries, the bulk model expanded to $k$-quadratic order is
	\begin{equation}
		\begin{aligned}
			\mathcal{H}_{eff}(\boldsymbol{k})=& E_{0}+\left(m_{0}-m_{1} k^{2}\right) \tau_{z}+v\left(k_{x} \sigma_{z}+k_{y} \sigma_{x}\right) \tau_{y} \\
			&+\left[\left(k_{x}^{2}-k_{y}^{2}\right) \sigma_{z}-2 k_{x} k_{y} \sigma_{x}\right]\left(\alpha_{1} \tau_{0}+\alpha_{2} \tau_{z}\right),
		\end{aligned}
	\end{equation}
	where $E_{0}=c_{0}+c_{1} k^{2}, k=|\boldsymbol{k}|$; $c_{i}, m_{i}, \alpha_{i}$ and $v$ are real parameters. The chiral symmetry is an approximate  symmetry and represented as $C = \tau_x$, when the first term and the $\alpha_1$ term can be ignored. 
	
	In the basis of the  two zero-energy edge modes, the edge Hamiltonian is given by \cite{SuppMater} $\mathcal{H}_{\text {edge }}(k)=v k s_{y}$, where $k$ is the wave vector along the edge, the Pauli matrices $s$ act on the space of $\left(\psi_{+}, \psi_{-}\right)^{T}$ and $v$ is Fermi velocity. In the basis, the symmetry operators can be represented by $\mathcal{M}_{y}=\mathcal{C}=s_{z}$. In the presence of $M_y$ and $C$ symmetry, the mass term is forced to be zero at edge Hamiltonian. However, for an edge that does not preserve $M_y$, the edge Hamiltonian will generally be gaped with the mass term $\Delta_{M}=m_{M} s_{x}$. Two edges related by $M_y$ must have opposite Dirac mass. Therefore, the protected 0D corner mode must exit at the intersection between the two edges, as shown in the finite-size flakes of the two mechanisms and candidate materials.

	\paragraph{\textcolor{blue}{Conclusion and discussion.}\textemdash{}}
	We put forward guidelines for designing the SOTI state in 2D light element materials with honeycomb lattice structure. Based on the guidelines, we have proposed a series of candidate materials with excellent performance, namely, hexagonal group IV materials hydrides and halides (graphane, halogenated graphenes, silicene hydride, silicene chloride, germanene hydride, stannene hydride), hexagonal group V materials (blue phosphorene, blue arsenene, black phosphorene, black arsenene), and hexagonal anti-Kekul\'e/Kekul\'e distortion group IV materials (anti-Kekul\'e/Kekul\'e distortion graphenes, silicenes, germanenes and stanenes). The second Stiefel-Whitney number $w_2$, characterizing the bulk topology of these candidate materials, is calculated by parity criterion and two Wilson loop methods. The hallmark corner states have been explicitly calculated from DFT and analyzed by effective model and edge theory. The fractional corner charge is obtained by the $C_n$-symmetry representations of the occupied bands and explicit numerical calculation.
	
	Experimentally, graphane and halogenated graphenes \cite{graphane, halide, halide2} were prepared shortly after the discovery of graphene. Anti-Kekul\'e or Kekul\'e ordered graphenes can also be induced experimentally through an external super-lattice potential \cite{kekuexp1, kekuexp2, kekuexp3} . Blue phosphorene/arsenene and black phosphorene/arsenene were synthesized with high quality \cite{blueph, blackP1, blackP2, blueP1, blackAs1, blackAs2, blueAs1}. In addition, silicene hydride has also been successfully prepared experimentally \cite{SiH}. These materials with giant gaps (up to 3.5 eV), which are larger than that of the 2D SOTI materials predicted previously \cite{graphdiyne1, graphdiyne2, graphyne1, graphyne2, note3}, will facilitate the experimental detection of the corner states as sharp peaks in the scanning tunneling spectroscopy measurement when the scanning tip moves close to the corners. As the majority of these materials are easy to grow and have large gaps, they are ideal candidates to explore the HOTI state and related novel properties.
	

	\begin{acknowledgments}
		S. Qian and C.-C. Liu are supported by the NSF of China (Grants No. 11922401, No. 11774028). Y.  Yao is supported by the National Key R\&D Program of China (Grant No. 2020YFA0308800), the NSF of China (Grants No. 11734003, No. 12061131002), the Strategic Priority Research Program of Chinese Academy of Sciences (Grant No. XDB30000000).
	\end{acknowledgments}

	\bibliographystyle{apsrev4-2}
	\bibliography{main}

\begin{thebibliography}{97}%
\makeatletter
\providecommand \@ifxundefined [1]{%
 \@ifx{#1\undefined}
}%
\providecommand \@ifnum [1]{%
 \ifnum #1\expandafter \@firstoftwo
 \else \expandafter \@secondoftwo
 \fi
}%
\providecommand \@ifx [1]{%
 \ifx #1\expandafter \@firstoftwo
 \else \expandafter \@secondoftwo
 \fi
}%
\providecommand \natexlab [1]{#1}%
\providecommand \enquote  [1]{``#1''}%
\providecommand \bibnamefont  [1]{#1}%
\providecommand \bibfnamefont [1]{#1}%
\providecommand \citenamefont [1]{#1}%
\providecommand \href@noop [0]{\@secondoftwo}%
\providecommand \href [0]{\begingroup \@sanitize@url \@href}%
\providecommand \@href[1]{\@@startlink{#1}\@@href}%
\providecommand \@@href[1]{\endgroup#1\@@endlink}%
\providecommand \@sanitize@url [0]{\catcode `\\12\catcode `\$12\catcode
  `\&12\catcode `\#12\catcode `\^12\catcode `\_12\catcode `\%12\relax}%
\providecommand \@@startlink[1]{}%
\providecommand \@@endlink[0]{}%
\providecommand \url  [0]{\begingroup\@sanitize@url \@url }%
\providecommand \@url [1]{\endgroup\@href {#1}{\urlprefix }}%
\providecommand \urlprefix  [0]{URL }%
\providecommand \Eprint [0]{\href }%
\providecommand \doibase [0]{https://doi.org/}%
\providecommand \selectlanguage [0]{\@gobble}%
\providecommand \bibinfo  [0]{\@secondoftwo}%
\providecommand \bibfield  [0]{\@secondoftwo}%
\providecommand \translation [1]{[#1]}%
\providecommand \BibitemOpen [0]{}%
\providecommand \bibitemStop [0]{}%
\providecommand \bibitemNoStop [0]{.\EOS\space}%
\providecommand \EOS [0]{\spacefactor3000\relax}%
\providecommand \BibitemShut  [1]{\csname bibitem#1\endcsname}%
\let\auto@bib@innerbib\@empty
\bibitem [{\citenamefont {Hasan}\ and\ \citenamefont {Kane}(2010)}]{TIRev1}%
  \BibitemOpen
  \bibfield  {author} {\bibinfo {author} {\bibfnamefont {M.~Z.}\ \bibnamefont
  {Hasan}}\ and\ \bibinfo {author} {\bibfnamefont {C.~L.}\ \bibnamefont
  {Kane}},\ }\href {https://doi.org/10.1103/revmodphys.82.3045} {\bibfield
  {journal} {\bibinfo  {journal} {Rev. Mod. Phys.}\ }\textbf {\bibinfo {volume}
  {82}},\ \bibinfo {pages} {3045} (\bibinfo {year} {2010})}\BibitemShut
  {NoStop}%
\bibitem [{\citenamefont {Qi}\ and\ \citenamefont {Zhang}(2011)}]{TIRev2}%
  \BibitemOpen
  \bibfield  {author} {\bibinfo {author} {\bibfnamefont {X.-L.}\ \bibnamefont
  {Qi}}\ and\ \bibinfo {author} {\bibfnamefont {S.-C.}\ \bibnamefont {Zhang}},\
  }\href {https://doi.org/10.1103/revmodphys.83.1057} {\bibfield  {journal}
  {\bibinfo  {journal} {Rev. Mod. Phys.}\ }\textbf {\bibinfo {volume} {83}},\
  \bibinfo {pages} {1057} (\bibinfo {year} {2011})}\BibitemShut {NoStop}%
\bibitem [{\citenamefont {Vergniory}\ \emph {et~al.}(2019)\citenamefont
  {Vergniory}, \citenamefont {Elcoro}, \citenamefont {Felser}, \citenamefont
  {Regnault}, \citenamefont {Bernevig},\ and\ \citenamefont {Wang}}]{fenlei1}%
  \BibitemOpen
  \bibfield  {author} {\bibinfo {author} {\bibfnamefont {M.}~\bibnamefont
  {Vergniory}}, \bibinfo {author} {\bibfnamefont {L.}~\bibnamefont {Elcoro}},
  \bibinfo {author} {\bibfnamefont {C.}~\bibnamefont {Felser}}, \bibinfo
  {author} {\bibfnamefont {N.}~\bibnamefont {Regnault}}, \bibinfo {author}
  {\bibfnamefont {B.~A.}\ \bibnamefont {Bernevig}},\ and\ \bibinfo {author}
  {\bibfnamefont {Z.}~\bibnamefont {Wang}},\ }\href
  {https://doi.org/https://doi.org/10.1038/s41586-019-0954-4} {\bibfield
  {journal} {\bibinfo  {journal} {Nature}\ }\textbf {\bibinfo {volume} {566}},\
  \bibinfo {pages} {480} (\bibinfo {year} {2019})}\BibitemShut {NoStop}%
\bibitem [{\citenamefont {Tang}\ \emph {et~al.}(2019)\citenamefont {Tang},
  \citenamefont {Po}, \citenamefont {Vishwanath},\ and\ \citenamefont
  {Wan}}]{fenlei2}%
  \BibitemOpen
  \bibfield  {author} {\bibinfo {author} {\bibfnamefont {F.}~\bibnamefont
  {Tang}}, \bibinfo {author} {\bibfnamefont {H.~C.}\ \bibnamefont {Po}},
  \bibinfo {author} {\bibfnamefont {A.}~\bibnamefont {Vishwanath}},\ and\
  \bibinfo {author} {\bibfnamefont {X.}~\bibnamefont {Wan}},\ }\href
  {https://doi.org/10.1038/s41586-019-0937-5} {\bibfield  {journal} {\bibinfo
  {journal} {Nature}\ }\textbf {\bibinfo {volume} {566}},\ \bibinfo {pages}
  {486} (\bibinfo {year} {2019})}\BibitemShut {NoStop}%
\bibitem [{\citenamefont {Zhang}\ \emph
  {et~al.}(2019{\natexlab{a}})\citenamefont {Zhang}, \citenamefont {Jiang},
  \citenamefont {Song}, \citenamefont {Huang}, \citenamefont {He},
  \citenamefont {Fang}, \citenamefont {Weng},\ and\ \citenamefont
  {Fang}}]{fenlei3}%
  \BibitemOpen
  \bibfield  {author} {\bibinfo {author} {\bibfnamefont {T.}~\bibnamefont
  {Zhang}}, \bibinfo {author} {\bibfnamefont {Y.}~\bibnamefont {Jiang}},
  \bibinfo {author} {\bibfnamefont {Z.}~\bibnamefont {Song}}, \bibinfo {author}
  {\bibfnamefont {H.}~\bibnamefont {Huang}}, \bibinfo {author} {\bibfnamefont
  {Y.}~\bibnamefont {He}}, \bibinfo {author} {\bibfnamefont {Z.}~\bibnamefont
  {Fang}}, \bibinfo {author} {\bibfnamefont {H.}~\bibnamefont {Weng}},\ and\
  \bibinfo {author} {\bibfnamefont {C.}~\bibnamefont {Fang}},\ }\href
  {https://doi.org/10.1038/s41586-019-0944-6} {\bibfield  {journal} {\bibinfo
  {journal} {Nature}\ }\textbf {\bibinfo {volume} {566}},\ \bibinfo {pages}
  {475} (\bibinfo {year} {2019}{\natexlab{a}})}\BibitemShut {NoStop}%
\bibitem [{\citenamefont {Benalcazar}\ \emph
  {et~al.}(2017{\natexlab{a}})\citenamefont {Benalcazar}, \citenamefont
  {Bernevig},\ and\ \citenamefont {Hughes}}]{HOTI2}%
  \BibitemOpen
  \bibfield  {author} {\bibinfo {author} {\bibfnamefont {W.~A.}\ \bibnamefont
  {Benalcazar}}, \bibinfo {author} {\bibfnamefont {B.~A.}\ \bibnamefont
  {Bernevig}},\ and\ \bibinfo {author} {\bibfnamefont {T.~L.}\ \bibnamefont
  {Hughes}},\ }\href {https://doi.org/10.1126/science.aah6442} {\bibfield
  {journal} {\bibinfo  {journal} {Science}\ }\textbf {\bibinfo {volume}
  {357}},\ \bibinfo {pages} {61} (\bibinfo {year}
  {2017}{\natexlab{a}})}\BibitemShut {NoStop}%
\bibitem [{\citenamefont {Langbehn}\ \emph {et~al.}(2017)\citenamefont
  {Langbehn}, \citenamefont {Peng}, \citenamefont {Trifunovic}, \citenamefont
  {von Oppen},\ and\ \citenamefont {Brouwer}}]{HOTI3}%
  \BibitemOpen
  \bibfield  {author} {\bibinfo {author} {\bibfnamefont {J.}~\bibnamefont
  {Langbehn}}, \bibinfo {author} {\bibfnamefont {Y.}~\bibnamefont {Peng}},
  \bibinfo {author} {\bibfnamefont {L.}~\bibnamefont {Trifunovic}}, \bibinfo
  {author} {\bibfnamefont {F.}~\bibnamefont {von Oppen}},\ and\ \bibinfo
  {author} {\bibfnamefont {P.~W.}\ \bibnamefont {Brouwer}},\ }\href
  {https://doi.org/10.1103/physrevlett.119.246401} {\bibfield  {journal}
  {\bibinfo  {journal} {Rev. Rev. Lett.}\ }\textbf {\bibinfo {volume} {119}},\
  \bibinfo {pages} {246401} (\bibinfo {year} {2017})}\BibitemShut {NoStop}%
\bibitem [{\citenamefont {Song}\ \emph {et~al.}(2017)\citenamefont {Song},
  \citenamefont {Fang},\ and\ \citenamefont {Fang}}]{HOTI4}%
  \BibitemOpen
  \bibfield  {author} {\bibinfo {author} {\bibfnamefont {Z.}~\bibnamefont
  {Song}}, \bibinfo {author} {\bibfnamefont {Z.}~\bibnamefont {Fang}},\ and\
  \bibinfo {author} {\bibfnamefont {C.}~\bibnamefont {Fang}},\ }\href
  {https://doi.org/10.1103/physrevlett.119.246402} {\bibfield  {journal}
  {\bibinfo  {journal} {Rev. Rev. Lett.}\ }\textbf {\bibinfo {volume} {119}},\
  \bibinfo {pages} {246402} (\bibinfo {year} {2017})}\BibitemShut {NoStop}%
\bibitem [{\citenamefont {Zhang}\ \emph {et~al.}(2013)\citenamefont {Zhang},
  \citenamefont {Kane},\ and\ \citenamefont {Mele}}]{HOTI1}%
  \BibitemOpen
  \bibfield  {author} {\bibinfo {author} {\bibfnamefont {F.}~\bibnamefont
  {Zhang}}, \bibinfo {author} {\bibfnamefont {C.~L.}\ \bibnamefont {Kane}},\
  and\ \bibinfo {author} {\bibfnamefont {E.~J.}\ \bibnamefont {Mele}},\ }\href
  {https://doi.org/10.1103/physrevlett.110.046404} {\bibfield  {journal}
  {\bibinfo  {journal} {Rev. Rev. Lett.}\ }\textbf {\bibinfo {volume} {110}},\
  \bibinfo {pages} {046404} (\bibinfo {year} {2013})}\BibitemShut {NoStop}%
\bibitem [{\citenamefont {Benalcazar}\ \emph
  {et~al.}(2017{\natexlab{b}})\citenamefont {Benalcazar}, \citenamefont
  {Bernevig},\ and\ \citenamefont {Hughes}}]{HOTI5}%
  \BibitemOpen
  \bibfield  {author} {\bibinfo {author} {\bibfnamefont {W.~A.}\ \bibnamefont
  {Benalcazar}}, \bibinfo {author} {\bibfnamefont {B.~A.}\ \bibnamefont
  {Bernevig}},\ and\ \bibinfo {author} {\bibfnamefont {T.~L.}\ \bibnamefont
  {Hughes}},\ }\href {https://doi.org/10.1103/physrevb.96.245115} {\bibfield
  {journal} {\bibinfo  {journal} {Rev. Rev. B}\ }\textbf {\bibinfo {volume}
  {96}},\ \bibinfo {pages} {245115} (\bibinfo {year}
  {2017}{\natexlab{b}})}\BibitemShut {NoStop}%
\bibitem [{\citenamefont {Xie}\ \emph {et~al.}(2021)\citenamefont {Xie},
  \citenamefont {Wang}, \citenamefont {Zhang}, \citenamefont {Zhan},
  \citenamefont {Jiang}, \citenamefont {Lu},\ and\ \citenamefont
  {Chen}}]{HOTIREV}%
  \BibitemOpen
  \bibfield  {author} {\bibinfo {author} {\bibfnamefont {B.}~\bibnamefont
  {Xie}}, \bibinfo {author} {\bibfnamefont {H.-X.}\ \bibnamefont {Wang}},
  \bibinfo {author} {\bibfnamefont {X.}~\bibnamefont {Zhang}}, \bibinfo
  {author} {\bibfnamefont {P.}~\bibnamefont {Zhan}}, \bibinfo {author}
  {\bibfnamefont {J.-H.}\ \bibnamefont {Jiang}}, \bibinfo {author}
  {\bibfnamefont {M.}~\bibnamefont {Lu}},\ and\ \bibinfo {author}
  {\bibfnamefont {Y.}~\bibnamefont {Chen}},\ }\href
  {https://doi.org/https://doi.org/10.1038/s42254-021-00323-4} {\bibfield
  {journal} {\bibinfo  {journal} {Nat. Rev. Phys.}\ ,\ \bibinfo {pages} {1}}
  (\bibinfo {year} {2021})}\BibitemShut {NoStop}%
\bibitem [{\citenamefont {Ezawa}(2018{\natexlab{a}})}]{BeathKogome}%
  \BibitemOpen
  \bibfield  {author} {\bibinfo {author} {\bibfnamefont {M.}~\bibnamefont
  {Ezawa}},\ }\href {https://doi.org/10.1103/PhysRevLett.120.026801} {\bibfield
   {journal} {\bibinfo  {journal} {Phys. Rev. Lett.}\ }\textbf {\bibinfo
  {volume} {120}},\ \bibinfo {pages} {026801} (\bibinfo {year}
  {2018}{\natexlab{a}})}\BibitemShut {NoStop}%
\bibitem [{\citenamefont {Franca}\ \emph {et~al.}(2018)\citenamefont {Franca},
  \citenamefont {van~den Brink},\ and\ \citenamefont {Fulga}}]{HOTIA}%
  \BibitemOpen
  \bibfield  {author} {\bibinfo {author} {\bibfnamefont {S.}~\bibnamefont
  {Franca}}, \bibinfo {author} {\bibfnamefont {J.}~\bibnamefont {van~den
  Brink}},\ and\ \bibinfo {author} {\bibfnamefont {I.~C.}\ \bibnamefont
  {Fulga}},\ }\href {https://doi.org/10.1103/PhysRevB.98.201114} {\bibfield
  {journal} {\bibinfo  {journal} {Phys. Rev. B}\ }\textbf {\bibinfo {volume}
  {98}},\ \bibinfo {pages} {201114} (\bibinfo {year} {2018})}\BibitemShut
  {NoStop}%
\bibitem [{\citenamefont {C\ifmmode \u{a}\else \u{a}\fi{}lug\ifmmode~\u{a}\else
  \u{a}\fi{}ru}\ \emph {et~al.}(2019)\citenamefont {C\ifmmode \u{a}\else
  \u{a}\fi{}lug\ifmmode~\u{a}\else \u{a}\fi{}ru}, \citenamefont {Juri\ifmmode
  \check{c}\else \v{c}\fi{}i\ifmmode~\acute{c}\else \'{c}\fi{}},\ and\
  \citenamefont {Roy}}]{HOTIGeneral}%
  \BibitemOpen
  \bibfield  {author} {\bibinfo {author} {\bibfnamefont {D.}~\bibnamefont
  {C\ifmmode \u{a}\else \u{a}\fi{}lug\ifmmode~\u{a}\else \u{a}\fi{}ru}},
  \bibinfo {author} {\bibfnamefont {V.}~\bibnamefont {Juri\ifmmode
  \check{c}\else \v{c}\fi{}i\ifmmode~\acute{c}\else \'{c}\fi{}}},\ and\
  \bibinfo {author} {\bibfnamefont {B.}~\bibnamefont {Roy}},\ }\href
  {https://doi.org/10.1103/PhysRevB.99.041301} {\bibfield  {journal} {\bibinfo
  {journal} {Phys. Rev. B}\ }\textbf {\bibinfo {volume} {99}},\ \bibinfo
  {pages} {041301} (\bibinfo {year} {2019})}\BibitemShut {NoStop}%
\bibitem [{\citenamefont {van Miert}\ and\ \citenamefont
  {Ortix}(2018)}]{HOTIPC}%
  \BibitemOpen
  \bibfield  {author} {\bibinfo {author} {\bibfnamefont {G.}~\bibnamefont {van
  Miert}}\ and\ \bibinfo {author} {\bibfnamefont {C.}~\bibnamefont {Ortix}},\
  }\href {https://doi.org/10.1103/PhysRevB.98.081110} {\bibfield  {journal}
  {\bibinfo  {journal} {Phys. Rev. B}\ }\textbf {\bibinfo {volume} {98}},\
  \bibinfo {pages} {081110} (\bibinfo {year} {2018})}\BibitemShut {NoStop}%
\bibitem [{\citenamefont {Liu}\ \emph {et~al.}(2019{\natexlab{a}})\citenamefont
  {Liu}, \citenamefont {Deng},\ and\ \citenamefont {Wakabayashi}}]{Fengliu}%
  \BibitemOpen
  \bibfield  {author} {\bibinfo {author} {\bibfnamefont {F.}~\bibnamefont
  {Liu}}, \bibinfo {author} {\bibfnamefont {H.-Y.}\ \bibnamefont {Deng}},\ and\
  \bibinfo {author} {\bibfnamefont {K.}~\bibnamefont {Wakabayashi}},\ }\href
  {https://doi.org/10.1103/PhysRevLett.122.086804} {\bibfield  {journal}
  {\bibinfo  {journal} {Phys. Rev. Lett.}\ }\textbf {\bibinfo {volume} {122}},\
  \bibinfo {pages} {086804} (\bibinfo {year} {2019}{\natexlab{a}})}\BibitemShut
  {NoStop}%
\bibitem [{\citenamefont {Liu}\ and\ \citenamefont
  {Wakabayashi}(2021)}]{Fengliugraphene}%
  \BibitemOpen
  \bibfield  {author} {\bibinfo {author} {\bibfnamefont {F.}~\bibnamefont
  {Liu}}\ and\ \bibinfo {author} {\bibfnamefont {K.}~\bibnamefont
  {Wakabayashi}},\ }\href {https://doi.org/10.1103/PhysRevResearch.3.023121}
  {\bibfield  {journal} {\bibinfo  {journal} {Phys. Rev. Research}\ }\textbf
  {\bibinfo {volume} {3}},\ \bibinfo {pages} {023121} (\bibinfo {year}
  {2021})}\BibitemShut {NoStop}%
\bibitem [{\citenamefont {Schindler}\ \emph
  {et~al.}(2018{\natexlab{a}})\citenamefont {Schindler}, \citenamefont {Wang},
  \citenamefont {Vergniory}, \citenamefont {Cook}, \citenamefont {Murani},
  \citenamefont {Sengupta}, \citenamefont {Kasumov}, \citenamefont {Deblock},
  \citenamefont {Jeon}, \citenamefont {Drozdov} \emph {et~al.}}]{HOTI_Bi}%
  \BibitemOpen
  \bibfield  {author} {\bibinfo {author} {\bibfnamefont {F.}~\bibnamefont
  {Schindler}}, \bibinfo {author} {\bibfnamefont {Z.}~\bibnamefont {Wang}},
  \bibinfo {author} {\bibfnamefont {M.~G.}\ \bibnamefont {Vergniory}}, \bibinfo
  {author} {\bibfnamefont {A.~M.}\ \bibnamefont {Cook}}, \bibinfo {author}
  {\bibfnamefont {A.}~\bibnamefont {Murani}}, \bibinfo {author} {\bibfnamefont
  {S.}~\bibnamefont {Sengupta}}, \bibinfo {author} {\bibfnamefont {A.~Y.}\
  \bibnamefont {Kasumov}}, \bibinfo {author} {\bibfnamefont {R.}~\bibnamefont
  {Deblock}}, \bibinfo {author} {\bibfnamefont {S.}~\bibnamefont {Jeon}},
  \bibinfo {author} {\bibfnamefont {I.}~\bibnamefont {Drozdov}}, \emph
  {et~al.},\ }\href {https://doi.org/10.1038/s41567-018-0224-7} {\bibfield
  {journal} {\bibinfo  {journal} {Nat. Phys.}\ }\textbf {\bibinfo {volume}
  {14}},\ \bibinfo {pages} {918} (\bibinfo {year}
  {2018}{\natexlab{a}})}\BibitemShut {NoStop}%
\bibitem [{\citenamefont {Serra-Garcia}\ \emph {et~al.}(2018)\citenamefont
  {Serra-Garcia}, \citenamefont {Peri}, \citenamefont {S{\"u}sstrunk},
  \citenamefont {Bilal}, \citenamefont {Larsen}, \citenamefont {Villanueva},\
  and\ \citenamefont {Huber}}]{mechanic}%
  \BibitemOpen
  \bibfield  {author} {\bibinfo {author} {\bibfnamefont {M.}~\bibnamefont
  {Serra-Garcia}}, \bibinfo {author} {\bibfnamefont {V.}~\bibnamefont {Peri}},
  \bibinfo {author} {\bibfnamefont {R.}~\bibnamefont {S{\"u}sstrunk}}, \bibinfo
  {author} {\bibfnamefont {O.~R.}\ \bibnamefont {Bilal}}, \bibinfo {author}
  {\bibfnamefont {T.}~\bibnamefont {Larsen}}, \bibinfo {author} {\bibfnamefont
  {L.~G.}\ \bibnamefont {Villanueva}},\ and\ \bibinfo {author} {\bibfnamefont
  {S.~D.}\ \bibnamefont {Huber}},\ }\href {https://doi.org/10.1038/nature25156}
  {\bibfield  {journal} {\bibinfo  {journal} {Nature}\ }\textbf {\bibinfo
  {volume} {555}},\ \bibinfo {pages} {342} (\bibinfo {year}
  {2018})}\BibitemShut {NoStop}%
\bibitem [{\citenamefont {Xue}\ \emph {et~al.}(2019)\citenamefont {Xue},
  \citenamefont {Yang}, \citenamefont {Gao}, \citenamefont {Chong},\ and\
  \citenamefont {Zhang}}]{acoustic1}%
  \BibitemOpen
  \bibfield  {author} {\bibinfo {author} {\bibfnamefont {H.}~\bibnamefont
  {Xue}}, \bibinfo {author} {\bibfnamefont {Y.}~\bibnamefont {Yang}}, \bibinfo
  {author} {\bibfnamefont {F.}~\bibnamefont {Gao}}, \bibinfo {author}
  {\bibfnamefont {Y.}~\bibnamefont {Chong}},\ and\ \bibinfo {author}
  {\bibfnamefont {B.}~\bibnamefont {Zhang}},\ }\href
  {https://doi.org/https://doi.org/10.1038/s41563-018-0251-x} {\bibfield
  {journal} {\bibinfo  {journal} {Nat. Mater.}\ }\textbf {\bibinfo {volume}
  {18}},\ \bibinfo {pages} {108} (\bibinfo {year} {2019})}\BibitemShut
  {NoStop}%
\bibitem [{\citenamefont {Ni}\ \emph {et~al.}(2019)\citenamefont {Ni},
  \citenamefont {Weiner}, \citenamefont {Alu},\ and\ \citenamefont
  {Khanikaev}}]{acoustic2}%
  \BibitemOpen
  \bibfield  {author} {\bibinfo {author} {\bibfnamefont {X.}~\bibnamefont
  {Ni}}, \bibinfo {author} {\bibfnamefont {M.}~\bibnamefont {Weiner}}, \bibinfo
  {author} {\bibfnamefont {A.}~\bibnamefont {Alu}},\ and\ \bibinfo {author}
  {\bibfnamefont {A.~B.}\ \bibnamefont {Khanikaev}},\ }\href
  {https://doi.org/https://doi.org/10.1038/s41563-018-0252-9} {\bibfield
  {journal} {\bibinfo  {journal} {Nat. Mater.}\ }\textbf {\bibinfo {volume}
  {18}},\ \bibinfo {pages} {113} (\bibinfo {year} {2019})}\BibitemShut
  {NoStop}%
\bibitem [{\citenamefont {Mittal}\ \emph {et~al.}(2019)\citenamefont {Mittal},
  \citenamefont {Orre}, \citenamefont {Zhu}, \citenamefont {Gorlach},
  \citenamefont {Poddubny},\ and\ \citenamefont {Hafezi}}]{photonic1}%
  \BibitemOpen
  \bibfield  {author} {\bibinfo {author} {\bibfnamefont {S.}~\bibnamefont
  {Mittal}}, \bibinfo {author} {\bibfnamefont {V.~V.}\ \bibnamefont {Orre}},
  \bibinfo {author} {\bibfnamefont {G.}~\bibnamefont {Zhu}}, \bibinfo {author}
  {\bibfnamefont {M.~A.}\ \bibnamefont {Gorlach}}, \bibinfo {author}
  {\bibfnamefont {A.}~\bibnamefont {Poddubny}},\ and\ \bibinfo {author}
  {\bibfnamefont {M.}~\bibnamefont {Hafezi}},\ }\href
  {https://doi.org/10.1038/s41566-019-0452-0} {\bibfield  {journal} {\bibinfo
  {journal} {Nat. Photon.}\ }\textbf {\bibinfo {volume} {13}},\ \bibinfo
  {pages} {692} (\bibinfo {year} {2019})}\BibitemShut {NoStop}%
\bibitem [{\citenamefont {Zhang}\ \emph
  {et~al.}(2020{\natexlab{a}})\citenamefont {Zhang}, \citenamefont {Xie},
  \citenamefont {Hao}, \citenamefont {Dang}, \citenamefont {Xiao},
  \citenamefont {Shi}, \citenamefont {Ni}, \citenamefont {Niu}, \citenamefont
  {Wang}, \citenamefont {Jin} \emph {et~al.}}]{photonic2}%
  \BibitemOpen
  \bibfield  {author} {\bibinfo {author} {\bibfnamefont {W.}~\bibnamefont
  {Zhang}}, \bibinfo {author} {\bibfnamefont {X.}~\bibnamefont {Xie}}, \bibinfo
  {author} {\bibfnamefont {H.}~\bibnamefont {Hao}}, \bibinfo {author}
  {\bibfnamefont {J.}~\bibnamefont {Dang}}, \bibinfo {author} {\bibfnamefont
  {S.}~\bibnamefont {Xiao}}, \bibinfo {author} {\bibfnamefont {S.}~\bibnamefont
  {Shi}}, \bibinfo {author} {\bibfnamefont {H.}~\bibnamefont {Ni}}, \bibinfo
  {author} {\bibfnamefont {Z.}~\bibnamefont {Niu}}, \bibinfo {author}
  {\bibfnamefont {C.}~\bibnamefont {Wang}}, \bibinfo {author} {\bibfnamefont
  {K.}~\bibnamefont {Jin}}, \emph {et~al.},\ }\href
  {https://doi.org/10.1038/s41377-020-00352-1} {\bibfield  {journal} {\bibinfo
  {journal} {Light Sci. Appl.}\ }\textbf {\bibinfo {volume} {9}},\ \bibinfo
  {pages} {1} (\bibinfo {year} {2020}{\natexlab{a}})}\BibitemShut {NoStop}%
\bibitem [{\citenamefont {Noh}\ \emph {et~al.}(2018)\citenamefont {Noh},
  \citenamefont {Benalcazar}, \citenamefont {Huang}, \citenamefont {Collins},
  \citenamefont {Chen}, \citenamefont {Hughes},\ and\ \citenamefont
  {Rechtsman}}]{photonic3}%
  \BibitemOpen
  \bibfield  {author} {\bibinfo {author} {\bibfnamefont {J.}~\bibnamefont
  {Noh}}, \bibinfo {author} {\bibfnamefont {W.~A.}\ \bibnamefont {Benalcazar}},
  \bibinfo {author} {\bibfnamefont {S.}~\bibnamefont {Huang}}, \bibinfo
  {author} {\bibfnamefont {M.~J.}\ \bibnamefont {Collins}}, \bibinfo {author}
  {\bibfnamefont {K.~P.}\ \bibnamefont {Chen}}, \bibinfo {author}
  {\bibfnamefont {T.~L.}\ \bibnamefont {Hughes}},\ and\ \bibinfo {author}
  {\bibfnamefont {M.~C.}\ \bibnamefont {Rechtsman}},\ }\href
  {https://doi.org/10.1038/s41566-018-0179-3} {\bibfield  {journal} {\bibinfo
  {journal} {Nat. Photon.}\ }\textbf {\bibinfo {volume} {12}},\ \bibinfo
  {pages} {408} (\bibinfo {year} {2018})}\BibitemShut {NoStop}%
\bibitem [{\citenamefont {Imhof}\ \emph {et~al.}(2018)\citenamefont {Imhof},
  \citenamefont {Berger}, \citenamefont {Bayer}, \citenamefont {Brehm},
  \citenamefont {Molenkamp}, \citenamefont {Kiessling}, \citenamefont
  {Schindler}, \citenamefont {Lee}, \citenamefont {Greiter}, \citenamefont
  {Neupert} \emph {et~al.}}]{elect1}%
  \BibitemOpen
  \bibfield  {author} {\bibinfo {author} {\bibfnamefont {S.}~\bibnamefont
  {Imhof}}, \bibinfo {author} {\bibfnamefont {C.}~\bibnamefont {Berger}},
  \bibinfo {author} {\bibfnamefont {F.}~\bibnamefont {Bayer}}, \bibinfo
  {author} {\bibfnamefont {J.}~\bibnamefont {Brehm}}, \bibinfo {author}
  {\bibfnamefont {L.~W.}\ \bibnamefont {Molenkamp}}, \bibinfo {author}
  {\bibfnamefont {T.}~\bibnamefont {Kiessling}}, \bibinfo {author}
  {\bibfnamefont {F.}~\bibnamefont {Schindler}}, \bibinfo {author}
  {\bibfnamefont {C.~H.}\ \bibnamefont {Lee}}, \bibinfo {author} {\bibfnamefont
  {M.}~\bibnamefont {Greiter}}, \bibinfo {author} {\bibfnamefont
  {T.}~\bibnamefont {Neupert}}, \emph {et~al.},\ }\href
  {https://doi.org/10.1038/s41567-018-0246-1} {\bibfield  {journal} {\bibinfo
  {journal} {Nat. Phys.}\ }\textbf {\bibinfo {volume} {14}},\ \bibinfo {pages}
  {925} (\bibinfo {year} {2018})}\BibitemShut {NoStop}%
\bibitem [{\citenamefont {Zangeneh-Nejad}\ and\ \citenamefont
  {Fleury}(2019)}]{elect2}%
  \BibitemOpen
  \bibfield  {author} {\bibinfo {author} {\bibfnamefont {F.}~\bibnamefont
  {Zangeneh-Nejad}}\ and\ \bibinfo {author} {\bibfnamefont {R.}~\bibnamefont
  {Fleury}},\ }\href {https://doi.org/10.1103/physrevlett.123.053902}
  {\bibfield  {journal} {\bibinfo  {journal} {Rev. Rev. Lett.}\ }\textbf
  {\bibinfo {volume} {123}},\ \bibinfo {pages} {053902} (\bibinfo {year}
  {2019})}\BibitemShut {NoStop}%
\bibitem [{\citenamefont {Zhang}\ \emph {et~al.}(2021)\citenamefont {Zhang},
  \citenamefont {Zou}, \citenamefont {Pei}, \citenamefont {He}, \citenamefont
  {Bao}, \citenamefont {Sun},\ and\ \citenamefont {Zhang}}]{elect3}%
  \BibitemOpen
  \bibfield  {author} {\bibinfo {author} {\bibfnamefont {W.}~\bibnamefont
  {Zhang}}, \bibinfo {author} {\bibfnamefont {D.}~\bibnamefont {Zou}}, \bibinfo
  {author} {\bibfnamefont {Q.}~\bibnamefont {Pei}}, \bibinfo {author}
  {\bibfnamefont {W.}~\bibnamefont {He}}, \bibinfo {author} {\bibfnamefont
  {J.}~\bibnamefont {Bao}}, \bibinfo {author} {\bibfnamefont {H.}~\bibnamefont
  {Sun}},\ and\ \bibinfo {author} {\bibfnamefont {X.}~\bibnamefont {Zhang}},\
  }\href {https://doi.org/10.1103/physrevlett.126.146802} {\bibfield  {journal}
  {\bibinfo  {journal} {Rev. Rev. Lett.}\ }\textbf {\bibinfo {volume} {126}},\
  \bibinfo {pages} {146802} (\bibinfo {year} {2021})}\BibitemShut {NoStop}%
\bibitem [{\citenamefont {Peterson}\ \emph {et~al.}(2018)\citenamefont
  {Peterson}, \citenamefont {Benalcazar}, \citenamefont {Hughes},\ and\
  \citenamefont {Bahl}}]{micro}%
  \BibitemOpen
  \bibfield  {author} {\bibinfo {author} {\bibfnamefont {C.~W.}\ \bibnamefont
  {Peterson}}, \bibinfo {author} {\bibfnamefont {W.~A.}\ \bibnamefont
  {Benalcazar}}, \bibinfo {author} {\bibfnamefont {T.~L.}\ \bibnamefont
  {Hughes}},\ and\ \bibinfo {author} {\bibfnamefont {G.}~\bibnamefont {Bahl}},\
  }\href {https://doi.org/10.1038/nature25777} {\bibfield  {journal} {\bibinfo
  {journal} {Nature}\ }\textbf {\bibinfo {volume} {555}},\ \bibinfo {pages}
  {346} (\bibinfo {year} {2018})}\BibitemShut {NoStop}%
\bibitem [{\citenamefont {Zhang}\ \emph
  {et~al.}(2019{\natexlab{b}})\citenamefont {Zhang}, \citenamefont {Xie},
  \citenamefont {Wang}, \citenamefont {Xu}, \citenamefont {Tian}, \citenamefont
  {Jiang}, \citenamefont {Lu},\ and\ \citenamefont {Chen}}]{HOTISonic1}%
  \BibitemOpen
  \bibfield  {author} {\bibinfo {author} {\bibfnamefont {X.}~\bibnamefont
  {Zhang}}, \bibinfo {author} {\bibfnamefont {B.-Y.}\ \bibnamefont {Xie}},
  \bibinfo {author} {\bibfnamefont {H.-F.}\ \bibnamefont {Wang}}, \bibinfo
  {author} {\bibfnamefont {X.}~\bibnamefont {Xu}}, \bibinfo {author}
  {\bibfnamefont {Y.}~\bibnamefont {Tian}}, \bibinfo {author} {\bibfnamefont
  {J.-H.}\ \bibnamefont {Jiang}}, \bibinfo {author} {\bibfnamefont {M.-H.}\
  \bibnamefont {Lu}},\ and\ \bibinfo {author} {\bibfnamefont {Y.-F.}\
  \bibnamefont {Chen}},\ }\href
  {https://doi.org/https://doi.org/10.1038/s41467-019-13333-9} {\bibfield
  {journal} {\bibinfo  {journal} {Nat. commun.}\ }\textbf {\bibinfo {volume}
  {10}},\ \bibinfo {pages} {1} (\bibinfo {year}
  {2019}{\natexlab{b}})}\BibitemShut {NoStop}%
\bibitem [{\citenamefont {Fan}\ \emph {et~al.}(2019)\citenamefont {Fan},
  \citenamefont {Xia}, \citenamefont {Tong}, \citenamefont {Zheng},\ and\
  \citenamefont {Yu}}]{HOTIElast}%
  \BibitemOpen
  \bibfield  {author} {\bibinfo {author} {\bibfnamefont {H.}~\bibnamefont
  {Fan}}, \bibinfo {author} {\bibfnamefont {B.}~\bibnamefont {Xia}}, \bibinfo
  {author} {\bibfnamefont {L.}~\bibnamefont {Tong}}, \bibinfo {author}
  {\bibfnamefont {S.}~\bibnamefont {Zheng}},\ and\ \bibinfo {author}
  {\bibfnamefont {D.}~\bibnamefont {Yu}},\ }\href
  {https://doi.org/10.1103/PhysRevLett.122.204301} {\bibfield  {journal}
  {\bibinfo  {journal} {Phys. Rev. Lett.}\ }\textbf {\bibinfo {volume} {122}},\
  \bibinfo {pages} {204301} (\bibinfo {year} {2019})}\BibitemShut {NoStop}%
\bibitem [{\citenamefont {Schindler}\ \emph {et~al.}(2019)\citenamefont
  {Schindler}, \citenamefont {Brzezi{\'n}ska}, \citenamefont {Benalcazar},
  \citenamefont {Iraola}, \citenamefont {Bouhon}, \citenamefont {Tsirkin},
  \citenamefont {Vergniory},\ and\ \citenamefont {Neupert}}]{V_Topo}%
  \BibitemOpen
  \bibfield  {author} {\bibinfo {author} {\bibfnamefont {F.}~\bibnamefont
  {Schindler}}, \bibinfo {author} {\bibfnamefont {M.}~\bibnamefont
  {Brzezi{\'n}ska}}, \bibinfo {author} {\bibfnamefont {W.~A.}\ \bibnamefont
  {Benalcazar}}, \bibinfo {author} {\bibfnamefont {M.}~\bibnamefont {Iraola}},
  \bibinfo {author} {\bibfnamefont {A.}~\bibnamefont {Bouhon}}, \bibinfo
  {author} {\bibfnamefont {S.~S.}\ \bibnamefont {Tsirkin}}, \bibinfo {author}
  {\bibfnamefont {M.~G.}\ \bibnamefont {Vergniory}},\ and\ \bibinfo {author}
  {\bibfnamefont {T.}~\bibnamefont {Neupert}},\ }\href
  {https://doi.org/10.1103/physrevresearch.1.033074} {\bibfield  {journal}
  {\bibinfo  {journal} {Phys. Rev. Research}\ }\textbf {\bibinfo {volume}
  {1}},\ \bibinfo {pages} {033074} (\bibinfo {year} {2019})}\BibitemShut
  {NoStop}%
\bibitem [{\citenamefont {Sheng}\ \emph {et~al.}(2019)\citenamefont {Sheng},
  \citenamefont {Chen}, \citenamefont {Liu}, \citenamefont {Chen},
  \citenamefont {Yu}, \citenamefont {Zhao},\ and\ \citenamefont
  {Yang}}]{graphdiyne1}%
  \BibitemOpen
  \bibfield  {author} {\bibinfo {author} {\bibfnamefont {X.-L.}\ \bibnamefont
  {Sheng}}, \bibinfo {author} {\bibfnamefont {C.}~\bibnamefont {Chen}},
  \bibinfo {author} {\bibfnamefont {H.}~\bibnamefont {Liu}}, \bibinfo {author}
  {\bibfnamefont {Z.}~\bibnamefont {Chen}}, \bibinfo {author} {\bibfnamefont
  {Z.-M.}\ \bibnamefont {Yu}}, \bibinfo {author} {\bibfnamefont
  {Y.}~\bibnamefont {Zhao}},\ and\ \bibinfo {author} {\bibfnamefont {S.~A.}\
  \bibnamefont {Yang}},\ }\href
  {https://doi.org/10.1103/physrevlett.123.256402} {\bibfield  {journal}
  {\bibinfo  {journal} {Rev. Rev. Lett.}\ }\textbf {\bibinfo {volume} {123}},\
  \bibinfo {pages} {256402} (\bibinfo {year} {2019})}\BibitemShut {NoStop}%
\bibitem [{\citenamefont {Lee}\ \emph {et~al.}(2020)\citenamefont {Lee},
  \citenamefont {Kim}, \citenamefont {Ahn},\ and\ \citenamefont
  {Yang}}]{graphdiyne2}%
  \BibitemOpen
  \bibfield  {author} {\bibinfo {author} {\bibfnamefont {E.}~\bibnamefont
  {Lee}}, \bibinfo {author} {\bibfnamefont {R.}~\bibnamefont {Kim}}, \bibinfo
  {author} {\bibfnamefont {J.}~\bibnamefont {Ahn}},\ and\ \bibinfo {author}
  {\bibfnamefont {B.-J.}\ \bibnamefont {Yang}},\ }\href
  {https://doi.org/10.1038/s41535-019-0206-8} {\bibfield  {journal} {\bibinfo
  {journal} {npj Quantum Mater.}\ }\textbf {\bibinfo {volume} {5}},\ \bibinfo
  {pages} {1} (\bibinfo {year} {2020})}\BibitemShut {NoStop}%
\bibitem [{\citenamefont {Liu}\ \emph {et~al.}(2019{\natexlab{b}})\citenamefont
  {Liu}, \citenamefont {Zhao}, \citenamefont {Liu},\ and\ \citenamefont
  {Wang}}]{graphyne1}%
  \BibitemOpen
  \bibfield  {author} {\bibinfo {author} {\bibfnamefont {B.}~\bibnamefont
  {Liu}}, \bibinfo {author} {\bibfnamefont {G.}~\bibnamefont {Zhao}}, \bibinfo
  {author} {\bibfnamefont {Z.}~\bibnamefont {Liu}},\ and\ \bibinfo {author}
  {\bibfnamefont {Z.}~\bibnamefont {Wang}},\ }\href
  {https://doi.org/10.1021/acs.nanolett.9b02719} {\bibfield  {journal}
  {\bibinfo  {journal} {Nano Lett.}\ }\textbf {\bibinfo {volume} {19}},\
  \bibinfo {pages} {6492} (\bibinfo {year} {2019}{\natexlab{b}})}\BibitemShut
  {NoStop}%
\bibitem [{\citenamefont {Chen}\ \emph {et~al.}(2020)\citenamefont {Chen},
  \citenamefont {Wu}, \citenamefont {Yu}, \citenamefont {Chen}, \citenamefont
  {Zhao}, \citenamefont {Sheng},\ and\ \citenamefont {Yang}}]{graphyne2}%
  \BibitemOpen
  \bibfield  {author} {\bibinfo {author} {\bibfnamefont {C.}~\bibnamefont
  {Chen}}, \bibinfo {author} {\bibfnamefont {W.}~\bibnamefont {Wu}}, \bibinfo
  {author} {\bibfnamefont {Z.-M.}\ \bibnamefont {Yu}}, \bibinfo {author}
  {\bibfnamefont {Z.}~\bibnamefont {Chen}}, \bibinfo {author} {\bibfnamefont
  {Y.}~\bibnamefont {Zhao}}, \bibinfo {author} {\bibfnamefont {X.-L.}\
  \bibnamefont {Sheng}},\ and\ \bibinfo {author} {\bibfnamefont {S.~A.}\
  \bibnamefont {Yang}},\ }\href {https://arxiv.org/abs/2011.14868} {\bibfield
  {journal} {\bibinfo  {journal} {arXiv:2011.14868}\ } (\bibinfo {year}
  {2020})}\BibitemShut {NoStop}%
\bibitem [{\citenamefont {Liu}\ \emph {et~al.}(2021)\citenamefont {Liu},
  \citenamefont {Xian}, \citenamefont {Mu}, \citenamefont {Zhao}, \citenamefont
  {Liu}, \citenamefont {Rubio},\ and\ \citenamefont {Wang}}]{TBG1}%
  \BibitemOpen
  \bibfield  {author} {\bibinfo {author} {\bibfnamefont {B.}~\bibnamefont
  {Liu}}, \bibinfo {author} {\bibfnamefont {L.}~\bibnamefont {Xian}}, \bibinfo
  {author} {\bibfnamefont {H.}~\bibnamefont {Mu}}, \bibinfo {author}
  {\bibfnamefont {G.}~\bibnamefont {Zhao}}, \bibinfo {author} {\bibfnamefont
  {Z.}~\bibnamefont {Liu}}, \bibinfo {author} {\bibfnamefont {A.}~\bibnamefont
  {Rubio}},\ and\ \bibinfo {author} {\bibfnamefont {Z.~F.}\ \bibnamefont
  {Wang}},\ }\href {https://doi.org/10.1103/PhysRevLett.126.066401} {\bibfield
  {journal} {\bibinfo  {journal} {Phys. Rev. Lett.}\ }\textbf {\bibinfo
  {volume} {126}},\ \bibinfo {pages} {066401} (\bibinfo {year}
  {2021})}\BibitemShut {NoStop}%
\bibitem [{\citenamefont {Park}\ \emph {et~al.}(2019)\citenamefont {Park},
  \citenamefont {Kim}, \citenamefont {Cho},\ and\ \citenamefont {Lee}}]{TBG2}%
  \BibitemOpen
  \bibfield  {author} {\bibinfo {author} {\bibfnamefont {M.~J.}\ \bibnamefont
  {Park}}, \bibinfo {author} {\bibfnamefont {Y.}~\bibnamefont {Kim}}, \bibinfo
  {author} {\bibfnamefont {G.~Y.}\ \bibnamefont {Cho}},\ and\ \bibinfo {author}
  {\bibfnamefont {S.}~\bibnamefont {Lee}},\ }\href
  {https://doi.org/10.1103/PhysRevLett.123.216803} {\bibfield  {journal}
  {\bibinfo  {journal} {Phys. Rev. Lett.}\ }\textbf {\bibinfo {volume} {123}},\
  \bibinfo {pages} {216803} (\bibinfo {year} {2019})}\BibitemShut {NoStop}%
\bibitem [{\citenamefont {Zhang}\ \emph
  {et~al.}(2020{\natexlab{b}})\citenamefont {Zhang}, \citenamefont {Wu},\ and\
  \citenamefont {Das~Sarma}}]{MnBiTe}%
  \BibitemOpen
  \bibfield  {author} {\bibinfo {author} {\bibfnamefont {R.-X.}\ \bibnamefont
  {Zhang}}, \bibinfo {author} {\bibfnamefont {F.}~\bibnamefont {Wu}},\ and\
  \bibinfo {author} {\bibfnamefont {S.}~\bibnamefont {Das~Sarma}},\ }\href
  {https://doi.org/10.1103/PhysRevLett.124.136407} {\bibfield  {journal}
  {\bibinfo  {journal} {Phys. Rev. Lett.}\ }\textbf {\bibinfo {volume} {124}},\
  \bibinfo {pages} {136407} (\bibinfo {year} {2020}{\natexlab{b}})}\BibitemShut
  {NoStop}%
\bibitem [{\citenamefont {Schindler}\ \emph
  {et~al.}(2018{\natexlab{b}})\citenamefont {Schindler}, \citenamefont {Cook},
  \citenamefont {Vergniory}, \citenamefont {Wang}, \citenamefont {Parkin},
  \citenamefont {Bernevig},\ and\ \citenamefont {Neupert}}]{HOTI}%
  \BibitemOpen
  \bibfield  {author} {\bibinfo {author} {\bibfnamefont {F.}~\bibnamefont
  {Schindler}}, \bibinfo {author} {\bibfnamefont {A.~M.}\ \bibnamefont {Cook}},
  \bibinfo {author} {\bibfnamefont {M.~G.}\ \bibnamefont {Vergniory}}, \bibinfo
  {author} {\bibfnamefont {Z.}~\bibnamefont {Wang}}, \bibinfo {author}
  {\bibfnamefont {S.~S.}\ \bibnamefont {Parkin}}, \bibinfo {author}
  {\bibfnamefont {B.~A.}\ \bibnamefont {Bernevig}},\ and\ \bibinfo {author}
  {\bibfnamefont {T.}~\bibnamefont {Neupert}},\ }\href
  {https://doi.org/10.1126/sciadv.aat0346} {\bibfield  {journal} {\bibinfo
  {journal} {Sci. adv.}\ }\textbf {\bibinfo {volume} {4}},\ \bibinfo {pages}
  {eaat0346} (\bibinfo {year} {2018}{\natexlab{b}})}\BibitemShut {NoStop}%
\bibitem [{\citenamefont {Xu}\ \emph {et~al.}(2019)\citenamefont {Xu},
  \citenamefont {Song}, \citenamefont {Wang}, \citenamefont {Weng},\ and\
  \citenamefont {Dai}}]{XiDai1}%
  \BibitemOpen
  \bibfield  {author} {\bibinfo {author} {\bibfnamefont {Y.}~\bibnamefont
  {Xu}}, \bibinfo {author} {\bibfnamefont {Z.}~\bibnamefont {Song}}, \bibinfo
  {author} {\bibfnamefont {Z.}~\bibnamefont {Wang}}, \bibinfo {author}
  {\bibfnamefont {H.}~\bibnamefont {Weng}},\ and\ \bibinfo {author}
  {\bibfnamefont {X.}~\bibnamefont {Dai}},\ }\href
  {https://doi.org/10.1103/physrevlett.122.256402} {\bibfield  {journal}
  {\bibinfo  {journal} {Phys. Rev. Lett.}\ }\textbf {\bibinfo {volume} {122}},\
  \bibinfo {pages} {256402} (\bibinfo {year} {2019})}\BibitemShut {NoStop}%
\bibitem [{\citenamefont {Yue}\ \emph {et~al.}(2019)\citenamefont {Yue},
  \citenamefont {Xu}, \citenamefont {Song}, \citenamefont {Weng}, \citenamefont
  {Lu}, \citenamefont {Fang},\ and\ \citenamefont {Dai}}]{XiDai2}%
  \BibitemOpen
  \bibfield  {author} {\bibinfo {author} {\bibfnamefont {C.}~\bibnamefont
  {Yue}}, \bibinfo {author} {\bibfnamefont {Y.}~\bibnamefont {Xu}}, \bibinfo
  {author} {\bibfnamefont {Z.}~\bibnamefont {Song}}, \bibinfo {author}
  {\bibfnamefont {H.}~\bibnamefont {Weng}}, \bibinfo {author} {\bibfnamefont
  {Y.-M.}\ \bibnamefont {Lu}}, \bibinfo {author} {\bibfnamefont
  {C.}~\bibnamefont {Fang}},\ and\ \bibinfo {author} {\bibfnamefont
  {X.}~\bibnamefont {Dai}},\ }\href
  {https://doi.org/https://doi.org/10.1038/s41567-019-0457-0} {\bibfield
  {journal} {\bibinfo  {journal} {Nat. Phys.}\ }\textbf {\bibinfo {volume}
  {15}},\ \bibinfo {pages} {577} (\bibinfo {year} {2019})}\BibitemShut
  {NoStop}%
\bibitem [{\citenamefont {Pesin}\ and\ \citenamefont
  {MacDonald}(2012)}]{Macdonald}%
  \BibitemOpen
  \bibfield  {author} {\bibinfo {author} {\bibfnamefont {D.}~\bibnamefont
  {Pesin}}\ and\ \bibinfo {author} {\bibfnamefont {A.~H.}\ \bibnamefont
  {MacDonald}},\ }\href {https://doi.org/https://doi.org/10.1038/nmat3305}
  {\bibfield  {journal} {\bibinfo  {journal} {Nature materials}\ }\textbf
  {\bibinfo {volume} {11}},\ \bibinfo {pages} {409} (\bibinfo {year}
  {2012})}\BibitemShut {NoStop}%
\bibitem [{\citenamefont {Xiao}\ \emph {et~al.}(2007)\citenamefont {Xiao},
  \citenamefont {Yao},\ and\ \citenamefont {Niu}}]{XiaoDi}%
  \BibitemOpen
  \bibfield  {author} {\bibinfo {author} {\bibfnamefont {D.}~\bibnamefont
  {Xiao}}, \bibinfo {author} {\bibfnamefont {W.}~\bibnamefont {Yao}},\ and\
  \bibinfo {author} {\bibfnamefont {Q.}~\bibnamefont {Niu}},\ }\href
  {https://doi.org/10.1103/PhysRevLett.99.236809} {\bibfield  {journal}
  {\bibinfo  {journal} {Phys. Rev. Lett.}\ }\textbf {\bibinfo {volume} {99}},\
  \bibinfo {pages} {236809} (\bibinfo {year} {2007})}\BibitemShut {NoStop}%
\bibitem [{\citenamefont {Pan}\ \emph {et~al.}(2014)\citenamefont {Pan},
  \citenamefont {Li}, \citenamefont {Liu}, \citenamefont {Zhu}, \citenamefont
  {Qiao},\ and\ \citenamefont {Yao}}]{yao1}%
  \BibitemOpen
  \bibfield  {author} {\bibinfo {author} {\bibfnamefont {H.}~\bibnamefont
  {Pan}}, \bibinfo {author} {\bibfnamefont {Z.}~\bibnamefont {Li}}, \bibinfo
  {author} {\bibfnamefont {C.-C.}\ \bibnamefont {Liu}}, \bibinfo {author}
  {\bibfnamefont {G.}~\bibnamefont {Zhu}}, \bibinfo {author} {\bibfnamefont
  {Z.}~\bibnamefont {Qiao}},\ and\ \bibinfo {author} {\bibfnamefont
  {Y.}~\bibnamefont {Yao}},\ }\href
  {https://doi.org/10.1103/PhysRevLett.112.106802} {\bibfield  {journal}
  {\bibinfo  {journal} {Phys. Rev. Lett.}\ }\textbf {\bibinfo {volume} {112}},\
  \bibinfo {pages} {106802} (\bibinfo {year} {2014})}\BibitemShut {NoStop}%
\bibitem [{\citenamefont {Xu}\ \emph {et~al.}(2013{\natexlab{a}})\citenamefont
  {Xu}, \citenamefont {Liang}, \citenamefont {Shi},\ and\ \citenamefont
  {Chen}}]{hexamater}%
  \BibitemOpen
  \bibfield  {author} {\bibinfo {author} {\bibfnamefont {M.}~\bibnamefont
  {Xu}}, \bibinfo {author} {\bibfnamefont {T.}~\bibnamefont {Liang}}, \bibinfo
  {author} {\bibfnamefont {M.}~\bibnamefont {Shi}},\ and\ \bibinfo {author}
  {\bibfnamefont {H.}~\bibnamefont {Chen}},\ }\href
  {https://doi.org/https://doi.org/10.1021/cr300263a} {\bibfield  {journal}
  {\bibinfo  {journal} {Chemical reviews}\ }\textbf {\bibinfo {volume} {113}},\
  \bibinfo {pages} {3766} (\bibinfo {year} {2013}{\natexlab{a}})}\BibitemShut
  {NoStop}%
\bibitem [{\citenamefont {Mak}\ \emph {et~al.}(2014)\citenamefont {Mak},
  \citenamefont {McGill}, \citenamefont {Park},\ and\ \citenamefont
  {McEuen}}]{valleyhall}%
  \BibitemOpen
  \bibfield  {author} {\bibinfo {author} {\bibfnamefont {K.~F.}\ \bibnamefont
  {Mak}}, \bibinfo {author} {\bibfnamefont {K.~L.}\ \bibnamefont {McGill}},
  \bibinfo {author} {\bibfnamefont {J.}~\bibnamefont {Park}},\ and\ \bibinfo
  {author} {\bibfnamefont {P.~L.}\ \bibnamefont {McEuen}},\ }\href
  {https://doi.org/10.1126/science.1250140} {\bibfield  {journal} {\bibinfo
  {journal} {Science}\ }\textbf {\bibinfo {volume} {344}},\ \bibinfo {pages}
  {1489} (\bibinfo {year} {2014})}\BibitemShut {NoStop}%
\bibitem [{\citenamefont {Song}\ \emph {et~al.}(2014)\citenamefont {Song},
  \citenamefont {Liu}, \citenamefont {Yang}, \citenamefont {Han}, \citenamefont
  {Ye}, \citenamefont {Fu}, \citenamefont {Yang}, \citenamefont {Niu},
  \citenamefont {Lu},\ and\ \citenamefont {Yao}}]{yao2}%
  \BibitemOpen
  \bibfield  {author} {\bibinfo {author} {\bibfnamefont {Z.}~\bibnamefont
  {Song}}, \bibinfo {author} {\bibfnamefont {C.-C.}\ \bibnamefont {Liu}},
  \bibinfo {author} {\bibfnamefont {J.}~\bibnamefont {Yang}}, \bibinfo {author}
  {\bibfnamefont {J.}~\bibnamefont {Han}}, \bibinfo {author} {\bibfnamefont
  {M.}~\bibnamefont {Ye}}, \bibinfo {author} {\bibfnamefont {B.}~\bibnamefont
  {Fu}}, \bibinfo {author} {\bibfnamefont {Y.}~\bibnamefont {Yang}}, \bibinfo
  {author} {\bibfnamefont {Q.}~\bibnamefont {Niu}}, \bibinfo {author}
  {\bibfnamefont {J.}~\bibnamefont {Lu}},\ and\ \bibinfo {author}
  {\bibfnamefont {Y.}~\bibnamefont {Yao}},\ }\href
  {https://doi.org/https://doi.org/10.1038/am.2014.113} {\bibfield  {journal}
  {\bibinfo  {journal} {NPG Asia Materials}\ }\textbf {\bibinfo {volume} {6}},\
  \bibinfo {pages} {e147} (\bibinfo {year} {2014})}\BibitemShut {NoStop}%
\bibitem [{\citenamefont {Wu}\ \emph {et~al.}(2007)\citenamefont {Wu},
  \citenamefont {Bergman}, \citenamefont {Balents},\ and\ \citenamefont
  {Das~Sarma}}]{congjun1}%
  \BibitemOpen
  \bibfield  {author} {\bibinfo {author} {\bibfnamefont {C.}~\bibnamefont
  {Wu}}, \bibinfo {author} {\bibfnamefont {D.}~\bibnamefont {Bergman}},
  \bibinfo {author} {\bibfnamefont {L.}~\bibnamefont {Balents}},\ and\ \bibinfo
  {author} {\bibfnamefont {S.}~\bibnamefont {Das~Sarma}},\ }\href
  {https://doi.org/10.1103/PhysRevLett.99.070401} {\bibfield  {journal}
  {\bibinfo  {journal} {Phys. Rev. Lett.}\ }\textbf {\bibinfo {volume} {99}},\
  \bibinfo {pages} {070401} (\bibinfo {year} {2007})}\BibitemShut {NoStop}%
\bibitem [{\citenamefont {Wu}(2008)}]{congjun2}%
  \BibitemOpen
  \bibfield  {author} {\bibinfo {author} {\bibfnamefont {C.}~\bibnamefont
  {Wu}},\ }\href {https://doi.org/10.1103/PhysRevLett.100.200406} {\bibfield
  {journal} {\bibinfo  {journal} {Phys. Rev. Lett.}\ }\textbf {\bibinfo
  {volume} {100}},\ \bibinfo {pages} {200406} (\bibinfo {year}
  {2008})}\BibitemShut {NoStop}%
\bibitem [{\citenamefont {Xu}\ \emph {et~al.}(2013{\natexlab{b}})\citenamefont
  {Xu}, \citenamefont {Yan}, \citenamefont {Zhang}, \citenamefont {Wang},
  \citenamefont {Xu}, \citenamefont {Tang}, \citenamefont {Duan},\ and\
  \citenamefont {Zhang}}]{yongxu}%
  \BibitemOpen
  \bibfield  {author} {\bibinfo {author} {\bibfnamefont {Y.}~\bibnamefont
  {Xu}}, \bibinfo {author} {\bibfnamefont {B.}~\bibnamefont {Yan}}, \bibinfo
  {author} {\bibfnamefont {H.-J.}\ \bibnamefont {Zhang}}, \bibinfo {author}
  {\bibfnamefont {J.}~\bibnamefont {Wang}}, \bibinfo {author} {\bibfnamefont
  {G.}~\bibnamefont {Xu}}, \bibinfo {author} {\bibfnamefont {P.}~\bibnamefont
  {Tang}}, \bibinfo {author} {\bibfnamefont {W.}~\bibnamefont {Duan}},\ and\
  \bibinfo {author} {\bibfnamefont {S.-C.}\ \bibnamefont {Zhang}},\ }\href
  {https://doi.org/10.1103/PhysRevLett.111.136804} {\bibfield  {journal}
  {\bibinfo  {journal} {Phys. Rev. Lett.}\ }\textbf {\bibinfo {volume} {111}},\
  \bibinfo {pages} {136804} (\bibinfo {year} {2013}{\natexlab{b}})}\BibitemShut
  {NoStop}%
\bibitem [{\citenamefont {Qiao}\ \emph {et~al.}(2010)\citenamefont {Qiao},
  \citenamefont {Yang}, \citenamefont {Feng}, \citenamefont {Tse},
  \citenamefont {Ding}, \citenamefont {Yao}, \citenamefont {Wang},\ and\
  \citenamefont {Niu}}]{yao3}%
  \BibitemOpen
  \bibfield  {author} {\bibinfo {author} {\bibfnamefont {Z.}~\bibnamefont
  {Qiao}}, \bibinfo {author} {\bibfnamefont {S.~A.}\ \bibnamefont {Yang}},
  \bibinfo {author} {\bibfnamefont {W.}~\bibnamefont {Feng}}, \bibinfo {author}
  {\bibfnamefont {W.-K.}\ \bibnamefont {Tse}}, \bibinfo {author} {\bibfnamefont
  {J.}~\bibnamefont {Ding}}, \bibinfo {author} {\bibfnamefont {Y.}~\bibnamefont
  {Yao}}, \bibinfo {author} {\bibfnamefont {J.}~\bibnamefont {Wang}},\ and\
  \bibinfo {author} {\bibfnamefont {Q.}~\bibnamefont {Niu}},\ }\href
  {https://doi.org/10.1103/PhysRevB.82.161414} {\bibfield  {journal} {\bibinfo
  {journal} {Phys. Rev. B}\ }\textbf {\bibinfo {volume} {82}},\ \bibinfo
  {pages} {161414} (\bibinfo {year} {2010})}\BibitemShut {NoStop}%
\bibitem [{\citenamefont {Das~Sarma}\ \emph {et~al.}(2011)\citenamefont
  {Das~Sarma}, \citenamefont {Adam}, \citenamefont {Hwang},\ and\ \citenamefont
  {Rossi}}]{Das}%
  \BibitemOpen
  \bibfield  {author} {\bibinfo {author} {\bibfnamefont {S.}~\bibnamefont
  {Das~Sarma}}, \bibinfo {author} {\bibfnamefont {S.}~\bibnamefont {Adam}},
  \bibinfo {author} {\bibfnamefont {E.~H.}\ \bibnamefont {Hwang}},\ and\
  \bibinfo {author} {\bibfnamefont {E.}~\bibnamefont {Rossi}},\ }\href
  {https://doi.org/10.1103/RevModPhys.83.407} {\bibfield  {journal} {\bibinfo
  {journal} {Rev. Mod. Phys.}\ }\textbf {\bibinfo {volume} {83}},\ \bibinfo
  {pages} {407} (\bibinfo {year} {2011})}\BibitemShut {NoStop}%
\bibitem [{\citenamefont {Liu}\ \emph {et~al.}(2014{\natexlab{a}})\citenamefont
  {Liu}, \citenamefont {Guan}, \citenamefont {Song}, \citenamefont {Yang},
  \citenamefont {Yang},\ and\ \citenamefont {Yao}}]{cclprb2014}%
  \BibitemOpen
  \bibfield  {author} {\bibinfo {author} {\bibfnamefont {C.-C.}\ \bibnamefont
  {Liu}}, \bibinfo {author} {\bibfnamefont {S.}~\bibnamefont {Guan}}, \bibinfo
  {author} {\bibfnamefont {Z.}~\bibnamefont {Song}}, \bibinfo {author}
  {\bibfnamefont {S.~A.}\ \bibnamefont {Yang}}, \bibinfo {author}
  {\bibfnamefont {J.}~\bibnamefont {Yang}},\ and\ \bibinfo {author}
  {\bibfnamefont {Y.}~\bibnamefont {Yao}},\ }\href
  {https://doi.org/10.1103/PhysRevB.90.085431} {\bibfield  {journal} {\bibinfo
  {journal} {Phys. Rev. B}\ }\textbf {\bibinfo {volume} {90}},\ \bibinfo
  {pages} {085431} (\bibinfo {year} {2014}{\natexlab{a}})}\BibitemShut
  {NoStop}%
\bibitem [{\citenamefont {Geim}\ and\ \citenamefont {Novoselov}(2007)}]{geim1}%
  \BibitemOpen
  \bibfield  {author} {\bibinfo {author} {\bibfnamefont {A.}~\bibnamefont
  {Geim}}\ and\ \bibinfo {author} {\bibfnamefont {K.}~\bibnamefont
  {Novoselov}},\ }\href {https://doi.org/https://doi.org/10.1038/nmat1849}
  {\bibfield  {journal} {\bibinfo  {journal} {Nat. Mater.}\ }\textbf {\bibinfo
  {volume} {6}},\ \bibinfo {pages} {183} (\bibinfo {year} {2007})}\BibitemShut
  {NoStop}%
\bibitem [{\citenamefont {Haldane}(1988)}]{haldane1}%
  \BibitemOpen
  \bibfield  {author} {\bibinfo {author} {\bibfnamefont {F.~D.~M.}\
  \bibnamefont {Haldane}},\ }\href
  {https://doi.org/10.1103/physrevlett.61.2015} {\bibfield  {journal} {\bibinfo
   {journal} {Rev. Rev. Lett.}\ }\textbf {\bibinfo {volume} {61}},\ \bibinfo
  {pages} {2015} (\bibinfo {year} {1988})}\BibitemShut {NoStop}%
\bibitem [{\citenamefont {Semenoff}\ and\ \citenamefont
  {Sodano}(1986)}]{semenoff}%
  \BibitemOpen
  \bibfield  {author} {\bibinfo {author} {\bibfnamefont {G.~W.}\ \bibnamefont
  {Semenoff}}\ and\ \bibinfo {author} {\bibfnamefont {P.}~\bibnamefont
  {Sodano}},\ }\href {https://doi.org/10.1103/physrevlett.57.1195} {\bibfield
  {journal} {\bibinfo  {journal} {Phys. Rev. Lett.}\ }\textbf {\bibinfo
  {volume} {57}},\ \bibinfo {pages} {1195} (\bibinfo {year}
  {1986})}\BibitemShut {NoStop}%
\bibitem [{\citenamefont {Kane}\ and\ \citenamefont {Mele}(2005)}]{kanemele}%
  \BibitemOpen
  \bibfield  {author} {\bibinfo {author} {\bibfnamefont {C.~L.}\ \bibnamefont
  {Kane}}\ and\ \bibinfo {author} {\bibfnamefont {E.~J.}\ \bibnamefont
  {Mele}},\ }\href {https://doi.org/10.1103/physrevlett.95.226801} {\bibfield
  {journal} {\bibinfo  {journal} {Rev. Rev. Lett.}\ }\textbf {\bibinfo {volume}
  {95}},\ \bibinfo {pages} {226801} (\bibinfo {year} {2005})}\BibitemShut
  {NoStop}%
\bibitem [{\citenamefont {Elias}\ \emph {et~al.}(2009)\citenamefont {Elias},
  \citenamefont {Nair}, \citenamefont {Mohiuddin}, \citenamefont {Morozov},
  \citenamefont {Blake}, \citenamefont {Halsall}, \citenamefont {Ferrari},
  \citenamefont {Boukhvalov}, \citenamefont {Katsnelson}, \citenamefont {Geim}
  \emph {et~al.}}]{graphane}%
  \BibitemOpen
  \bibfield  {author} {\bibinfo {author} {\bibfnamefont {D.~C.}\ \bibnamefont
  {Elias}}, \bibinfo {author} {\bibfnamefont {R.~R.}\ \bibnamefont {Nair}},
  \bibinfo {author} {\bibfnamefont {T.}~\bibnamefont {Mohiuddin}}, \bibinfo
  {author} {\bibfnamefont {S.}~\bibnamefont {Morozov}}, \bibinfo {author}
  {\bibfnamefont {P.}~\bibnamefont {Blake}}, \bibinfo {author} {\bibfnamefont
  {M.}~\bibnamefont {Halsall}}, \bibinfo {author} {\bibfnamefont {A.~C.}\
  \bibnamefont {Ferrari}}, \bibinfo {author} {\bibfnamefont {D.}~\bibnamefont
  {Boukhvalov}}, \bibinfo {author} {\bibfnamefont {M.}~\bibnamefont
  {Katsnelson}}, \bibinfo {author} {\bibfnamefont {A.}~\bibnamefont {Geim}},
  \emph {et~al.},\ }\href {https://doi.org/10.1126/science.1167130} {\bibfield
  {journal} {\bibinfo  {journal} {Science}\ }\textbf {\bibinfo {volume}
  {323}},\ \bibinfo {pages} {610} (\bibinfo {year} {2009})}\BibitemShut
  {NoStop}%
\bibitem [{\citenamefont {Karlicky}\ \emph {et~al.}(2013)\citenamefont
  {Karlicky}, \citenamefont {Kumara Ramanatha~Datta}, \citenamefont {Otyepka},\
  and\ \citenamefont {Zboril}}]{halide}%
  \BibitemOpen
  \bibfield  {author} {\bibinfo {author} {\bibfnamefont {F.}~\bibnamefont
  {Karlicky}}, \bibinfo {author} {\bibfnamefont {K.}~\bibnamefont {Kumara
  Ramanatha~Datta}}, \bibinfo {author} {\bibfnamefont {M.}~\bibnamefont
  {Otyepka}},\ and\ \bibinfo {author} {\bibfnamefont {R.}~\bibnamefont
  {Zboril}},\ }\href {https://doi.org/10.1021/nn4024027} {\bibfield  {journal}
  {\bibinfo  {journal} {ACS nano}\ }\textbf {\bibinfo {volume} {7}},\ \bibinfo
  {pages} {6434} (\bibinfo {year} {2013})}\BibitemShut {NoStop}%
\bibitem [{\citenamefont {Zbo{\v{r}}il}\ \emph {et~al.}(2010)\citenamefont
  {Zbo{\v{r}}il}, \citenamefont {Karlick{\`y}}, \citenamefont {Bourlinos},
  \citenamefont {Steriotis}, \citenamefont {Stubos}, \citenamefont
  {Georgakilas}, \citenamefont {{\v{S}}af{\'a}{\v{r}}ov{\'a}}, \citenamefont
  {Jan{\v{c}}{\'\i}k}, \citenamefont {Trapalis},\ and\ \citenamefont
  {Otyepka}}]{halide2}%
  \BibitemOpen
  \bibfield  {author} {\bibinfo {author} {\bibfnamefont {R.}~\bibnamefont
  {Zbo{\v{r}}il}}, \bibinfo {author} {\bibfnamefont {F.}~\bibnamefont
  {Karlick{\`y}}}, \bibinfo {author} {\bibfnamefont {A.~B.}\ \bibnamefont
  {Bourlinos}}, \bibinfo {author} {\bibfnamefont {T.~A.}\ \bibnamefont
  {Steriotis}}, \bibinfo {author} {\bibfnamefont {A.~K.}\ \bibnamefont
  {Stubos}}, \bibinfo {author} {\bibfnamefont {V.}~\bibnamefont {Georgakilas}},
  \bibinfo {author} {\bibfnamefont {K.}~\bibnamefont
  {{\v{S}}af{\'a}{\v{r}}ov{\'a}}}, \bibinfo {author} {\bibfnamefont
  {D.}~\bibnamefont {Jan{\v{c}}{\'\i}k}}, \bibinfo {author} {\bibfnamefont
  {C.}~\bibnamefont {Trapalis}},\ and\ \bibinfo {author} {\bibfnamefont
  {M.}~\bibnamefont {Otyepka}},\ }\href
  {https://doi.org/10.1002/smll.201001401} {\bibfield  {journal} {\bibinfo
  {journal} {Small}\ }\textbf {\bibinfo {volume} {6}},\ \bibinfo {pages} {2885}
  (\bibinfo {year} {2010})}\BibitemShut {NoStop}%
\bibitem [{\citenamefont {Qiu}\ \emph {et~al.}(2015)\citenamefont {Qiu},
  \citenamefont {Fu}, \citenamefont {Xu}, \citenamefont {Oreshkin},
  \citenamefont {Shao}, \citenamefont {Li}, \citenamefont {Meng}, \citenamefont
  {Chen},\ and\ \citenamefont {Wu}}]{SiH}%
  \BibitemOpen
  \bibfield  {author} {\bibinfo {author} {\bibfnamefont {J.}~\bibnamefont
  {Qiu}}, \bibinfo {author} {\bibfnamefont {H.}~\bibnamefont {Fu}}, \bibinfo
  {author} {\bibfnamefont {Y.}~\bibnamefont {Xu}}, \bibinfo {author}
  {\bibfnamefont {A.~I.}\ \bibnamefont {Oreshkin}}, \bibinfo {author}
  {\bibfnamefont {T.}~\bibnamefont {Shao}}, \bibinfo {author} {\bibfnamefont
  {H.}~\bibnamefont {Li}}, \bibinfo {author} {\bibfnamefont {S.}~\bibnamefont
  {Meng}}, \bibinfo {author} {\bibfnamefont {L.}~\bibnamefont {Chen}},\ and\
  \bibinfo {author} {\bibfnamefont {K.}~\bibnamefont {Wu}},\ }\href
  {https://doi.org/10.1103/PhysRevLett.114.126101} {\bibfield  {journal}
  {\bibinfo  {journal} {Phys. Rev. Lett.}\ }\textbf {\bibinfo {volume} {114}},\
  \bibinfo {pages} {126101} (\bibinfo {year} {2015})}\BibitemShut {NoStop}%
\bibitem [{\citenamefont {Zhang}\ \emph {et~al.}(2018)\citenamefont {Zhang},
  \citenamefont {Enriquez}, \citenamefont {Tong}, \citenamefont {Bendounan},
  \citenamefont {Kara}, \citenamefont {Seitsonen}, \citenamefont {Mayne},
  \citenamefont {Dujardin},\ and\ \citenamefont {Oughaddou}}]{blueph}%
  \BibitemOpen
  \bibfield  {author} {\bibinfo {author} {\bibfnamefont {W.}~\bibnamefont
  {Zhang}}, \bibinfo {author} {\bibfnamefont {H.}~\bibnamefont {Enriquez}},
  \bibinfo {author} {\bibfnamefont {Y.}~\bibnamefont {Tong}}, \bibinfo {author}
  {\bibfnamefont {A.}~\bibnamefont {Bendounan}}, \bibinfo {author}
  {\bibfnamefont {A.}~\bibnamefont {Kara}}, \bibinfo {author} {\bibfnamefont
  {A.~P.}\ \bibnamefont {Seitsonen}}, \bibinfo {author} {\bibfnamefont {A.~J.}\
  \bibnamefont {Mayne}}, \bibinfo {author} {\bibfnamefont {G.}~\bibnamefont
  {Dujardin}},\ and\ \bibinfo {author} {\bibfnamefont {H.}~\bibnamefont
  {Oughaddou}},\ }\href {https://doi.org/10.1002/smll.201804066} {\bibfield
  {journal} {\bibinfo  {journal} {Small}\ }\textbf {\bibinfo {volume} {14}},\
  \bibinfo {pages} {1804066} (\bibinfo {year} {2018})}\BibitemShut {NoStop}%
\bibitem [{\citenamefont {Li}\ \emph {et~al.}(2014)\citenamefont {Li},
  \citenamefont {Yu}, \citenamefont {Ye}, \citenamefont {Ge}, \citenamefont
  {Ou}, \citenamefont {Wu}, \citenamefont {Feng}, \citenamefont {Chen},\ and\
  \citenamefont {Zhang}}]{blackP1}%
  \BibitemOpen
  \bibfield  {author} {\bibinfo {author} {\bibfnamefont {L.}~\bibnamefont
  {Li}}, \bibinfo {author} {\bibfnamefont {Y.}~\bibnamefont {Yu}}, \bibinfo
  {author} {\bibfnamefont {G.~J.}\ \bibnamefont {Ye}}, \bibinfo {author}
  {\bibfnamefont {Q.}~\bibnamefont {Ge}}, \bibinfo {author} {\bibfnamefont
  {X.}~\bibnamefont {Ou}}, \bibinfo {author} {\bibfnamefont {H.}~\bibnamefont
  {Wu}}, \bibinfo {author} {\bibfnamefont {D.}~\bibnamefont {Feng}}, \bibinfo
  {author} {\bibfnamefont {X.~H.}\ \bibnamefont {Chen}},\ and\ \bibinfo
  {author} {\bibfnamefont {Y.}~\bibnamefont {Zhang}},\ }\href
  {https://doi.org/https://doi.org/10.1038/nnano.2014.35} {\bibfield  {journal}
  {\bibinfo  {journal} {Nat. nanotechnol.}\ }\textbf {\bibinfo {volume} {9}},\
  \bibinfo {pages} {372} (\bibinfo {year} {2014})}\BibitemShut {NoStop}%
\bibitem [{\citenamefont {Liu}\ \emph {et~al.}(2014{\natexlab{b}})\citenamefont
  {Liu}, \citenamefont {Neal}, \citenamefont {Zhu}, \citenamefont {Luo},
  \citenamefont {Xu}, \citenamefont {Tom{\'a}nek},\ and\ \citenamefont
  {Ye}}]{blackP2}%
  \BibitemOpen
  \bibfield  {author} {\bibinfo {author} {\bibfnamefont {H.}~\bibnamefont
  {Liu}}, \bibinfo {author} {\bibfnamefont {A.~T.}\ \bibnamefont {Neal}},
  \bibinfo {author} {\bibfnamefont {Z.}~\bibnamefont {Zhu}}, \bibinfo {author}
  {\bibfnamefont {Z.}~\bibnamefont {Luo}}, \bibinfo {author} {\bibfnamefont
  {X.}~\bibnamefont {Xu}}, \bibinfo {author} {\bibfnamefont {D.}~\bibnamefont
  {Tom{\'a}nek}},\ and\ \bibinfo {author} {\bibfnamefont {P.~D.}\ \bibnamefont
  {Ye}},\ }\href {https://doi.org/https://doi.org/10.1021/nn501226z} {\bibfield
   {journal} {\bibinfo  {journal} {ACS nano}\ }\textbf {\bibinfo {volume}
  {8}},\ \bibinfo {pages} {4033} (\bibinfo {year}
  {2014}{\natexlab{b}})}\BibitemShut {NoStop}%
\bibitem [{\citenamefont {Zhang}\ \emph {et~al.}(2016)\citenamefont {Zhang},
  \citenamefont {Zhao}, \citenamefont {Han}, \citenamefont {Wang},
  \citenamefont {Zhong}, \citenamefont {Sun}, \citenamefont {Guo},
  \citenamefont {Zhou}, \citenamefont {Gu}, \citenamefont {Yuan} \emph
  {et~al.}}]{blueP1}%
  \BibitemOpen
  \bibfield  {author} {\bibinfo {author} {\bibfnamefont {J.~L.}\ \bibnamefont
  {Zhang}}, \bibinfo {author} {\bibfnamefont {S.}~\bibnamefont {Zhao}},
  \bibinfo {author} {\bibfnamefont {C.}~\bibnamefont {Han}}, \bibinfo {author}
  {\bibfnamefont {Z.}~\bibnamefont {Wang}}, \bibinfo {author} {\bibfnamefont
  {S.}~\bibnamefont {Zhong}}, \bibinfo {author} {\bibfnamefont
  {S.}~\bibnamefont {Sun}}, \bibinfo {author} {\bibfnamefont {R.}~\bibnamefont
  {Guo}}, \bibinfo {author} {\bibfnamefont {X.}~\bibnamefont {Zhou}}, \bibinfo
  {author} {\bibfnamefont {C.~D.}\ \bibnamefont {Gu}}, \bibinfo {author}
  {\bibfnamefont {K.~D.}\ \bibnamefont {Yuan}}, \emph {et~al.},\ }\href
  {https://doi.org/https://doi.org/10.1021/acs.nanolett.6b01459} {\bibfield
  {journal} {\bibinfo  {journal} {Nano lett.}\ }\textbf {\bibinfo {volume}
  {16}},\ \bibinfo {pages} {4903} (\bibinfo {year} {2016})}\BibitemShut
  {NoStop}%
\bibitem [{\citenamefont {Zhong}\ \emph {et~al.}(2018)\citenamefont {Zhong},
  \citenamefont {Xia}, \citenamefont {Pan}, \citenamefont {Liu}, \citenamefont
  {Chen}, \citenamefont {Deng}, \citenamefont {Li},\ and\ \citenamefont
  {Wei}}]{blackAs1}%
  \BibitemOpen
  \bibfield  {author} {\bibinfo {author} {\bibfnamefont {M.}~\bibnamefont
  {Zhong}}, \bibinfo {author} {\bibfnamefont {Q.}~\bibnamefont {Xia}}, \bibinfo
  {author} {\bibfnamefont {L.}~\bibnamefont {Pan}}, \bibinfo {author}
  {\bibfnamefont {Y.}~\bibnamefont {Liu}}, \bibinfo {author} {\bibfnamefont
  {Y.}~\bibnamefont {Chen}}, \bibinfo {author} {\bibfnamefont {H.-X.}\
  \bibnamefont {Deng}}, \bibinfo {author} {\bibfnamefont {J.}~\bibnamefont
  {Li}},\ and\ \bibinfo {author} {\bibfnamefont {Z.}~\bibnamefont {Wei}},\
  }\href {https://doi.org/https://doi.org/10.1002/adfm.201802581} {\bibfield
  {journal} {\bibinfo  {journal} {Adv. Funct. Mater.}\ }\textbf {\bibinfo
  {volume} {28}},\ \bibinfo {pages} {1802581} (\bibinfo {year}
  {2018})}\BibitemShut {NoStop}%
\bibitem [{\citenamefont {Chen}\ \emph {et~al.}(2018)\citenamefont {Chen},
  \citenamefont {Chen}, \citenamefont {Kealhofer}, \citenamefont {Liu},
  \citenamefont {Yuan}, \citenamefont {Jiang}, \citenamefont {Suh},
  \citenamefont {Park}, \citenamefont {Ko}, \citenamefont {Choe}, \citenamefont
  {Avila}, \citenamefont {Zhong}, \citenamefont {Wei}, \citenamefont {Li},
  \citenamefont {Li}, \citenamefont {Gao}, \citenamefont {Liu}, \citenamefont
  {Analytis}, \citenamefont {Xia}, \citenamefont {Asensio},\ and\ \citenamefont
  {Wu}}]{blackAs2}%
  \BibitemOpen
  \bibfield  {author} {\bibinfo {author} {\bibfnamefont {Y.}~\bibnamefont
  {Chen}}, \bibinfo {author} {\bibfnamefont {C.}~\bibnamefont {Chen}}, \bibinfo
  {author} {\bibfnamefont {R.}~\bibnamefont {Kealhofer}}, \bibinfo {author}
  {\bibfnamefont {H.}~\bibnamefont {Liu}}, \bibinfo {author} {\bibfnamefont
  {Z.}~\bibnamefont {Yuan}}, \bibinfo {author} {\bibfnamefont {L.}~\bibnamefont
  {Jiang}}, \bibinfo {author} {\bibfnamefont {J.}~\bibnamefont {Suh}}, \bibinfo
  {author} {\bibfnamefont {J.}~\bibnamefont {Park}}, \bibinfo {author}
  {\bibfnamefont {C.}~\bibnamefont {Ko}}, \bibinfo {author} {\bibfnamefont
  {H.~S.}\ \bibnamefont {Choe}}, \bibinfo {author} {\bibfnamefont
  {J.}~\bibnamefont {Avila}}, \bibinfo {author} {\bibfnamefont
  {M.}~\bibnamefont {Zhong}}, \bibinfo {author} {\bibfnamefont
  {Z.}~\bibnamefont {Wei}}, \bibinfo {author} {\bibfnamefont {J.}~\bibnamefont
  {Li}}, \bibinfo {author} {\bibfnamefont {S.}~\bibnamefont {Li}}, \bibinfo
  {author} {\bibfnamefont {H.}~\bibnamefont {Gao}}, \bibinfo {author}
  {\bibfnamefont {Y.}~\bibnamefont {Liu}}, \bibinfo {author} {\bibfnamefont
  {J.}~\bibnamefont {Analytis}}, \bibinfo {author} {\bibfnamefont
  {Q.}~\bibnamefont {Xia}}, \bibinfo {author} {\bibfnamefont {M.~C.}\
  \bibnamefont {Asensio}},\ and\ \bibinfo {author} {\bibfnamefont
  {J.}~\bibnamefont {Wu}},\ }\href
  {https://doi.org/https://doi.org/10.1002/adma.201800754} {\bibfield
  {journal} {\bibinfo  {journal} {Adv. Mater.}\ }\textbf {\bibinfo {volume}
  {30}},\ \bibinfo {pages} {1800754} (\bibinfo {year} {2018})}\BibitemShut
  {NoStop}%
\bibitem [{\citenamefont {Shah}\ \emph {et~al.}(2020)\citenamefont {Shah},
  \citenamefont {Wang}, \citenamefont {Sohail},\ and\ \citenamefont
  {Uhrberg}}]{blueAs1}%
  \BibitemOpen
  \bibfield  {author} {\bibinfo {author} {\bibfnamefont {J.}~\bibnamefont
  {Shah}}, \bibinfo {author} {\bibfnamefont {W.}~\bibnamefont {Wang}}, \bibinfo
  {author} {\bibfnamefont {H.~M.}\ \bibnamefont {Sohail}},\ and\ \bibinfo
  {author} {\bibfnamefont {R.}~\bibnamefont {Uhrberg}},\ }\href
  {https://doi.org/https://doi.org/10.1088/2053-1583/ab64fb} {\bibfield
  {journal} {\bibinfo  {journal} {2D Mater.}\ }\textbf {\bibinfo {volume}
  {7}},\ \bibinfo {pages} {025013} (\bibinfo {year} {2020})}\BibitemShut
  {NoStop}%
\bibitem [{\citenamefont {Bao}\ \emph {et~al.}(2021)\citenamefont {Bao},
  \citenamefont {Zhang}, \citenamefont {Zhang}, \citenamefont {Wu},
  \citenamefont {Luo}, \citenamefont {Zhou}, \citenamefont {Li}, \citenamefont
  {Hou}, \citenamefont {Yao}, \citenamefont {Liu}, \citenamefont {Yu},
  \citenamefont {Li}, \citenamefont {Duan}, \citenamefont {Yao}, \citenamefont
  {Wang},\ and\ \citenamefont {Zhou}}]{Kekule1}%
  \BibitemOpen
  \bibfield  {author} {\bibinfo {author} {\bibfnamefont {C.}~\bibnamefont
  {Bao}}, \bibinfo {author} {\bibfnamefont {H.}~\bibnamefont {Zhang}}, \bibinfo
  {author} {\bibfnamefont {T.}~\bibnamefont {Zhang}}, \bibinfo {author}
  {\bibfnamefont {X.}~\bibnamefont {Wu}}, \bibinfo {author} {\bibfnamefont
  {L.}~\bibnamefont {Luo}}, \bibinfo {author} {\bibfnamefont {S.}~\bibnamefont
  {Zhou}}, \bibinfo {author} {\bibfnamefont {Q.}~\bibnamefont {Li}}, \bibinfo
  {author} {\bibfnamefont {Y.}~\bibnamefont {Hou}}, \bibinfo {author}
  {\bibfnamefont {W.}~\bibnamefont {Yao}}, \bibinfo {author} {\bibfnamefont
  {L.}~\bibnamefont {Liu}}, \bibinfo {author} {\bibfnamefont {P.}~\bibnamefont
  {Yu}}, \bibinfo {author} {\bibfnamefont {J.}~\bibnamefont {Li}}, \bibinfo
  {author} {\bibfnamefont {W.}~\bibnamefont {Duan}}, \bibinfo {author}
  {\bibfnamefont {H.}~\bibnamefont {Yao}}, \bibinfo {author} {\bibfnamefont
  {Y.}~\bibnamefont {Wang}},\ and\ \bibinfo {author} {\bibfnamefont
  {S.}~\bibnamefont {Zhou}},\ }\href
  {https://doi.org/10.1103/PhysRevLett.126.206804} {\bibfield  {journal}
  {\bibinfo  {journal} {Phys. Rev. Lett.}\ }\textbf {\bibinfo {volume} {126}},\
  \bibinfo {pages} {206804} (\bibinfo {year} {2021})}\BibitemShut {NoStop}%
\bibitem [{\citenamefont {Gomes}\ \emph {et~al.}(2012)\citenamefont {Gomes},
  \citenamefont {Mar}, \citenamefont {Ko}, \citenamefont {Guinea},\ and\
  \citenamefont {Manoharan}}]{kekuexp1}%
  \BibitemOpen
  \bibfield  {author} {\bibinfo {author} {\bibfnamefont {K.~K.}\ \bibnamefont
  {Gomes}}, \bibinfo {author} {\bibfnamefont {W.}~\bibnamefont {Mar}}, \bibinfo
  {author} {\bibfnamefont {W.}~\bibnamefont {Ko}}, \bibinfo {author}
  {\bibfnamefont {F.}~\bibnamefont {Guinea}},\ and\ \bibinfo {author}
  {\bibfnamefont {H.~C.}\ \bibnamefont {Manoharan}},\ }\href
  {https://doi.org/https://doi.org/10.1038/nature10941} {\bibfield  {journal}
  {\bibinfo  {journal} {Nature}\ }\textbf {\bibinfo {volume} {483}},\ \bibinfo
  {pages} {306} (\bibinfo {year} {2012})}\BibitemShut {NoStop}%
\bibitem [{\citenamefont {Giovannetti}\ \emph {et~al.}(2015)\citenamefont
  {Giovannetti}, \citenamefont {Capone}, \citenamefont {van~den Brink},\ and\
  \citenamefont {Ortix}}]{kekuexp2}%
  \BibitemOpen
  \bibfield  {author} {\bibinfo {author} {\bibfnamefont {G.}~\bibnamefont
  {Giovannetti}}, \bibinfo {author} {\bibfnamefont {M.}~\bibnamefont {Capone}},
  \bibinfo {author} {\bibfnamefont {J.}~\bibnamefont {van~den Brink}},\ and\
  \bibinfo {author} {\bibfnamefont {C.}~\bibnamefont {Ortix}},\ }\href
  {https://doi.org/10.1103/PhysRevB.91.121417} {\bibfield  {journal} {\bibinfo
  {journal} {Phys. Rev. B}\ }\textbf {\bibinfo {volume} {91}},\ \bibinfo
  {pages} {121417} (\bibinfo {year} {2015})}\BibitemShut {NoStop}%
\bibitem [{\citenamefont {Lin}\ \emph {et~al.}(2017)\citenamefont {Lin},
  \citenamefont {Qin}, \citenamefont {Zeng}, \citenamefont {Chen},
  \citenamefont {Cui}, \citenamefont {Cho}, \citenamefont {Qiao},\ and\
  \citenamefont {Zhang}}]{kekuexp3}%
  \BibitemOpen
  \bibfield  {author} {\bibinfo {author} {\bibfnamefont {Z.}~\bibnamefont
  {Lin}}, \bibinfo {author} {\bibfnamefont {W.}~\bibnamefont {Qin}}, \bibinfo
  {author} {\bibfnamefont {J.}~\bibnamefont {Zeng}}, \bibinfo {author}
  {\bibfnamefont {W.}~\bibnamefont {Chen}}, \bibinfo {author} {\bibfnamefont
  {P.}~\bibnamefont {Cui}}, \bibinfo {author} {\bibfnamefont {J.-H.}\
  \bibnamefont {Cho}}, \bibinfo {author} {\bibfnamefont {Z.}~\bibnamefont
  {Qiao}},\ and\ \bibinfo {author} {\bibfnamefont {Z.}~\bibnamefont {Zhang}},\
  }\href {https://doi.org/https://doi.org/10.1021/acs.nanolett.6b05354}
  {\bibfield  {journal} {\bibinfo  {journal} {Nano lett.}\ }\textbf {\bibinfo
  {volume} {17}},\ \bibinfo {pages} {4013} (\bibinfo {year}
  {2017})}\BibitemShut {NoStop}%
\bibitem [{\citenamefont {Zhao}\ \emph {et~al.}(2016)\citenamefont {Zhao},
  \citenamefont {Schnyder},\ and\ \citenamefont {Wang}}]{PT}%
  \BibitemOpen
  \bibfield  {author} {\bibinfo {author} {\bibfnamefont {Y.}~\bibnamefont
  {Zhao}}, \bibinfo {author} {\bibfnamefont {A.~P.}\ \bibnamefont {Schnyder}},\
  and\ \bibinfo {author} {\bibfnamefont {Z.}~\bibnamefont {Wang}},\ }\href
  {https://doi.org/https://doi.org/10.1103/PhysRevLett.116.156402} {\bibfield
  {journal} {\bibinfo  {journal} {Rev. Rev. Lett.}\ }\textbf {\bibinfo {volume}
  {116}},\ \bibinfo {pages} {156402} (\bibinfo {year} {2016})}\BibitemShut
  {NoStop}%
\bibitem [{\citenamefont {Zhao}\ and\ \citenamefont {Lu}(2017)}]{PT1}%
  \BibitemOpen
  \bibfield  {author} {\bibinfo {author} {\bibfnamefont {Y.}~\bibnamefont
  {Zhao}}\ and\ \bibinfo {author} {\bibfnamefont {Y.}~\bibnamefont {Lu}},\
  }\href {https://doi.org/10.1103/physrevlett.118.056401} {\bibfield  {journal}
  {\bibinfo  {journal} {Rev. Rev. Lett.}\ }\textbf {\bibinfo {volume} {118}},\
  \bibinfo {pages} {056401} (\bibinfo {year} {2017})}\BibitemShut {NoStop}%
\bibitem [{\citenamefont {Ahn}\ \emph {et~al.}(2019)\citenamefont {Ahn},
  \citenamefont {Park}, \citenamefont {Kim}, \citenamefont {Kim},\ and\
  \citenamefont {Yang}}]{PT2}%
  \BibitemOpen
  \bibfield  {author} {\bibinfo {author} {\bibfnamefont {J.}~\bibnamefont
  {Ahn}}, \bibinfo {author} {\bibfnamefont {S.}~\bibnamefont {Park}}, \bibinfo
  {author} {\bibfnamefont {D.}~\bibnamefont {Kim}}, \bibinfo {author}
  {\bibfnamefont {Y.}~\bibnamefont {Kim}},\ and\ \bibinfo {author}
  {\bibfnamefont {B.-J.}\ \bibnamefont {Yang}},\ }\href
  {https://doi.org/10.1088/1674-1056/ab4d3b} {\bibfield  {journal} {\bibinfo
  {journal} {Chin. Phys. B}\ }\textbf {\bibinfo {volume} {28}},\ \bibinfo
  {pages} {117101} (\bibinfo {year} {2019})}\BibitemShut {NoStop}%
\bibitem [{\citenamefont {Ahn}\ \emph {et~al.}(2018)\citenamefont {Ahn},
  \citenamefont {Kim}, \citenamefont {Kim},\ and\ \citenamefont {Yang}}]{PT3}%
  \BibitemOpen
  \bibfield  {author} {\bibinfo {author} {\bibfnamefont {J.}~\bibnamefont
  {Ahn}}, \bibinfo {author} {\bibfnamefont {D.}~\bibnamefont {Kim}}, \bibinfo
  {author} {\bibfnamefont {Y.}~\bibnamefont {Kim}},\ and\ \bibinfo {author}
  {\bibfnamefont {B.-J.}\ \bibnamefont {Yang}},\ }\href
  {https://doi.org/10.1103/physrevlett.121.106403} {\bibfield  {journal}
  {\bibinfo  {journal} {Rev. Rev. Lett.}\ }\textbf {\bibinfo {volume} {121}},\
  \bibinfo {pages} {106403} (\bibinfo {year} {2018})}\BibitemShut {NoStop}%
\bibitem [{\citenamefont {Fu}\ and\ \citenamefont {Kane}(2007)}]{parity}%
  \BibitemOpen
  \bibfield  {author} {\bibinfo {author} {\bibfnamefont {L.}~\bibnamefont
  {Fu}}\ and\ \bibinfo {author} {\bibfnamefont {C.~L.}\ \bibnamefont {Kane}},\
  }\href {https://doi.org/10.1103/physrevb.76.045302} {\bibfield  {journal}
  {\bibinfo  {journal} {Phys. Rev. B}\ }\textbf {\bibinfo {volume} {76}},\
  \bibinfo {pages} {045302} (\bibinfo {year} {2007})}\BibitemShut {NoStop}%
\bibitem [{\citenamefont {Wieder}\ and\ \citenamefont
  {Bernevig}(2018)}]{axion1}%
  \BibitemOpen
  \bibfield  {author} {\bibinfo {author} {\bibfnamefont {B.~J.}\ \bibnamefont
  {Wieder}}\ and\ \bibinfo {author} {\bibfnamefont {B.~A.}\ \bibnamefont
  {Bernevig}},\ }\href {https://arxiv.org/abs/1810.02373} {\bibfield  {journal}
  {\bibinfo  {journal} {arXiv:1810.02373}\ } (\bibinfo {year}
  {2018})}\BibitemShut {NoStop}%
\bibitem [{Sup()}]{SuppMater}%
  \BibitemOpen
  \href@noop {} {\bibinfo  {journal} {See Supplemental Material for (I)
  Calculation methods, (II) Projection band structures of graphane(CH) and
  graphene fluoride(CF), (III) Model I: Eight-band TB model for planar or
  buckled hexagonal lattices, (IV) Model II: Anti-Kekul\'e and Kekul\'e
  distortion hexagonal lattice model, (V) Wilson loop and nested Wilson loop,
  (VI) The geometry structures, bulk bands, edge states, Wilson loop spectra,
  energy spectra of hexagonal finite-size flakes, and charge spatial
  distribution of corner states of the SOTI material candidates (2D hexagonal
  group IV hydrides/halides, group V materials), (VII) The $p_z$ orbitals and
  $sp^2$ orbitals decomposition of SOTI anti-Kekul\'e and Kekul\'e distortion
  graphenes, (VIII) The geometry structures, bulk bands, edge states, Wilson
  loop spectra, energy spectra of hexagonal finite-size flakes, and charge
  spatial distribution of corner states of the SOTI material candidates
  (anti-Kekul\'e and Kekul\'e distortion group IV materials), (IX) $k\cdot p$
  Hamiltonian and edge theory, (X) Corner charge, second Stiefel-Whitney
  number, nested Wilson loop and bulk gaps of all the SOTI material candidates,
  (XI) Charge spatial distribution of corner states, (XII) Numerical
  calculation of corner charge, which includes Refs. \cite{PBE, VASP, wannier1,
  wannier2, wannier3, qe, irvsp, silicene1, silicene2, skpara, JRmodel}}\
  }\BibitemShut {NoStop}%
\bibitem [{\citenamefont {Perdew}\ \emph {et~al.}(1996)\citenamefont {Perdew},
  \citenamefont {Burke},\ and\ \citenamefont {Ernzerhof}}]{PBE}%
  \BibitemOpen
  \bibfield  {author} {\bibinfo {author} {\bibfnamefont {J.~P.}\ \bibnamefont
  {Perdew}}, \bibinfo {author} {\bibfnamefont {K.}~\bibnamefont {Burke}},\ and\
  \bibinfo {author} {\bibfnamefont {M.}~\bibnamefont {Ernzerhof}},\ }\href
  {https://doi.org/10.1103/physrevlett.77.3865} {\bibfield  {journal} {\bibinfo
   {journal} {Rev. Rev. Lett.}\ }\textbf {\bibinfo {volume} {77}},\ \bibinfo
  {pages} {3865} (\bibinfo {year} {1996})}\BibitemShut {NoStop}%
\bibitem [{\citenamefont {Kresse}\ and\ \citenamefont
  {Furthm{\"u}ller}(1996)}]{VASP}%
  \BibitemOpen
  \bibfield  {author} {\bibinfo {author} {\bibfnamefont {G.}~\bibnamefont
  {Kresse}}\ and\ \bibinfo {author} {\bibfnamefont {J.}~\bibnamefont
  {Furthm{\"u}ller}},\ }\href {https://doi.org/10.1103/physrevb.54.11169}
  {\bibfield  {journal} {\bibinfo  {journal} {Rev. Rev. B}\ }\textbf {\bibinfo
  {volume} {54}},\ \bibinfo {pages} {11169} (\bibinfo {year}
  {1996})}\BibitemShut {NoStop}%
\bibitem [{\citenamefont {Mostofi}\ \emph {et~al.}(2008)\citenamefont
  {Mostofi}, \citenamefont {Yates}, \citenamefont {Lee}, \citenamefont {Souza},
  \citenamefont {Vanderbilt},\ and\ \citenamefont {Marzari}}]{wannier1}%
  \BibitemOpen
  \bibfield  {author} {\bibinfo {author} {\bibfnamefont {A.~A.}\ \bibnamefont
  {Mostofi}}, \bibinfo {author} {\bibfnamefont {J.~R.}\ \bibnamefont {Yates}},
  \bibinfo {author} {\bibfnamefont {Y.-S.}\ \bibnamefont {Lee}}, \bibinfo
  {author} {\bibfnamefont {I.}~\bibnamefont {Souza}}, \bibinfo {author}
  {\bibfnamefont {D.}~\bibnamefont {Vanderbilt}},\ and\ \bibinfo {author}
  {\bibfnamefont {N.}~\bibnamefont {Marzari}},\ }\href
  {https://doi.org/10.1016/j.cpc.2007.11.016} {\bibfield  {journal} {\bibinfo
  {journal} {Comput. Phys. Commun.}\ }\textbf {\bibinfo {volume} {178}},\
  \bibinfo {pages} {685} (\bibinfo {year} {2008})}\BibitemShut {NoStop}%
\bibitem [{\citenamefont {Marzari}\ and\ \citenamefont
  {Vanderbilt}(1997)}]{wannier2}%
  \BibitemOpen
  \bibfield  {author} {\bibinfo {author} {\bibfnamefont {N.}~\bibnamefont
  {Marzari}}\ and\ \bibinfo {author} {\bibfnamefont {D.}~\bibnamefont
  {Vanderbilt}},\ }\href {https://doi.org/10.1103/physrevb.56.12847} {\bibfield
   {journal} {\bibinfo  {journal} {Rev. Rev. B}\ }\textbf {\bibinfo {volume}
  {56}},\ \bibinfo {pages} {12847} (\bibinfo {year} {1997})}\BibitemShut
  {NoStop}%
\bibitem [{\citenamefont {Souza}\ \emph {et~al.}(2001)\citenamefont {Souza},
  \citenamefont {Marzari},\ and\ \citenamefont {Vanderbilt}}]{wannier3}%
  \BibitemOpen
  \bibfield  {author} {\bibinfo {author} {\bibfnamefont {I.}~\bibnamefont
  {Souza}}, \bibinfo {author} {\bibfnamefont {N.}~\bibnamefont {Marzari}},\
  and\ \bibinfo {author} {\bibfnamefont {D.}~\bibnamefont {Vanderbilt}},\
  }\href {https://doi.org/10.1103/physrevb.65.035109} {\bibfield  {journal}
  {\bibinfo  {journal} {Rev. Rev. B}\ }\textbf {\bibinfo {volume} {65}},\
  \bibinfo {pages} {035109} (\bibinfo {year} {2001})}\BibitemShut {NoStop}%
\bibitem [{\citenamefont {Giannozzi}\ \emph {et~al.}(2009)\citenamefont
  {Giannozzi}, \citenamefont {Baroni}, \citenamefont {Bonini}, \citenamefont
  {Calandra}, \citenamefont {Car}, \citenamefont {Cavazzoni}, \citenamefont
  {Ceresoli}, \citenamefont {Chiarotti}, \citenamefont {Cococcioni},
  \citenamefont {Dabo} \emph {et~al.}}]{qe}%
  \BibitemOpen
  \bibfield  {author} {\bibinfo {author} {\bibfnamefont {P.}~\bibnamefont
  {Giannozzi}}, \bibinfo {author} {\bibfnamefont {S.}~\bibnamefont {Baroni}},
  \bibinfo {author} {\bibfnamefont {N.}~\bibnamefont {Bonini}}, \bibinfo
  {author} {\bibfnamefont {M.}~\bibnamefont {Calandra}}, \bibinfo {author}
  {\bibfnamefont {R.}~\bibnamefont {Car}}, \bibinfo {author} {\bibfnamefont
  {C.}~\bibnamefont {Cavazzoni}}, \bibinfo {author} {\bibfnamefont
  {D.}~\bibnamefont {Ceresoli}}, \bibinfo {author} {\bibfnamefont {G.~L.}\
  \bibnamefont {Chiarotti}}, \bibinfo {author} {\bibfnamefont {M.}~\bibnamefont
  {Cococcioni}}, \bibinfo {author} {\bibfnamefont {I.}~\bibnamefont {Dabo}},
  \emph {et~al.},\ }\href {https://doi.org/10.1088/0953-8984/21/39/395502}
  {\bibfield  {journal} {\bibinfo  {journal} {J. Phys. Condens. Matter.}\
  }\textbf {\bibinfo {volume} {21}},\ \bibinfo {pages} {395502} (\bibinfo
  {year} {2009})}\BibitemShut {NoStop}%
\bibitem [{\citenamefont {Gao}\ \emph {et~al.}(2021)\citenamefont {Gao},
  \citenamefont {Wu}, \citenamefont {Persson},\ and\ \citenamefont
  {Wang}}]{irvsp}%
  \BibitemOpen
  \bibfield  {author} {\bibinfo {author} {\bibfnamefont {J.}~\bibnamefont
  {Gao}}, \bibinfo {author} {\bibfnamefont {Q.}~\bibnamefont {Wu}}, \bibinfo
  {author} {\bibfnamefont {C.}~\bibnamefont {Persson}},\ and\ \bibinfo {author}
  {\bibfnamefont {Z.}~\bibnamefont {Wang}},\ }\href
  {https://doi.org/10.1016/j.cpc.2020.107760} {\bibfield  {journal} {\bibinfo
  {journal} {Comput. Phys. Commun.}\ }\textbf {\bibinfo {volume} {261}},\
  \bibinfo {pages} {107760} (\bibinfo {year} {2021})}\BibitemShut {NoStop}%
\bibitem [{\citenamefont {Liu}\ \emph {et~al.}(2011{\natexlab{a}})\citenamefont
  {Liu}, \citenamefont {Feng},\ and\ \citenamefont {Yao}}]{silicene1}%
  \BibitemOpen
  \bibfield  {author} {\bibinfo {author} {\bibfnamefont {C.-C.}\ \bibnamefont
  {Liu}}, \bibinfo {author} {\bibfnamefont {W.}~\bibnamefont {Feng}},\ and\
  \bibinfo {author} {\bibfnamefont {Y.}~\bibnamefont {Yao}},\ }\href
  {https://doi.org/10.1103/physrevlett.107.076802} {\bibfield  {journal}
  {\bibinfo  {journal} {Rev. Rev. Lett.}\ }\textbf {\bibinfo {volume} {107}},\
  \bibinfo {pages} {076802} (\bibinfo {year} {2011}{\natexlab{a}})}\BibitemShut
  {NoStop}%
\bibitem [{\citenamefont {Liu}\ \emph {et~al.}(2011{\natexlab{b}})\citenamefont
  {Liu}, \citenamefont {Jiang},\ and\ \citenamefont {Yao}}]{silicene2}%
  \BibitemOpen
  \bibfield  {author} {\bibinfo {author} {\bibfnamefont {C.-C.}\ \bibnamefont
  {Liu}}, \bibinfo {author} {\bibfnamefont {H.}~\bibnamefont {Jiang}},\ and\
  \bibinfo {author} {\bibfnamefont {Y.}~\bibnamefont {Yao}},\ }\href
  {https://doi.org/10.1103/physrevb.84.195430} {\bibfield  {journal} {\bibinfo
  {journal} {Rev. Rev. B}\ }\textbf {\bibinfo {volume} {84}},\ \bibinfo {pages}
  {195430} (\bibinfo {year} {2011}{\natexlab{b}})}\BibitemShut {NoStop}%
\bibitem [{\citenamefont {Dresselhaus}\ \emph {et~al.}(1998)\citenamefont
  {Dresselhaus}, \citenamefont {Dresselhaus},\ and\ \citenamefont
  {Saito}}]{skpara}%
  \BibitemOpen
  \bibfield  {author} {\bibinfo {author} {\bibfnamefont {G.}~\bibnamefont
  {Dresselhaus}}, \bibinfo {author} {\bibfnamefont {M.~S.}\ \bibnamefont
  {Dresselhaus}},\ and\ \bibinfo {author} {\bibfnamefont {R.}~\bibnamefont
  {Saito}},\ }\href@noop {} {\emph {\bibinfo {title} {Physical properties of
  carbon nanotubes}}}\ (\bibinfo  {publisher} {World scientific},\ \bibinfo
  {year} {1998})\BibitemShut {NoStop}%
\bibitem [{\citenamefont {Jackiw}\ and\ \citenamefont {Rebbi}(1976)}]{JRmodel}%
  \BibitemOpen
  \bibfield  {author} {\bibinfo {author} {\bibfnamefont {R.}~\bibnamefont
  {Jackiw}}\ and\ \bibinfo {author} {\bibfnamefont {C.}~\bibnamefont {Rebbi}},\
  }\href {https://doi.org/10.1103/PhysRevD.13.3398} {\bibfield  {journal}
  {\bibinfo  {journal} {Phys. Rev. D}\ }\textbf {\bibinfo {volume} {13}},\
  \bibinfo {pages} {3398} (\bibinfo {year} {1976})}\BibitemShut {NoStop}%
\bibitem [{\citenamefont {Ezawa}(2018{\natexlab{b}})}]{blackPmodel1}%
  \BibitemOpen
\bibfield  {journal} {  }\bibfield  {author} {\bibinfo {author} {\bibfnamefont
  {M.}~\bibnamefont {Ezawa}},\ }\href
  {https://doi.org/10.1103/PhysRevB.98.045125} {\bibfield  {journal} {\bibinfo
  {journal} {Phys. Rev. B}\ }\textbf {\bibinfo {volume} {98}},\ \bibinfo
  {pages} {045125} (\bibinfo {year} {2018}{\natexlab{b}})}\BibitemShut
  {NoStop}%
\bibitem [{\citenamefont {Hitomi}\ \emph {et~al.}(2021)\citenamefont {Hitomi},
  \citenamefont {Kawakami},\ and\ \citenamefont {Koshino}}]{blackPmodel2}%
  \BibitemOpen
  \bibfield  {author} {\bibinfo {author} {\bibfnamefont {M.}~\bibnamefont
  {Hitomi}}, \bibinfo {author} {\bibfnamefont {T.}~\bibnamefont {Kawakami}},\
  and\ \bibinfo {author} {\bibfnamefont {M.}~\bibnamefont {Koshino}},\ }\href
  {https://arxiv.org/abs/2105.05402} {\bibfield  {journal} {\bibinfo  {journal}
  {arXiv: 2105.05402}\ } (\bibinfo {year} {2021})}\BibitemShut {NoStop}%
\bibitem [{\citenamefont {Benalcazar}\ \emph {et~al.}(2019)\citenamefont
  {Benalcazar}, \citenamefont {Li},\ and\ \citenamefont
  {Hughes}}]{kekulemodel3}%
  \BibitemOpen
  \bibfield  {author} {\bibinfo {author} {\bibfnamefont {W.~A.}\ \bibnamefont
  {Benalcazar}}, \bibinfo {author} {\bibfnamefont {T.}~\bibnamefont {Li}},\
  and\ \bibinfo {author} {\bibfnamefont {T.~L.}\ \bibnamefont {Hughes}},\
  }\href {https://doi.org/10.1103/PhysRevB.99.245151} {\bibfield  {journal}
  {\bibinfo  {journal} {Phys. Rev. B}\ }\textbf {\bibinfo {volume} {99}},\
  \bibinfo {pages} {245151} (\bibinfo {year} {2019})}\BibitemShut {NoStop}%
\bibitem [{\citenamefont {Bradlyn}\ \emph {et~al.}(2017)\citenamefont
  {Bradlyn}, \citenamefont {Elcoro}, \citenamefont {Cano}, \citenamefont
  {Vergniory}, \citenamefont {Wang}, \citenamefont {Felser}, \citenamefont
  {Aroyo},\ and\ \citenamefont {Bernevig}}]{ebrs}%
  \BibitemOpen
  \bibfield  {author} {\bibinfo {author} {\bibfnamefont {B.}~\bibnamefont
  {Bradlyn}}, \bibinfo {author} {\bibfnamefont {L.}~\bibnamefont {Elcoro}},
  \bibinfo {author} {\bibfnamefont {J.}~\bibnamefont {Cano}}, \bibinfo {author}
  {\bibfnamefont {M.}~\bibnamefont {Vergniory}}, \bibinfo {author}
  {\bibfnamefont {Z.}~\bibnamefont {Wang}}, \bibinfo {author} {\bibfnamefont
  {C.}~\bibnamefont {Felser}}, \bibinfo {author} {\bibfnamefont
  {M.}~\bibnamefont {Aroyo}},\ and\ \bibinfo {author} {\bibfnamefont {B.~A.}\
  \bibnamefont {Bernevig}},\ }\href {https://doi.org/10.1038/nature23268}
  {\bibfield  {journal} {\bibinfo  {journal} {Nature}\ }\textbf {\bibinfo
  {volume} {547}},\ \bibinfo {pages} {298} (\bibinfo {year}
  {2017})}\BibitemShut {NoStop}%
\bibitem [{not({\natexlab{a}})}]{note1}%
  \BibitemOpen
  \href@noop {} {\bibfield  {journal} {\bibinfo  {journal} {The
  Anti-Kekul\'eSi/Ge/Sn have trivial second Stiefel-Whitney number but still
  have corner states \cite{SuppMater}}\ } }\BibitemShut
  {NoStop}%
\bibitem [{not({\natexlab{b}})}]{note2}%
  \BibitemOpen
  \href@noop {} {\bibfield  {journal} {\bibinfo  {journal} {The Kekul\'eGe/Sn
  have nontrivial second Stiefel-Whitney number. However, since the bulk band
  gaps are too small and the dispersion of edge states are too large, the
  corner states are mixed with bulk states or edge states \cite{SuppMater}}\ }
  }\BibitemShut {NoStop}%
\bibitem [{not({\natexlab{c}})}]{note3}%
  \BibitemOpen
  \href@noop {} {\bibfield  {journal} {\bibinfo  {journal} {Recently proposed
  SOTI materials graphdiyne \cite{graphdiyne1, graphdiyne2} and graphyne
  \cite{graphyne1, graphyne2}, whose gaps are induced by the band folding and
  alternate equivalent C-C bond lengths, can be considered as belonging to
  Mechanism II of anti-Kekul\'e and Kekul\'e distortion}\ }
  }\BibitemShut {NoStop}%
\end{thebibliography}%
	
	\clearpage 
	\renewcommand\thefigure{S\arabic{figure}}
	\renewcommand\thetable{S\arabic{table}}
	\setcounter{equation}{0}
	\setcounter{figure}{0}
	\setcounter{table}{0}
	\onecolumngrid
	\begin{center}
		\textbf{\large Supplemental Material \textemdash{} Second Order Topological Insulator State in Hexagonal Lattices and its Material Realization}
	\end{center}
	~\\
	~\\
	~\\
	~\\
	~\\
	\twocolumngrid
	\tableofcontents
	
	\section{Calculation methods}
	The calculations of the band structures were performed using DFT in the Perdew-Becke-Ernzerhof (PBE) generalized gradient approximation (GGA) \cite{PBE} implemented in the Vienna $ab$ $initio$ simulation package (VASP) \cite{VASP}. The plane-wave energy cutoff is set to 600 eV and the Brillouin zone is sampled by a 12 $\times$ 12 $\times$ 1 mesh. The vacuum layer is up to 20 \AA\ to avoid the interactions between the layers. The symmetry adapted Wannier functions are constructed using the WANNIER90 code \cite{wannier1, wannier2, wannier3} and quantum espresso \cite{qe}. Based on the constructed Wannier functions, we obtain the edge states, corner states and Wilson loop. The parity eigenvalues and irreducible representations of electronic states from DFT results were calculated using irvsp code \cite{irvsp}, which relies on the space-group character tables published on the Bilbao Crystallographic server. 
	
	\section{Projection band structures of graphane (CH) and graphene fluoride (CF)}
	Due to the different relative electronegativity, the $p_z$ orbitals of hexagonal group IV hydrides are pushed down the Fermi level, while those of hexagonal group IV halides are pushed up.  Figure \ref{fig:1} shows the projection band structures of graphane and graphene fluoride. The hydrogen has weaker electronegativity than that of carbon which induce the $p_z$ orbitals of carbon shift down the Fermi level, while the halide has stronger electronegativity which induce the $p_z$ orbitals shift up the Fermi level.

	\begin{figure}[H]
		\begin{center}
			\includegraphics[width=1\linewidth]{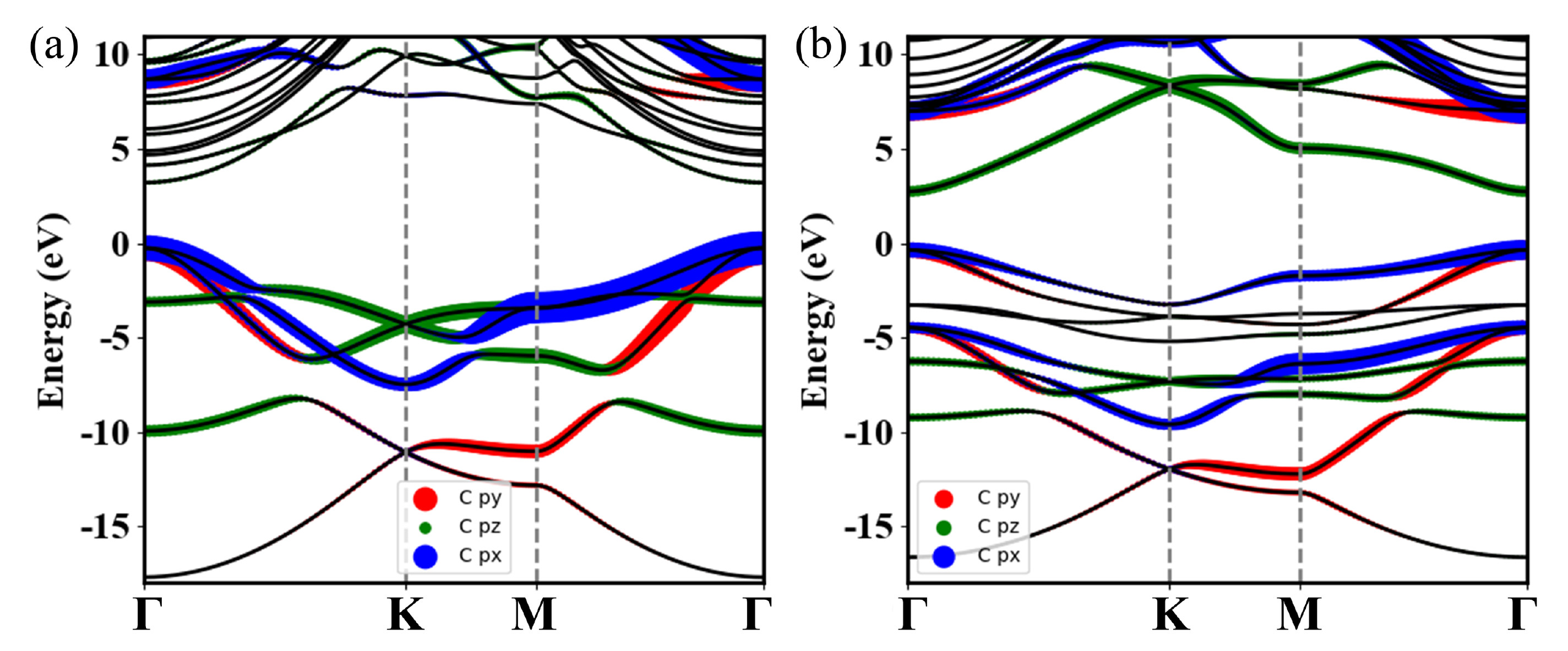}
		\end{center}
		\caption{Projection band structures of graphane (CH) and graphene fluoride (CF).} \label{fig:1}
	\end{figure}

	\section{Model I: Eight-band TB model for planar or buckled hexagonal lattice}
	For Mechanism I, we build an eight-band tight binding (TB) model with one $s$ orbital and three $p$ orbitals per site \cite{silicene1, silicene2}. In the basis of $\left\{\left|p_{z}^{A}\right\rangle,\left|p_{z}^{B}\right\rangle,\left|p_{y}^{A}\right\rangle,\left|p_{x}^{A}\right\rangle,\left|s^{A}\right\rangle,\left|p_{y}^{B}\right\rangle,\left|p_{x}^{B}\right\rangle,\left|s^{B}\right\rangle\right\}$ with  A and B labeling two sublattices, the Hamiltonian has the form
	\begin{equation}
		H_{I}=\left(\begin{array}{cc}
			\mathbf{H}_{\pi} & \mathbf{H}_{{n}} \\
			\mathbf{H}_{{n}}{ }^{\dagger} & \mathbf{H}_{\sigma}
		\end{array}\right),
	\end{equation}
	where 
	\begin{equation}
		\mathbf{H}_{\sigma}=\left(\begin{array}{cc}
			\mathbf{E} & \mathbf{T} \\
			\mathbf{T}^{\dagger} & \mathbf{E}
		\end{array}\right).
	\end{equation}
	The 2 $\times$ 2 diagonal block $\mathbf{H}_{\pi}$ and 6 $\times$ 6  $\mathbf{H}_{\sigma}$ describe the Hamiltonian for $p_z$ and  $sp^2$ ($s$, $p_x$, $p_y$), respectively. The nondiagonal block $\mathbf{H}_{n}$ is a 2 $\times$ 6 matrix which couples the $\mathbf{H}_{\pi}$ and $\mathbf{H}_{\sigma}$. The matrix elements of $\mathbf{H}_{n}$ are nonzero in the buckled structures. The matrix $\mathbf{E}$  describes the on-site energy of $\left\{p_y, p_x, s\right\}$ orbitals and can be written as
	\begin{equation}
		\mathbf{E}=\left(\begin{array}{lll}
			\epsilon_{p} & 0 & 0 \\
			0 & \epsilon_{p} & 0 \\
			0 & 0 & \epsilon_{s}
		\end{array}\right).
	\end{equation}
	$\mathbf{T}$ matrix describes the nearest neighbour hopping between A and B sublattices. Therefore, the matrix $\mathbf{T}$ and $\mathbf{H}_{n}$ can be written as 
	\begin{equation} 
		\mathbf{T}=\begin{bmatrix}t_{p_{y}^{A},p_{y}^{B}} & t_{p_{y}^{A},p_{x}^{B}} & t_{p_{y}^{A},s^{B}}\\
			t_{p_{x}^{A},p_{y}^{B}} & t_{p_{x}^{A},p_{x}^{B}} & t_{p_{x}^{A},s^{B}}\\
			t_{s^{A},p_{y}^{B}} & t_{s^{A},p_{x}^{B}} & t_{s^{A},s^{B}}
		\end{bmatrix},
	\end{equation}
	\begin{equation}
		\mathbf{H}_{n}=\begin{bmatrix}0 & 0 & 0 & t_{p_{z}^{A},p_{y}^{B}} & t_{p_{z}^{A},p_{x}^{B}} & t_{p_{z}^{A},s^{B}}\\
			t_{p_{z}^{B},p_{y}^{A}} & t_{p_{z}^{B},p_{x}^{A}} & t_{p_{z}^{B},s^{A}} & 0 & 0 & 0
		\end{bmatrix},
	\end{equation} 
	where the $t_{\alpha, \beta}$ represents the $\alpha$ orbital hopping to $\beta$ and can be written as 
	\begin{equation} \label{hopping}
		t_{\alpha, \beta}=\sum_{i=1}^{3} t_{\alpha, \beta}\left(\vec{d}_{i}\right) e^{i \vec{k} \cdot \vec{d}_{i}},
	\end{equation}
	where $\alpha$ and $\beta$ are the orbitals in different sublattices and $\vec{d}_{i}$ is the vector between the two sublattices. 
	We use Slater-Koster parameters for the TB model and represent the hopping integrals  $t_{\alpha, \beta}\left(\vec{d}_{i}\right)$ using direction cosine ($l, m, n$) and bond integral ($V$), which are given in Table. \ref{table: sk}.  
	The diagonal block $\mathbf{H}_{\pi}$ can be written as 
	\begin{equation}
		H_{\pi}=\begin{bmatrix}\ensuremath{\epsilon_{p}} & t_{p_{z}^{A},p_{z}^{B}} \\
			t_{p_{z}^{B},p_{z}^{A}} & \ensuremath{\epsilon_{p}}
		\end{bmatrix},
	\end{equation}
	where the diagonal elements $\epsilon_{p}$ are the on-site energy of $p_z$ orbitals and the non-diagonal elements represent hopping between the $p_z$ orbitals given in the form of Eq. (\ref{hopping}).  For Figs. 1 (b-d) in the text, the parameters $V_{s s \sigma}$, $V_{s p \sigma}$, $V_{p p \sigma}$ and $V_{p p \pi}$ are -2$t_1$, 2$t_1$, 2$t_1$ and -$t_1$, respectively. $t_1$ is about 3 eV for graphene \cite{skpara}. The onsite energy $\epsilon_{s}$ and $\epsilon_{p}$ of $s$ and $p$ orbitals are -3$t_1$ and 0, respectively. We can change the on-site energy of $p_z$ orbitals to push the $p_z$ orbitals up or down the Fermi level. In Figs. 1(c, d) in the text, the onsite energy energy of $p_z$ orbitals is -5$t_1$.
	
	\begin{table}[htbp]
		\centering
		{\tabcolsep0.2in
			\caption{The hopping integrals between $s$ and $p$ orbitals are described as functions of the direction cosine (${l, m, n}$).}
			\label{table: sk}
			\begin{tabular}{lc}
				\hline \hline
				$t_{s, s}$ & $V_{s s \sigma}$ \\
				$t_{x, x}$ & $l^{2} V_{p p \sigma}+\left(1-l^{2}\right) V_{p p \pi}$ \\
				$t_{s, x}$ & $l V_{s p \sigma}$ \\
				$t_{x, y}$ & $\operatorname{lm}\left(V_{p p \sigma^{-}} V_{p p \pi}\right)$ \\
				$t_{x, s}$ & $-l V_{s p \sigma}$ \\
				$t_{y, z}$ & $m n\left(V_{p p \sigma^{-}} V_{p p \pi}\right)$ \\
				\hline \hline
			\end{tabular}
		}
		\label{tab: sk}%
	\end{table}%
	
	\section{Model II: Anti-Kekul\'e and Kekul\'e distortion hexagonal lattice model}
	For Mechanism II, we consider anti-Kekul\'e and Kekul\'e distortion hexagonal lattice model. The unit cell of anti-Kekul\'e or Kekul\'e distortion hexagonal lattice is $\sqrt{3}\times\sqrt{3}$ supercell of hexagonal unit cell, as shown in Fig. 4(a) in the text. The TB Hamiltonian with one $p_z$ orbital per site reads
	\begin{equation}
		H_{II}(\mathbf{k})=\left(\begin{array}{cccccc}
			0 & t_{3} & 0 & t_{2}e^{i\mathbf{k}\cdot\mathbf{a}_{2}} & 0 & t_{3}\\
			& 0 & t_{3} & 0 & t_{2}e^{-i\mathbf{k}\cdot\mathbf{a}_{3}} & 0\\
			&  & 0 & t_{3} & 0 & t_{2}e^{-i\mathbf{k}\cdot\mathbf{a}_{1}}\\
			& \dagger &  & 0 & t_{3} & 0\\
			&  &  &  & 0 & t_{3}\\
			&  &  &  &  & 0
		\end{array}\right),
	\end{equation}
	where $\mathbf{a}_{1}$ and $\mathbf{a}_{2}$ are the unit lattice vectors, and  $\mathbf{a}_{3} = \mathbf{a}_{1}-\mathbf{a}_{2}$. $t_{2}$ and $t_{3}$ represent the inter-cell hopping and intra-cell hopping, respectively. For anti-Kekul\'e distortion, $t_3/t_2 < 1$, while $t_3/t_2 > 1$ for Kekul\'e distortion. Figure 1(e) shows band structures of the anti-Kekul\'e distortion hexagonal lattice with $t_3/t_2 = 0.3$.
	
	\section{Wilson loop and nested Wilson loop}
	One can use  Wilson loop and nested Wilson loop to calculate the second Stiefel-Whitney number $w_{2}$.
	The Wilson loop operator of $k_x$ is defined by 
	\begin{equation}
		W_{1 \left(k_{x}, k_{y}\right) \rightarrow\left(k_{x}+2 \pi, k_{y}\right)}=\lim _{N \rightarrow \infty} F_{0} F_{1} \cdots F_{N-2} F_{N-1},
	\end{equation}
	where $F_{j}$ is the overlap matrix whose matrix elements are given by 
	\begin{equation}
		\left[F_{i}\right]_{m n}=\left\langle u_{m}\left(2 \pi(i)/N, k_{y}\right) \mid u_{n}\left(2 \pi (i+1)/N, k_{y}\right)\right\rangle,
	\end{equation}
	and its spectrum is gauge-invariant. The topological information is encoded in the phase factors $\theta_{m}\left(k_{y}\right) \in(-\pi, \pi]$ of the eigenvalues $\lambda_{m}\left(k_{y}\right)$ of Wilson operator:
	\begin{equation} 
		\theta_{m}\left(k_{y}\right)=\operatorname{Im}\left[\log \lambda_{m}\left(k_{y}\right)\right]. 
	\end{equation}
	The spectrum in the $\theta-k_{y}$ diagram represents evolution of the Wanner centers. The topological classification is indicated by the number of linear crossing points on $\theta=\pi$. A spectrum corresponds to $w_{2} = 0$ ($w_{2} = 1$) when it has even (odd) linear crossing points on $\theta=\pi$. 
	
	The calculation of nested Wilson loop is similar to Wilson loop, by projecting onto the eigenstates $\left|\nu_{i}\left(k_y\right)\right\rangle$ of $W_{1 k_y}$. We use the set of eigenvectors corresponding to the sub-band centered at $\theta=\pi$, the nested Wilson loop operator $W_2$ along the $k_y$ is given by
	\begin{equation}
		W_{2 \left(k_{y}\right) \rightarrow \left(k_{y}+2 \pi\right) }=\lim _{N \rightarrow \infty} \tilde{F}_{0} \tilde{F}_{2} \cdots \tilde{F}_{N-2} \tilde{F}_{N-1},
	\end{equation}
	where 
	\begin{equation}
		\left[\tilde{F}_{i}\right]_{m n}=\left\langle\nu_{m}\left(2 \pi (i)/N\right) \mid \nu_{n}\left(2 \pi (i+1)/N\right)\right\rangle. 
	\end{equation}
	The determinant of the nested Wilson loop operator det($W_2$) is identical to $(-1)^{w_{2}}$.

	\section{The geometry structures, bulk bands, edge states, Wilson loop spectra, energy spectra of hexagonal finite-size flakes, and charge spatial distribution of corner states of the SOTI material candidates (2D hexagonal group IV hydrides/halides, group V materials)}
	Here, we demonstrate that other 2D hexagonal group IV hydrides/halides, group V materials are also SOTI material candidates with global gaps at Fermi level. Figures S2-S11 show the geometry structures, bulk bands, edge states, Wilson loop spectra, discrete energy spectra and charge distribution of corner states of these SOTI material candidates, which include graphene fluoride (CF), graphene chloride (CCl), silicene chloride (SiCl), silicene hydride (SiH), germanene hydride (GeH), stanene hydride (SnH), and blue phosphorene (BlueP), black phosphene (BlackP), blue arsenene (BlueAs), black arsenene (BlackAs). These materials all have nontrivial second Stiefel-Whitney number and hallmark corner states.
	
	\begin{figure}[H]
		\begin{center}
			\includegraphics[width=1\linewidth]{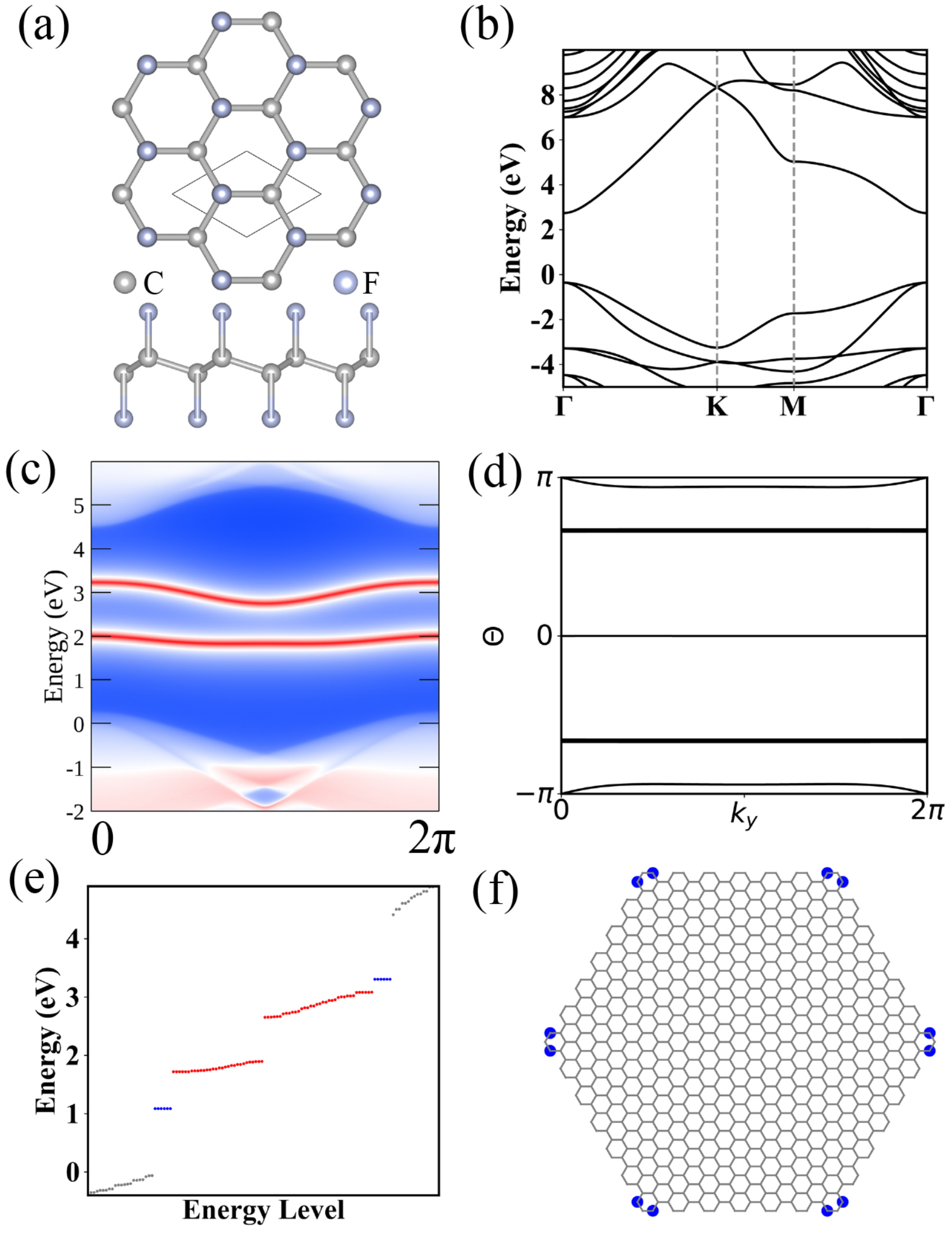}
		\end{center}
		\caption{(a) Top view and side view of geometry structures of graphene fluoride (CF). (b) (c) Bulk bands and edge states of CF from DFT calculation. (d) Wilson loop of CF. The number of crossing on $\theta=\pi$ is 1 and therefore the second Stiefel-Whitney number $w_2$ is 1. (e) Energy spectrum of hexagonal finite-size flake shown in (f). Blue, red and grey dots represent the corner, edge and bulk states, respectively. (f) Charge spatial distribution of blue states in (e).}
	\end{figure}
	
	\begin{figure}[H]
		\begin{center}
			\includegraphics[width=1\linewidth]{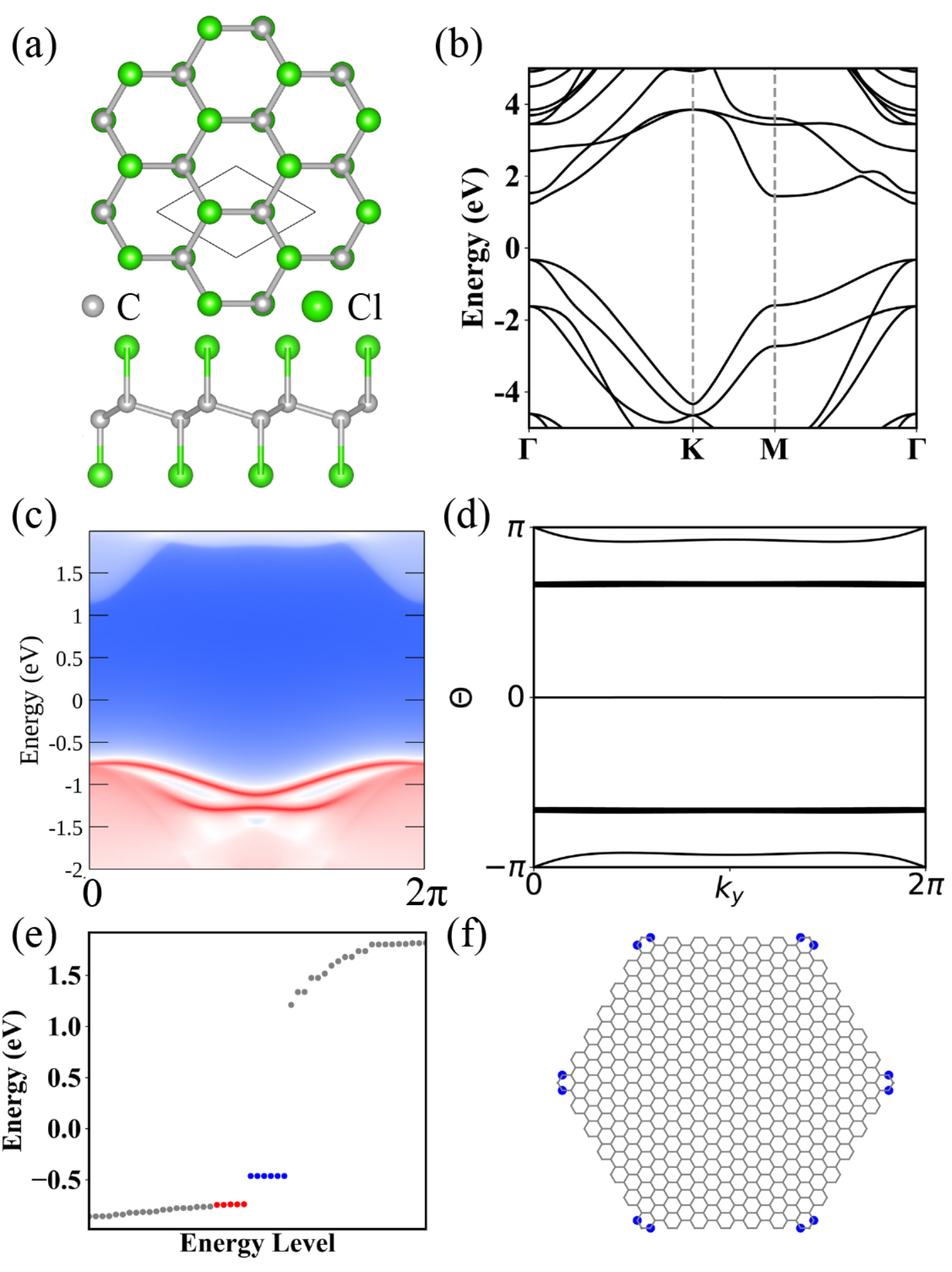}
		\end{center}
		\caption{Top view and side view of geometry structures of graphene chloride (CCl). (b) (c) Bulk bands and edge states of CCl from DFT calculation. (d) Wilson loop of CCl. The number of crossing on $\theta=\pi$ is 1 and therefore the second Stiefel-Whitney number $w_2$ is 1. (e) Energy spectrum of hexagonal finite-size flake shown in (f). Blue, red and grey dots represent the corner, edge and bulk states, respectively. (f) Charge spatial distribution of blue states in (e).}
	\end{figure}
	
	\begin{figure}[H]
		\begin{center}
			\includegraphics[width=1\linewidth]{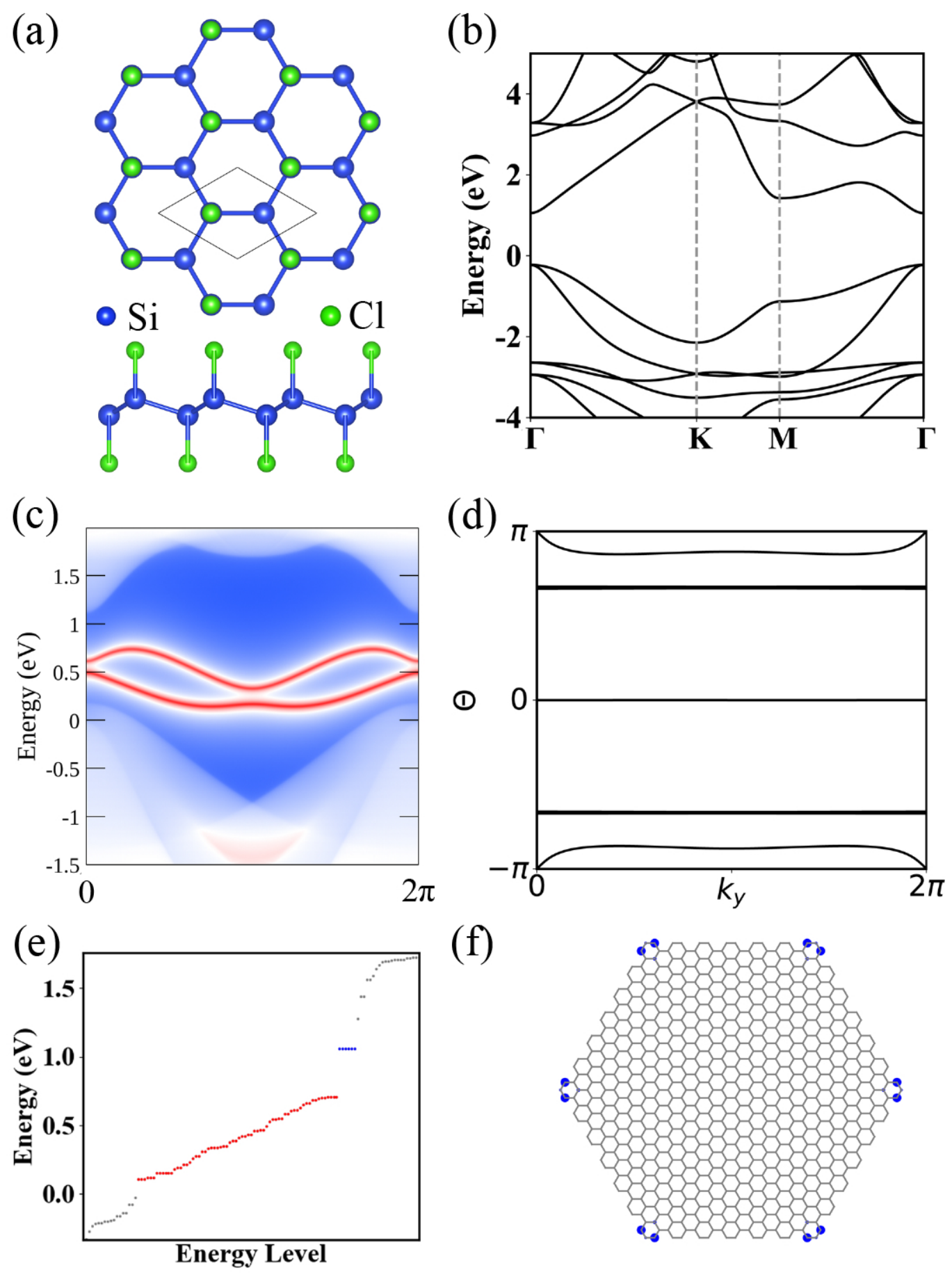}
		\end{center}
		\caption{Top view and side view of geometry structures of silicene chloride (SiCl). (b) (c) Bulk bands and edge states of SiCl from DFT calculation. (d) Wilson loop of SiCl. The number of crossing on $\theta=\pi$ is 1 and therefore the second Stiefel-Whitney number $w_2$ is 1. (e) Energy spectrum of hexagonal finite-size flake shown in (f). Blue, red and grey dots represent the corner, edge and bulk states, respectively. (f) Charge spatial distribution of blue states in (e).}
	\end{figure}
	
	\begin{figure}[H]
		\begin{center}
			\includegraphics[width=1\linewidth]{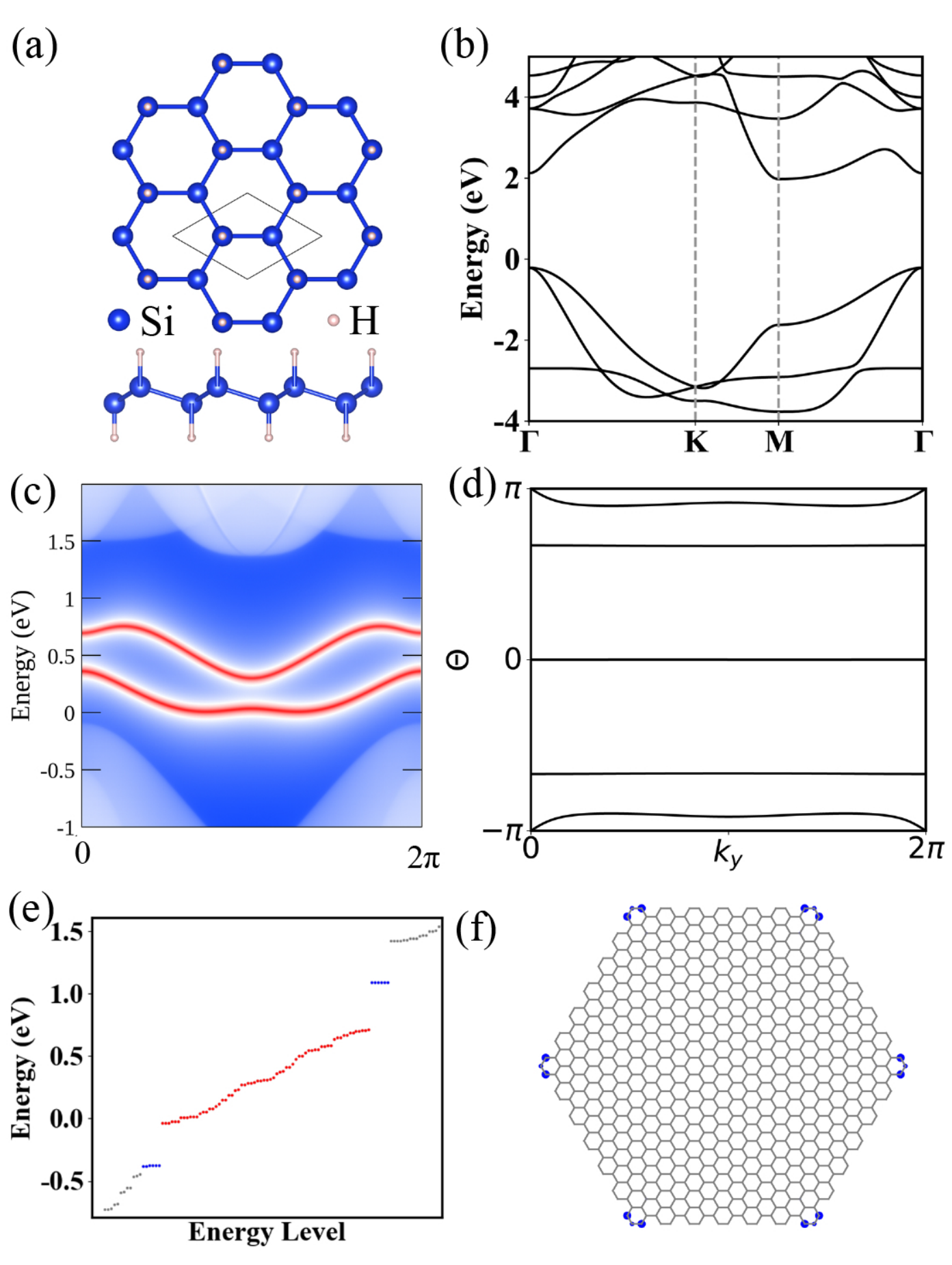}
		\end{center}
		\caption{Top view and side view of geometry structures of silicene hydride (SiH). (b) (c) Bulk bands and edge states of SiH from DFT calculation. (d) Wilson loop of SiH. The number of crossing on $\theta=\pi$ is 1 and therefore the second Stiefel-Whitney number $w_2$ is 1. (e) Energy spectrum of hexagonal finite-size flake shown in (f). Blue, red and grey dots represent the corner, edge and bulk states, respectively. (f)  Charge spatial distribution of blue states in (e).}
	\end{figure}
	
	\begin{figure}[H]
		\begin{center}
			\includegraphics[width=1\linewidth]{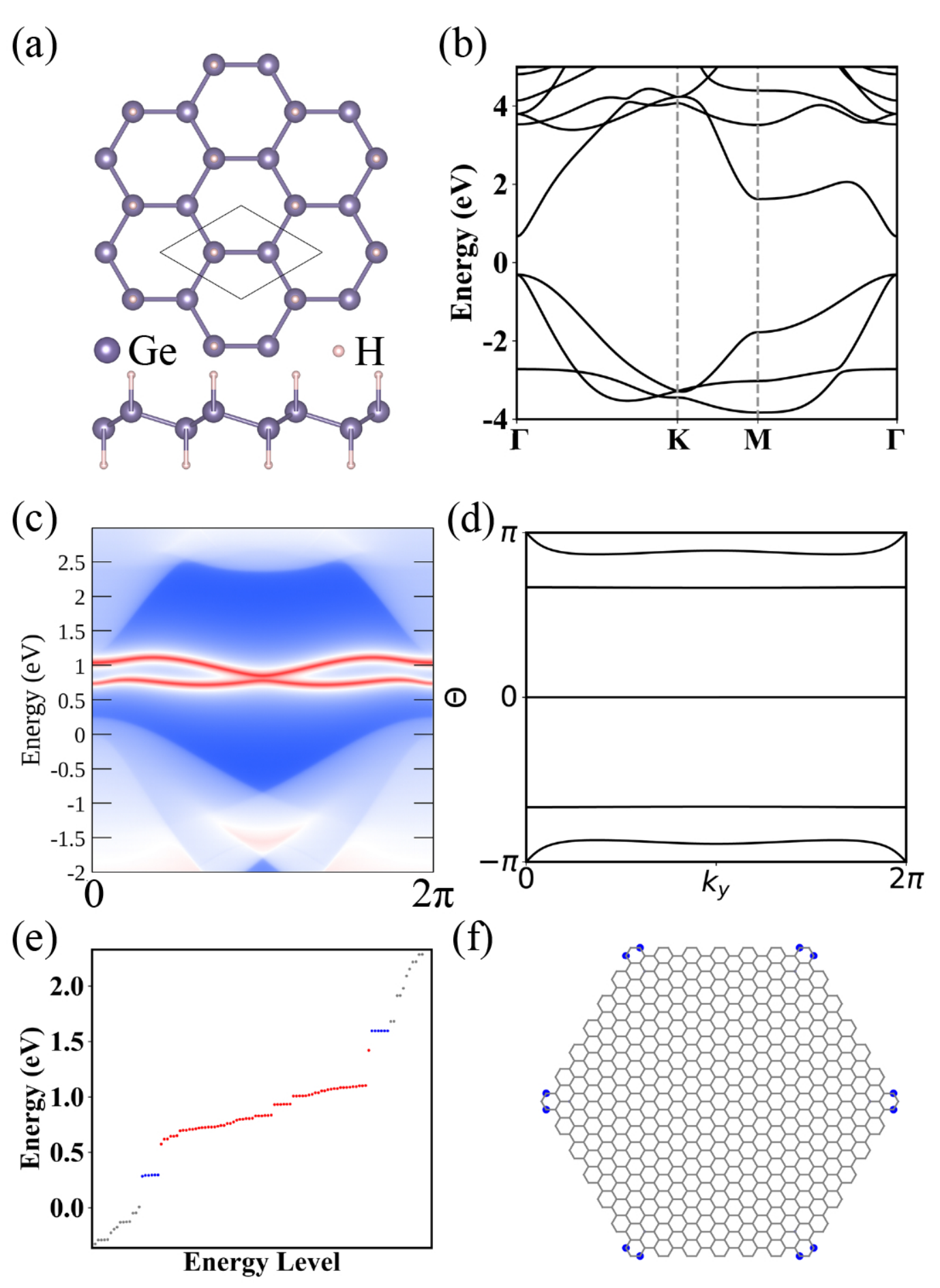}
		\end{center}
		\caption{Top view and side view of geometry structures of germanene hydride (GeH). (b) (c) Bulk bands and edge states of GeH from DFT calculation. (d) Wilson loop of GeH. The number of crossing on $\theta=\pi$ is 1 and therefore the second Stiefel-Whitney number $w_2$ is 1. (e) Energy spectrum of hexagonal finite-size flake shown in (f). Blue, red and grey dots represent the corner, edge and bulk states, respectively. (f)  Charge spatial distribution of blue states in (e).}
	\end{figure}
	
	\begin{figure}[H]
		\begin{center}
			\includegraphics[width=1\linewidth]{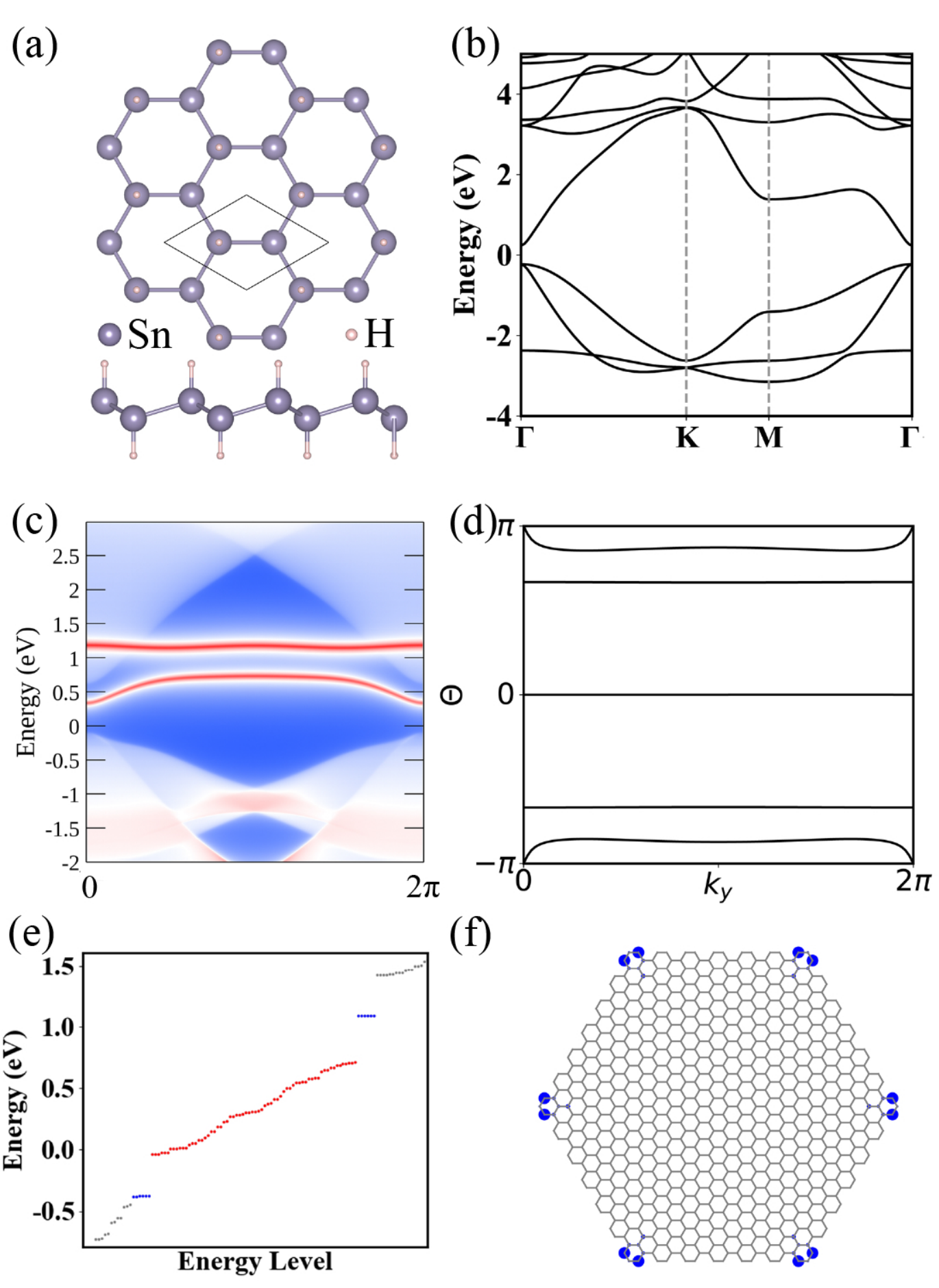}
		\end{center}
		\caption{Top view and side view of geometry structures of stanene hydride (SnH). (b) (c) Bulk bands and edge states of SnH from DFT calculation. (d) Wilson loop of SnH. The number of crossing on $\theta=\pi$ is 1 and therefore the second Stiefel-Whitney number $w_2$ is 1. (e) Energy spectrum of hexagonal finite-size flake shown in (f). Blue, red and grey dots represent the corner states, edge states and bulk states, respectively. (f)  Charge spatial distribution of blue states in (e).}
	\end{figure}
	
	\begin{figure}[H]
		\begin{center}
			\includegraphics[width=1\linewidth]{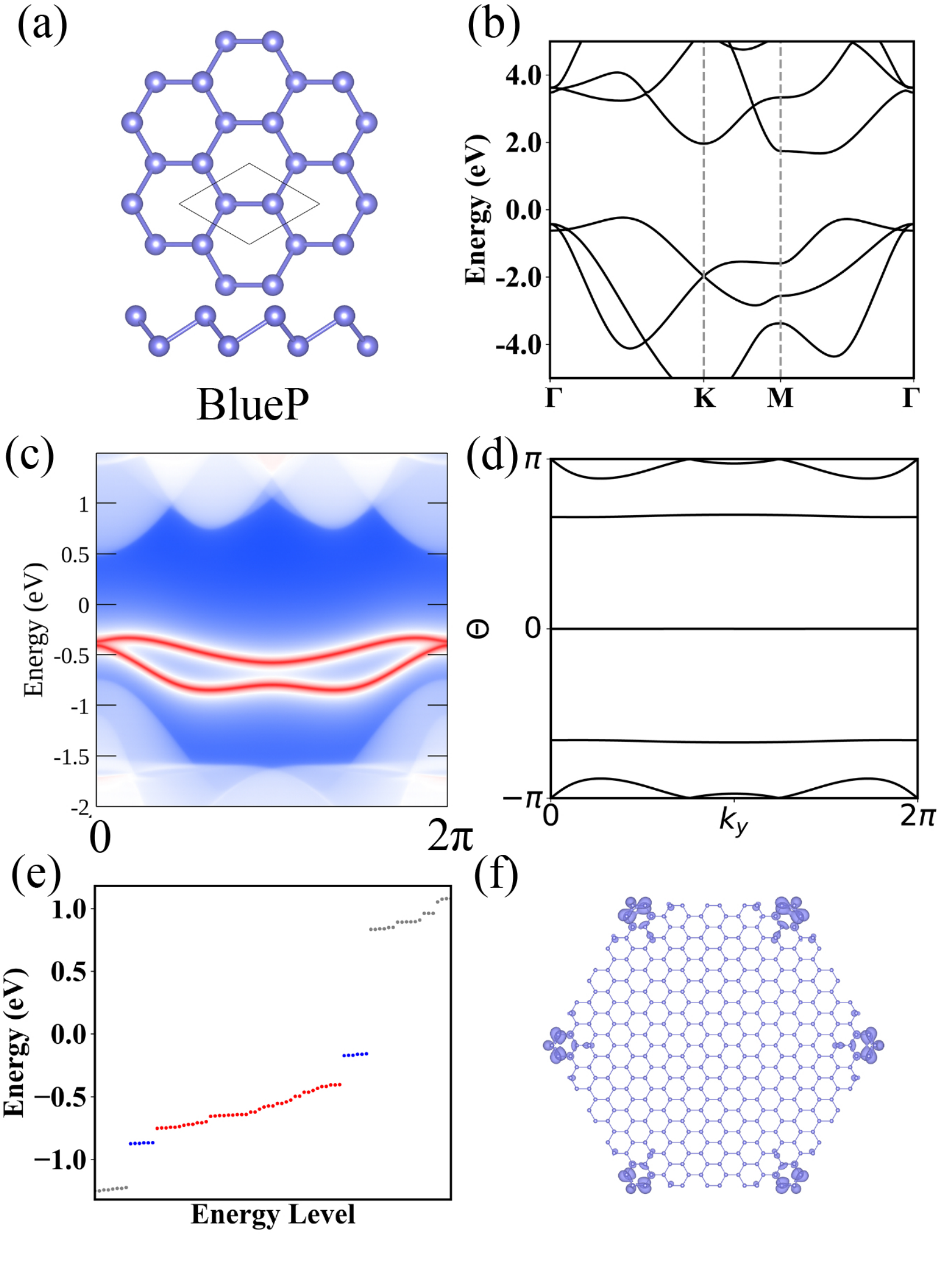}
		\end{center}
		\caption{Top view and side view of geometry structures of blue phosphorene (BlueP). (b) (c) Bulk bands and edge states of BlueP from DFT calculation. (d) Wilson loop of BlueP. The number of crossing on $\theta=\pi$ is 3 and therefore the second Stiefel-Whitney number $w_2$ is 1. (e) Energy spectrum of hexagonal finite-size flake shown in (f). Blue, red and grey dots represent the corner, edge and bulk states, respectively. (f) Charge spatial distribution of blue states in (e).}
	\end{figure}
	
	\begin{figure}[H]
		\begin{center}
			\includegraphics[width=1\linewidth]{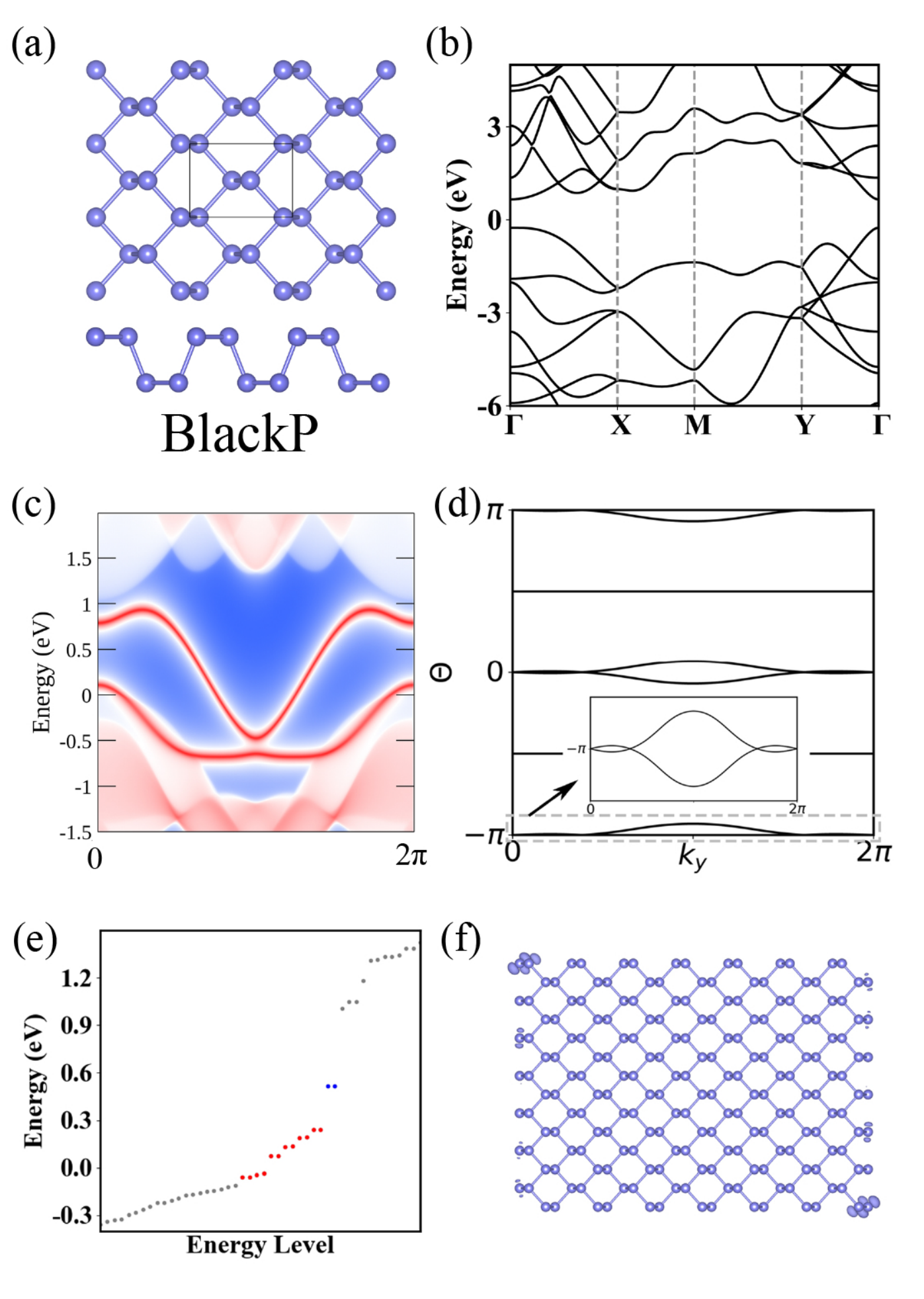}
		\end{center}
		\caption{Top view and side view of geometry structures of black phosphene (BlackP). (b) (c) Bulk bands and edge states of BlackP from DFT calculation. (d) Wilson loop of BlackP. The number of crossing on $\theta=\pi$ is 3 and therefore the second Stiefel-Whitney number $w_2$ is 1. (e) Energy spectrum of rectangle finite-size flake shown in (f). Blue, red and grey dots represent the corner, edge and bulk states, respectively. (f) Charge spatial distribution of blue states in (e).}
	\end{figure}

	\begin{figure}[H]
		\begin{center}
			\includegraphics[width=1\linewidth]{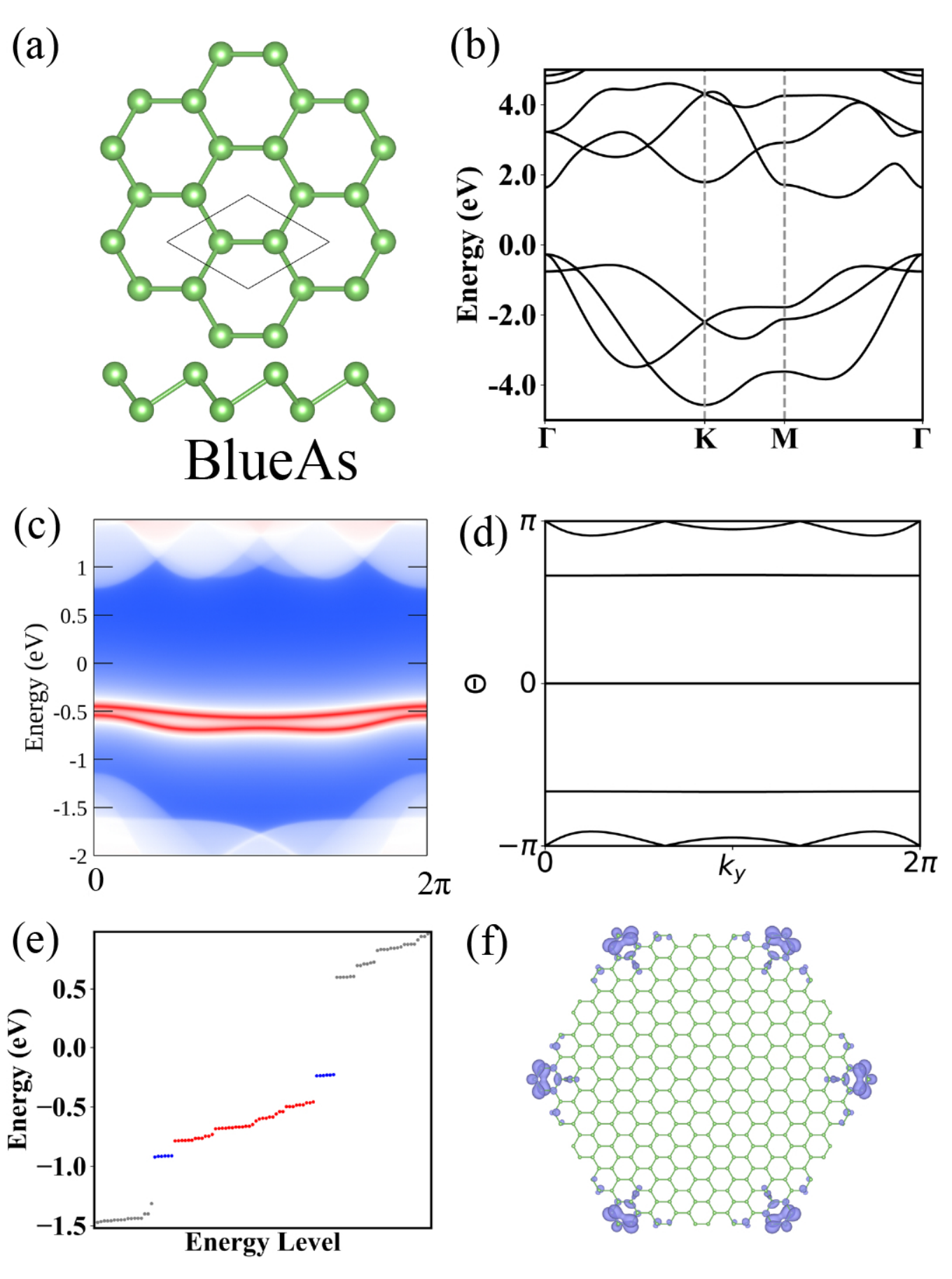}
		\end{center}
		\caption{Top view and side view of geometry structures of blue arsenene (BlueAs). (b) (c) Bulk bands and edge states of BlueAs from DFT calculation. (d) Wilson loop of BlueAs. The number of crossing on $\theta=\pi$ is 3 and therefore the second Stiefel-Whitney number $w_2$ is 1. (e) Energy spectrum of hexagonal finite-size flake shown in (f). Blue, red and grey dots represent the corner, edge and bulk states, respectively. (f) Charge spatial distribution of blue states in (e).}
	\end{figure}
	
	\begin{figure}[H]
		\begin{center}
			\includegraphics[width=1\linewidth]{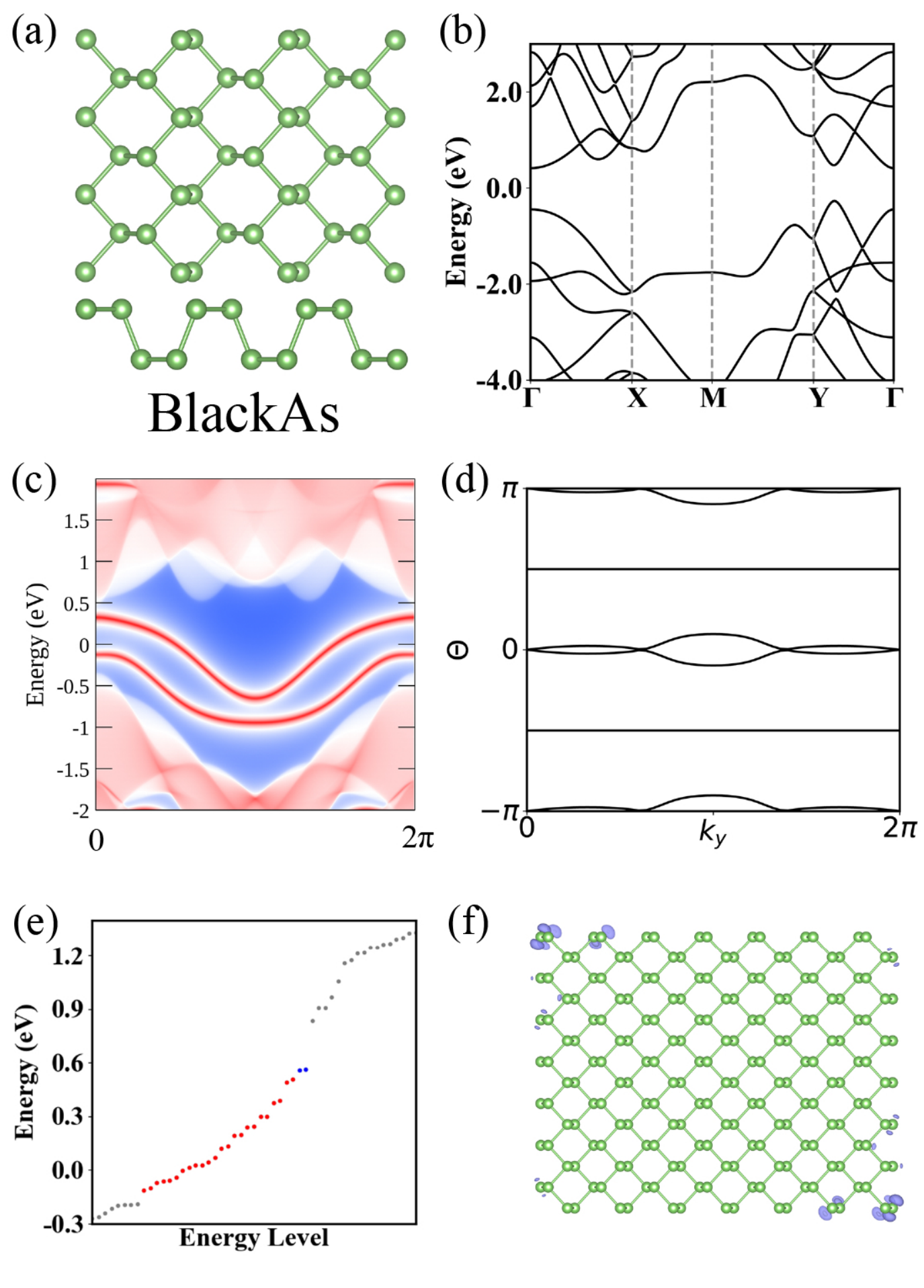}
		\end{center}
		\caption{Top view and side view of geometry structures of black arsenene (BlackAs). (b) (c) Bulk bands and edge states of BlackAs from DFT calculation. (d) Wilson loop of BlackAs. The number of crossing on $\theta=\pi$ is 3 and thus the second Stiefel-Whitney number $w_2$ is 1. (e) Energy spectrum of rectangle finite-size flake shown in (f). Blue, red and grey dots represent the corner, edge and bulk states, respectively. (f) Charge spatial distribution of blue states in (e).}
	\end{figure}

	\section{The $p_z$ orbitals and $sp^2$ orbitals decomposition of SOTI anti-Kekul\'e and Kekul\'e distortion graphenes}
\begin{figure*}
		\begin{center}
			\includegraphics[width=1\linewidth]{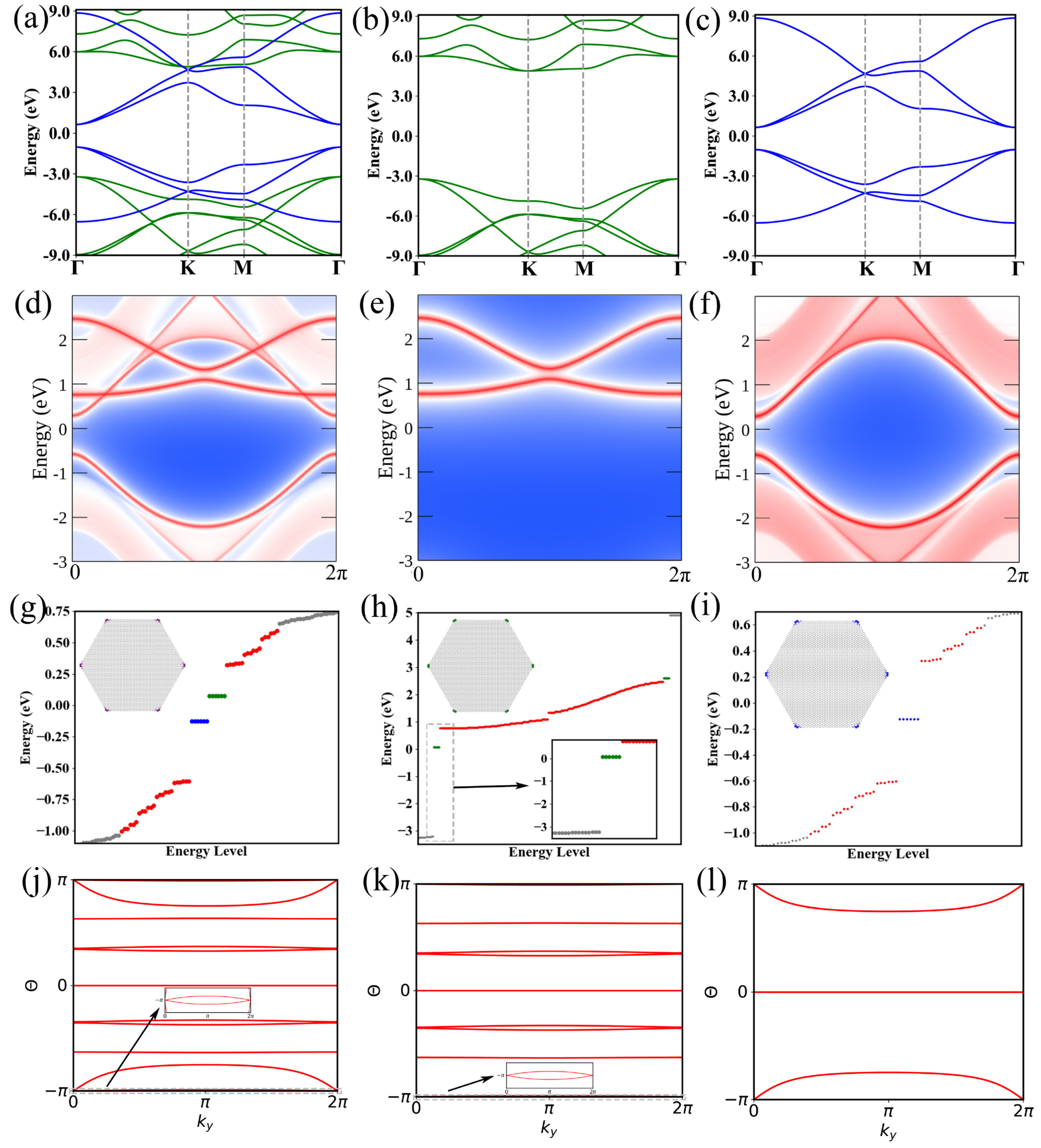}
		\end{center}
		\caption{Bulk bands, edge statets, corner states and Wilson loop of anti-Kekul\'e distortion graphene based on DFT calculations and Wannier functions. Band structures of anti-Kekul\'e graphene with all $s$ and $p$ orbitals (a), only $s$, $p_x$ and $p_y$ ($sp^2$) orbitals (b) and only $p_z$ orbitals (c). Green lines represent the band structures contributed by $sp^2$ orbitals while blue lines represent the band structures contributed by $p_z$ orbitals. Edge states of anti-Kekul\'e graphene with all $s$ and $p$ orbitals (d), only $sp^2$ orbitals (e) and only $p_z$ orbitals (f). Energy discrete spectra of hexagonal finite-size flake with all $s$ and $p$ orbitals (g), only $sp^2$ orbitals (h) and only $p_z$ orbitals (i). Green dots represent the corner states with only $sp^2$ orbitals, blue dots represent the corner states with only $p_z$,  red dots represent the edge states and grey dots represent the bulk states. The corresponding charge spatial distribution of blue states and green states in (g), (h) and (i) plotted in their insets are well localized at the corners, i.e., the corner states. Wilson loop of anti-Kekul\'e graphene with all $s$ and $p$ orbitals (j), only $sp^2$ orbitals (k) and only $p_z$ orbitals (l). The number of crossing on $\theta=\pi$ of anti-Kekul\'e graphene with only $sp^2$ orbitals (k) and only $p_z$ orbitals (l) are both 1.} \label{fig: anti}
	\end{figure*}

	The $p_z$ orbitals and $sp^2$ orbitals ($s$, $p_x$ and $p_y$) of anti-Kekul\'e distortion graphene are decoupled, because the anti-Kekul\'e distortion graphene is fully planar. The $w_2$ of Wilson loop is 0 if we consider all occupied states [Fig. \ref{fig: anti} (j)], which indicates the anti-Kekul\'e distortion graphene is topological trivial. However, since the $p_z$ orbitals and $sp^2$ orbitals are independent, one can consider them separately. Both of them are topological nontrivial with $w_2 = 1$ [Figs. \ref{fig: anti} (k, l)]. We also calculated the edge states and corner states of anti-Kekul\'e distortion graphene with only $p_z$ orbitals and $sp^2$ orbitals [Figs. \ref{fig: anti}(e, h, f, i)]. The anti-Kekul\'e distortion graphene with all occupied states can be seen as the direct summation of two independent topological phases. Therefore, anti-Kekul\'e order graphene is topological nontrivial.
	
\begin{figure*}
		\begin{center}
			\includegraphics[width=1\linewidth]{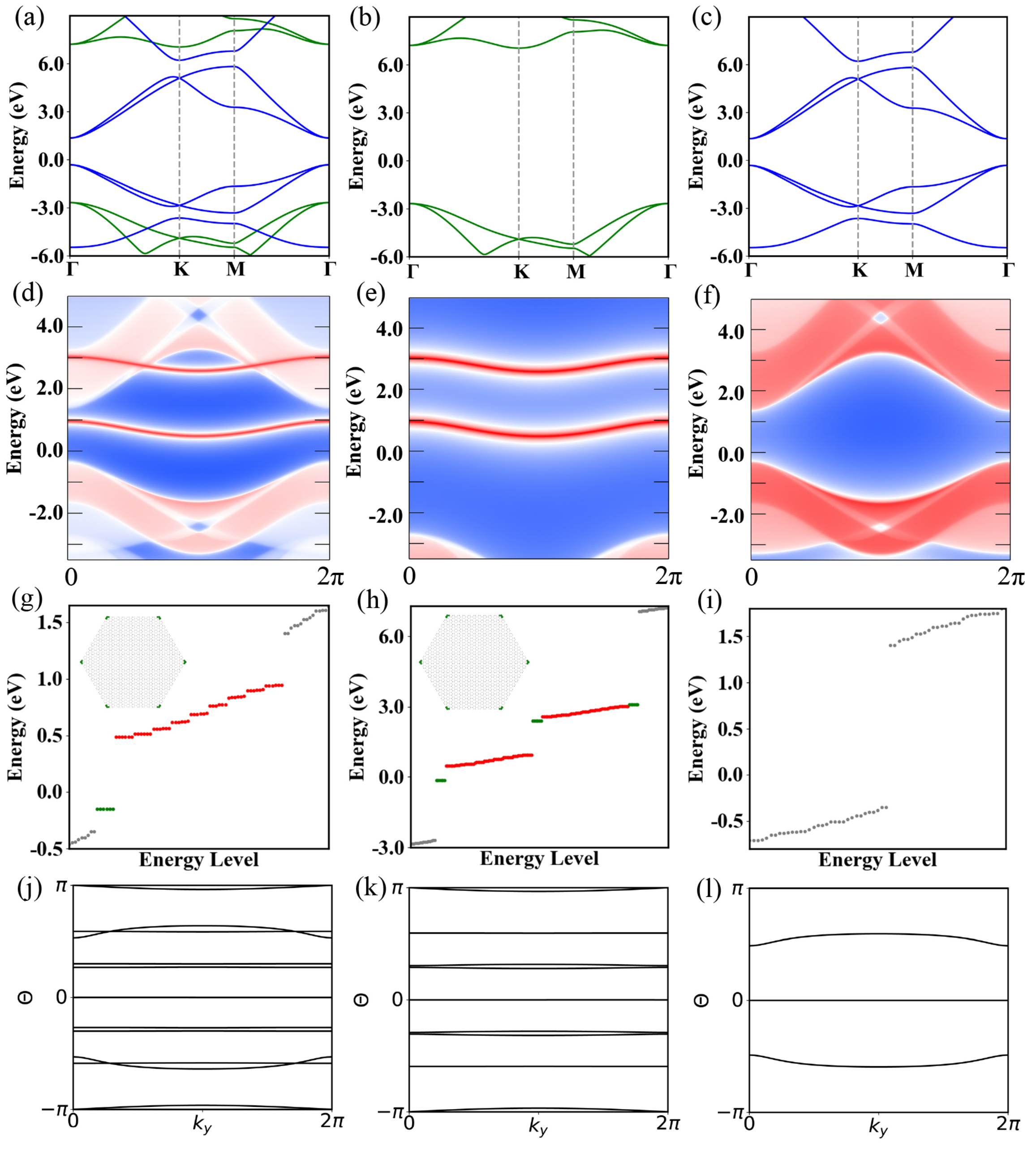}
		\end{center}
		\caption{Bulk bands, edge statets, corner states and Wilson loop of Kekul\'e distortion graphene based on DFT calculations and Wannier functions. Band structures of Kekul\'e graphene with all $s$ and $p$ orbitals (a), only $s$, $p_x$ and $p_y$ ($sp^2$) orbitals (b) and only $p_z$ orbitals (c). Green lines represent the band structures contributed by $sp^2$ orbitals while  blue lines represent the band structures contributed by $p_z$ orbitals. Edge states of Kekul\'e graphene with all $s$ and $p$ orbitals (d), only $sp^2$ orbitals (e) and only $p_z$ orbitals (f). Energy discrete spectra of hexagonal finite-size flake with all $s$ and $p$ orbitals (g), only $sp^2$ orbitals (h) and only $p_z$ orbitals (i). Green dots represent the corner states with $sp^2$ orbitals, red dots represent the edge states and grey dots represent the bulk states. The corresponding charge spatial distribution of green states in (g), (h) and (i) plotted in their insets are well localized at the corners, i.e., the corner states. Wilson loop of Kekul\'e graphene with all $s$ and $p$ orbitals (j), only $sp^2$ orbitals (k) and only $p_z$ orbitals (l). The number of crossing on $\theta=\pi$ of Kekul\'e graphene with only $sp^2$ orbitals (k) and only $p_z$ orbitals are 1 and 0, respectively.} \label{fig: keku}
	\end{figure*}

	For Kekul\'e distortion graphene, the $p_z$ orbitals and $sp^2$ orbitals are also decoupled. The $p_z$ orbitals are trivial and the $sp^2$ orbitals are nontrivial according to the Wilson loop [Figs. \ref{fig: keku} (k, l)], edge states [Figs. \ref{fig: keku} (h, i)] and corner states [Figs. \ref{fig: keku} (h, i)]. As a result, Kekul\'e order graphene is also topological nontrivial with corner states.
	
	\section{The geometry structures, bulk bands, edge states, Wilson loop spectra, energy spectra of hexagonal finite-size flakes, and charge spatial distribution of corner states of the SOTI material candidates (anti-Kekul\'e and Kekul\'e distortion group IV materials)}
	The anti-Kekul\'e/Kekul\'e distortion group IV materials are also SOTI material candidates with global gaps at Fermi level. Figures S14-S19 show the geometry structures, bulk bands, edge states, Wilson loop spectra, discrete energy spectra and charge distribution of corner states of these SOTI material candidates, which include anti-Kekul\'e/Kekul\'e distortion graphenes (anti-Kekul\'eGr/Kekul\'eGr) and silicenes (anti-Kekul\'eSi/Kekul\'eSi), anti-Kekul\'e distortion germanene (anti-Kekul\'eGe), anti-Kekul\'e distortion stanene (anti-Kekul\'eSn). It should be noticed that anti-Kekul\'eSi/Ge/Sn have the trivial second Stiefel-Whitney number $w_{2}=0$, different from the anti-Kekul\'eGr due to their buckled geometry. The buckling of the anti-Kekul\'eSi/Ge/Sn structures makes the $p_z$ orbitals couple with $sp^2$ orbitals and therefore one should consider both $p_z$ and $sp^2$ orbitals together, which results in a trivial second Stiefel-Whitney number. However, the buckling do not close the gaps and one can still observe the robust corner states in the anti-Kekul\'eSi/Ge/Sn, as shown in Fig. S16, Fig. S18, and Fig. S19. Figures S20-S21 show the geometry structures, bulk bands, edge states, and Wilson loops of Kekul\'e distortion germanene (Kekul\'eGe) and Kekul\'e  distortion stanene (Kekul\'eSn). The Kekul\'eGe/Sn have nontrivial second Stiefel-Whitney number $w_{2}=1$. However, since the bulk band gaps are too small and the dispersion of edge states are too large, the corner states are mixed bulk states or edge states.
	\begin{figure}[H]
		\begin{center}
			\includegraphics[width=1\linewidth]{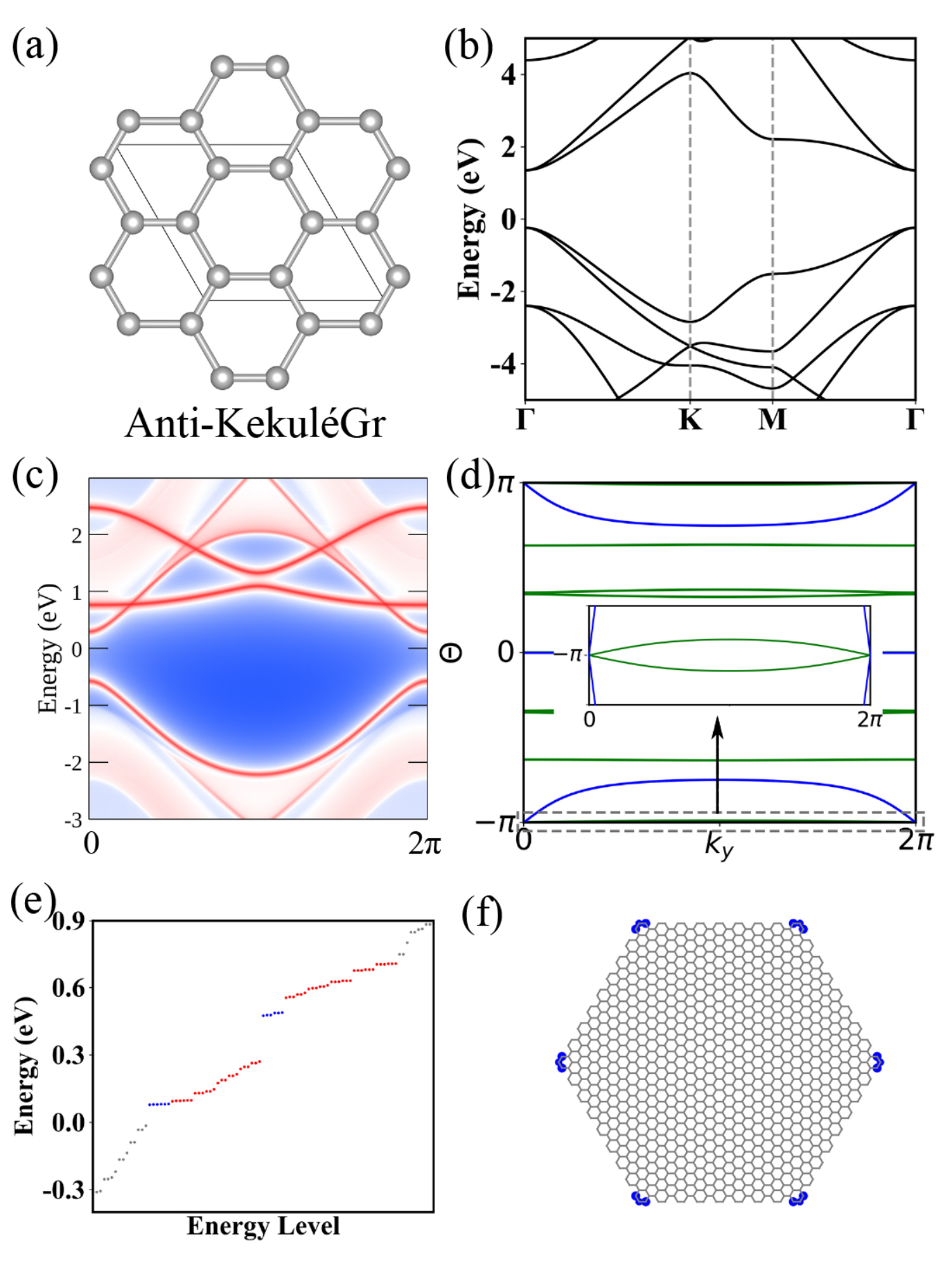}
		\end{center}
		\caption{Top view of geometry structures of anti-Kekul\'e distortion graphene (anti-Kekul\'eGr). In order to reduce the computational burden, we set the weak (strong) bond length 1.6 (1.4) \AA\ and thus get a larger band gap at $\Gamma$ than the experimental value but with the same $w_2$. (b) (c) Bulk bands and edge states of anti-Kekul\'eGr from DFT calculation. (d) Wilson loop of $sp^2$ orbitals (green) and $p_z$ orbitals (blue) of anti-Kekul\'eGr. The number of crossing on $\theta=\pi$ is 2 and therefore the second Stiefel-Whitney number $w_2$ is 0. (e) Energy spectrum of hexagonal finite-size flake shown in (f). Blue, red and grey dots represent the corner states, edge states and bulk states, respectively. (f)  Charge spatial distribution of blue states in (e).}
	\end{figure}
	
	\begin{figure}[H]
		\begin{center}
			\includegraphics[width=1\linewidth]{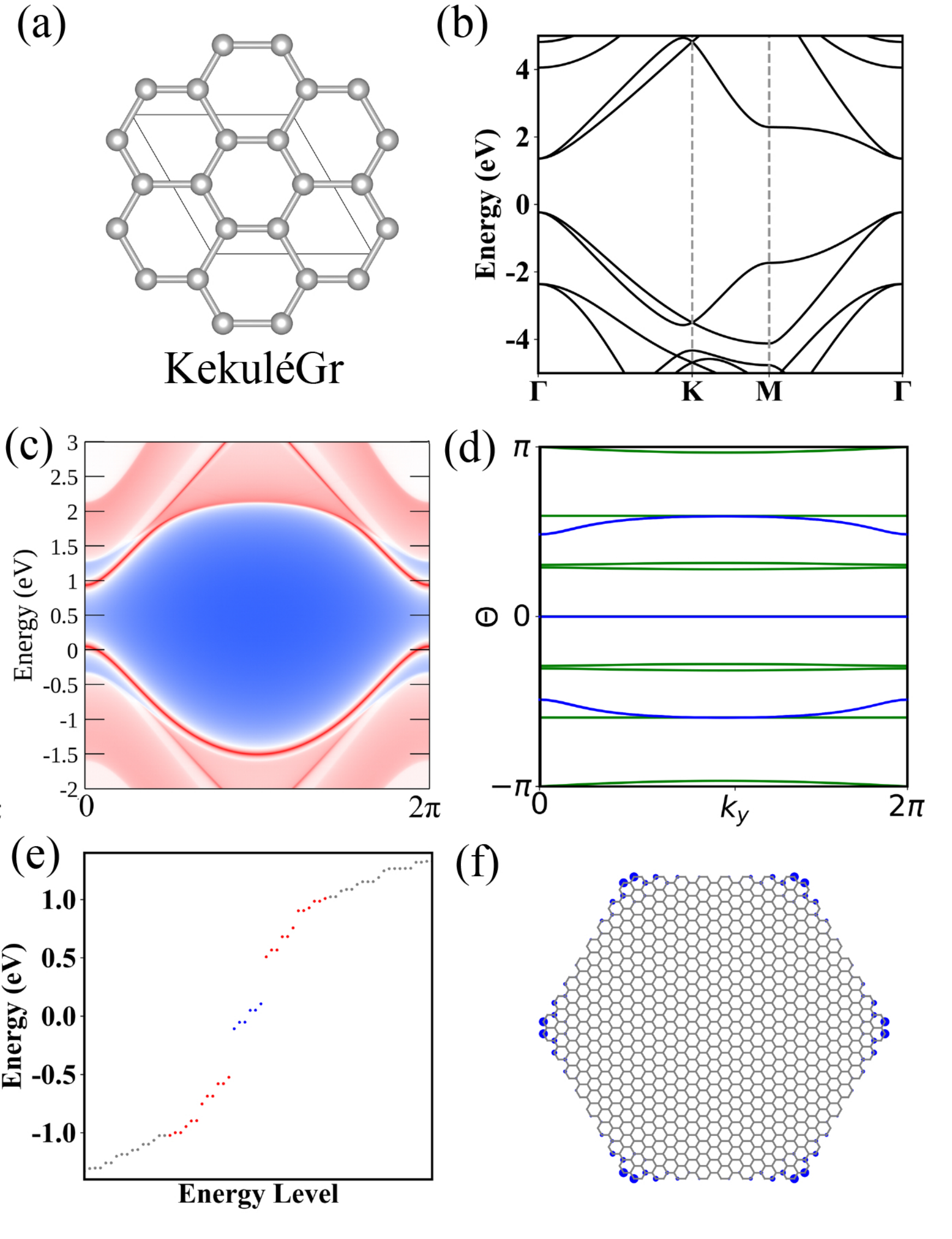}
		\end{center}
		\caption{Top view of geometry structures of Kekul\'e distortion graphene (Kekul\'eGr). In order to reduce the computational burden, we set the weak (strong) bond length 1.6 (1.4) \AA\ and thus get a larger band gap at $\Gamma$ than the experimental value but with the same $w_2$. (b) (c) Bulk bands and edge states of Kekul\'eGr from DFT calculation. (d) Wilson loop of $sp^2$ orbitals (green) and $p_z$ orbitals (blue) of Kekul\'eGr. The number of crossing on $\theta=\pi$ is 1 and therefore the second Stiefel-Whitney number $w_2$ is 1. (e) Energy spectrum of hexagonal finite-size flake shown in (f). Blue, red and grey dots represent the corner, edge and bulk states, respectively. (f)  Charge spatial distribution of blue states in (e).}
	\end{figure}
	
	\begin{figure}[H]
		\begin{center}
			\includegraphics[width=1\linewidth]{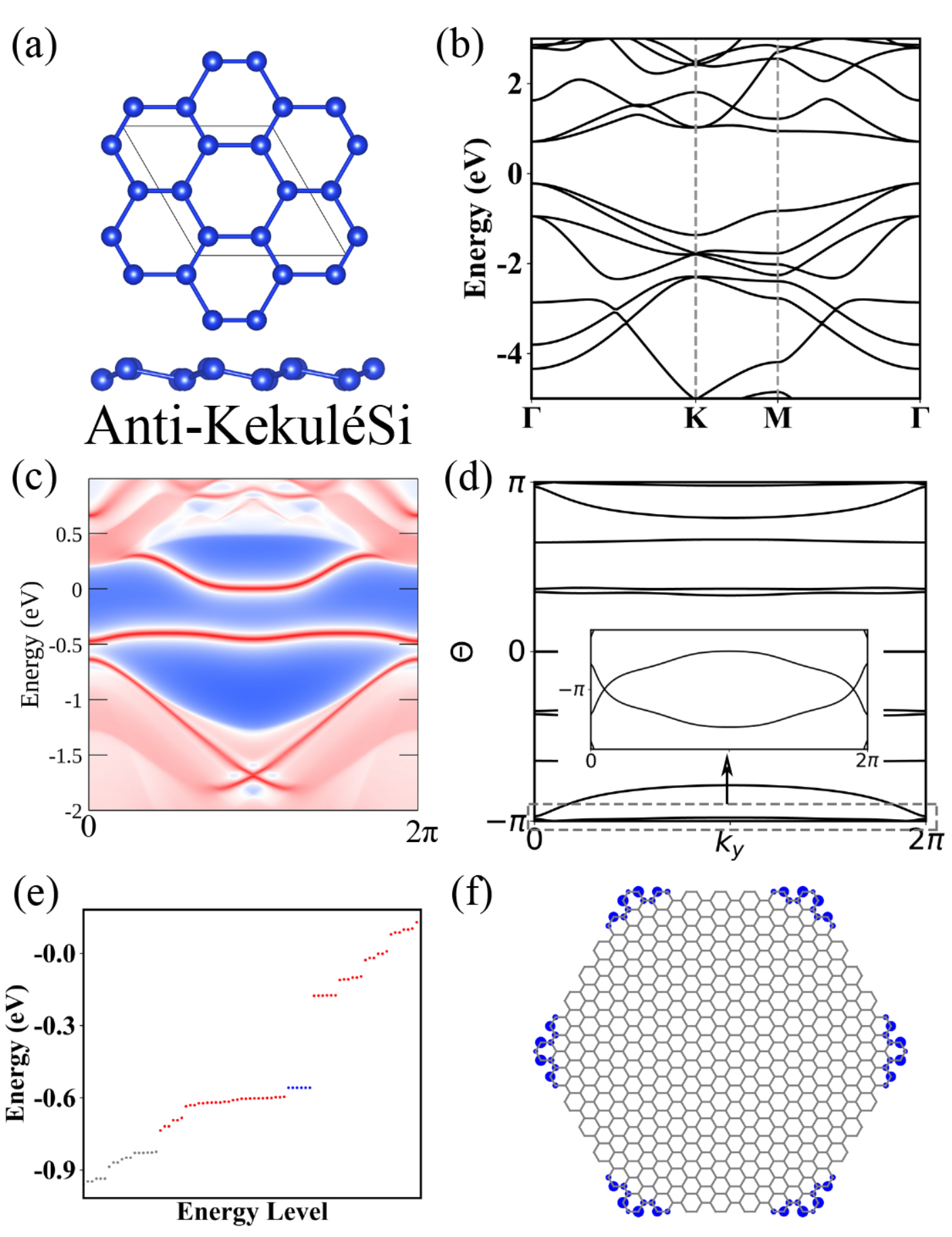}
		\end{center}
		\caption{Top view and side view of geometry structures of anti-Kekul\'e distortion silicene (anti-Kekul\'eSi). The weak (strong) bond length is set as 2.44 (2.05) \AA . (b) (c) Bulk bands and edge states of anti-Kekul\'eSi from DFT calculation. (d) Wilson loop of anti-Kekul\'eSi. The number of crossing on $\theta=\pi$ is 2 and therefore the second Stiefel-Whitney number $w_2$ is 0. (e) Energy spectrum of hexagonal finite-size flake shown in (f). Blue, red and grey dots represent the corner states, edge states and bulk states, respectively. (f)  Charge spatial distribution of blue states in (e).}
	\end{figure}
	
	\begin{figure}[H]
		\begin{center}
			\includegraphics[width=1\linewidth]{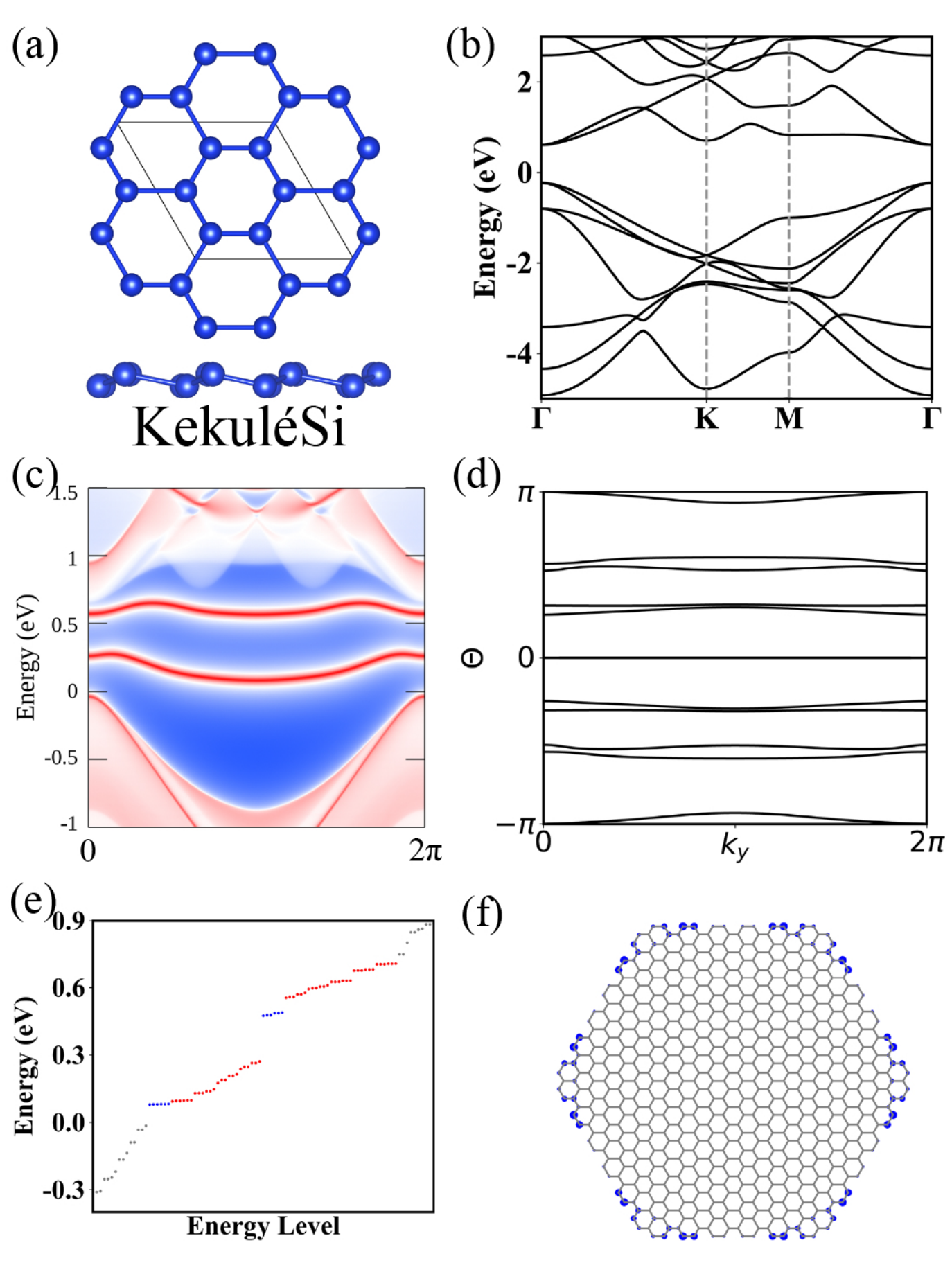}
		\end{center}
		\caption{Top view and side view of geometry structures of Kekul\'e distortion silicene (Kekul\'eSi). The weak (strong) bond length is 2.44 (2.05) \AA . (b) (c) Bulk bands and edge states of Kekul\'eSi from DFT calculation. (d) Wilson loop of Kekul\'eSi. The number of crossing on $\theta=\pi$ is 1 and therefore the second Stiefel-Whitney number $w_2$ is 1. (e) Energy spectrum of hexagonal finite-size flake shown in (f). Blue, red and grey dots represent the corner states, edge states and bulk states, respectively. (f)  Charge spatial distribution of blue states in (e).}
	\end{figure}

	\begin{figure}[H]
		\begin{center}
			\includegraphics[width=1\linewidth]{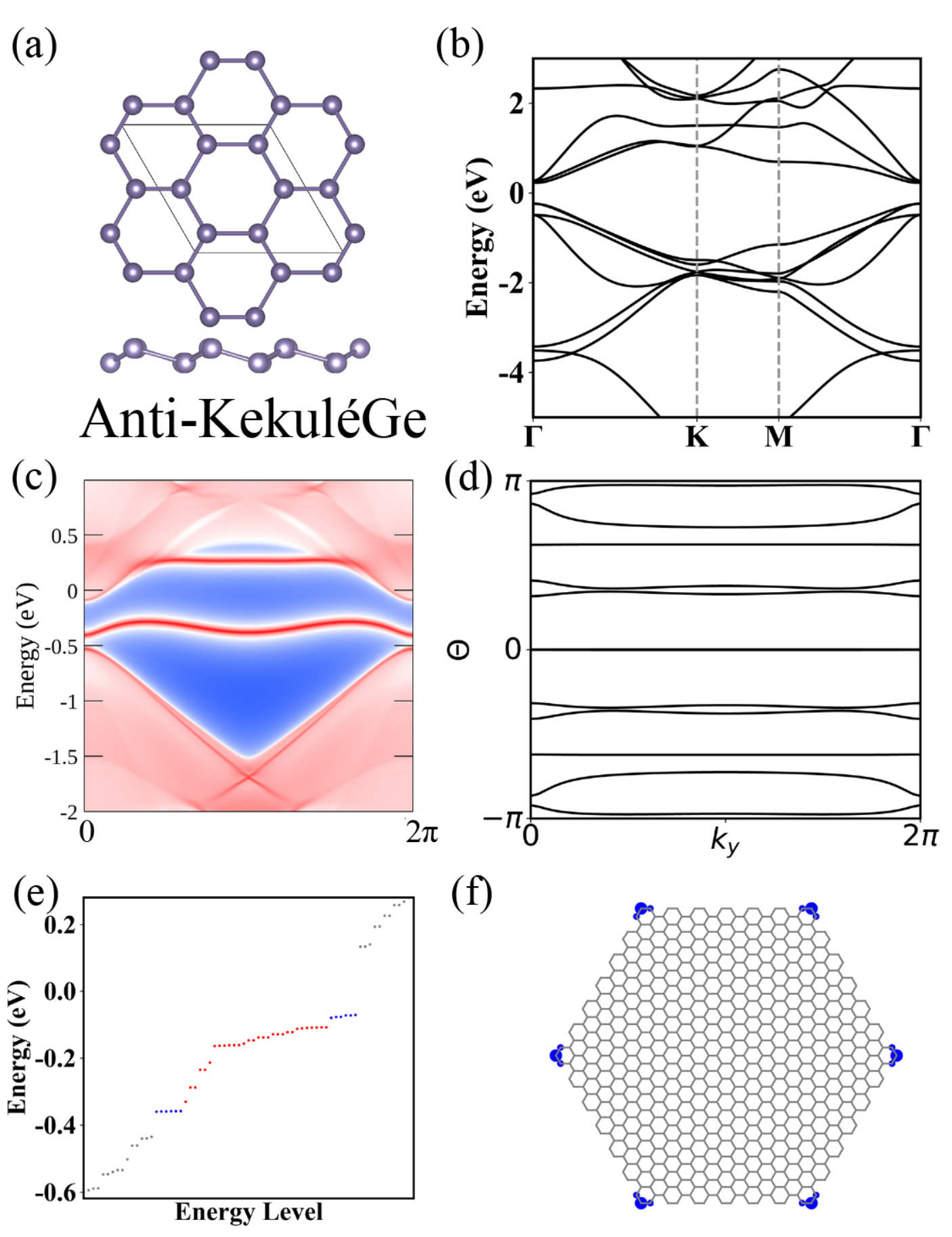}
		\end{center}
		\caption{Top view and side view of geometry structures of anti-Kekul\'e distortion germanene (anti-Kekul\'eGe). The weak (strong) bond length is 2.54 (2.30) \AA . (b) (c) Bulk bands and edge states of anti-Kekul\'eGe from DFT calculation. (d) Wilson loop of anti-Kekul\'eGe. The number of crossing on $\theta=\pi$ is 0 and therefore the second Stiefel-Whitney number $w_2$ is 0. (e) Energy spectrum of hexagonal finite-size flake shown in (f). Blue, red and grey dots represent the corner states, edge states and bulk states, respectively. (f)  Charge spatial distribution of blue states in (e).}
	\end{figure}

	\begin{figure}[H]
		\begin{center}
			\includegraphics[width=1\linewidth]{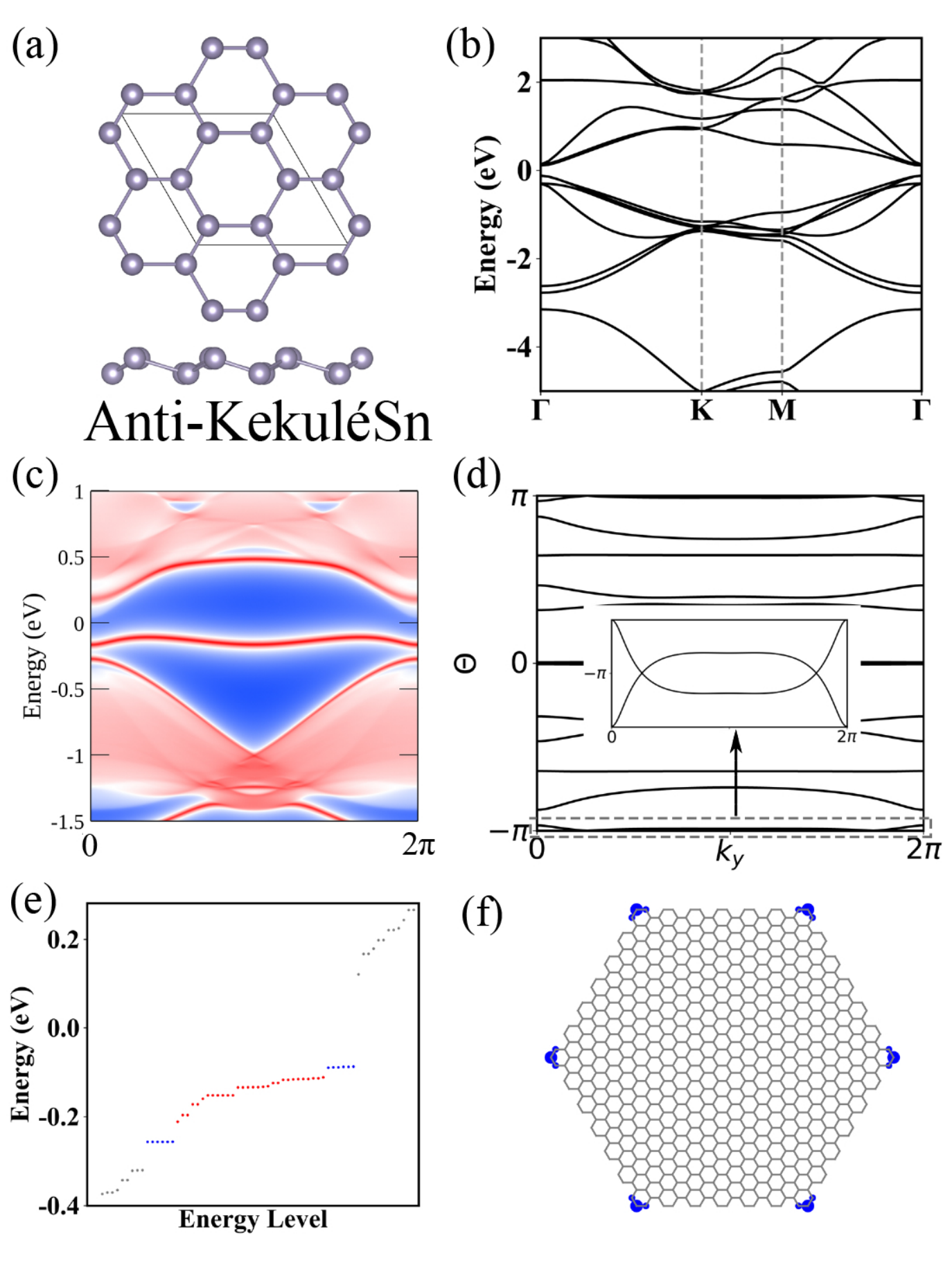}
		\end{center}
		\caption{Top view and side view of geometry structures of anti-Kekul\'e distortion stanene (anti-Kekul\'eSn). The weak (strong) bond length is 2.93 (2.70) \AA . (b) (c) Bulk bands and edge states of anti-Kekul\'eSn from DFT calculation. (d) Wilson loop of anti-Kekul\'eSn. The number of crossing on $\theta=\pi$ is 2 and therefore the second Stiefel-Whitney number $w_2$ is 0. (e) Energy spectrum of hexagonal finite-size flake shown in (f). Blue, red and grey dots represent the corner states, edge states and bulk states, respectively. (f)  Charge spatial distribution of blue states in (e).}
	\end{figure}
	
	\begin{figure}[H]
		\begin{center}
			\includegraphics[width=1\linewidth]{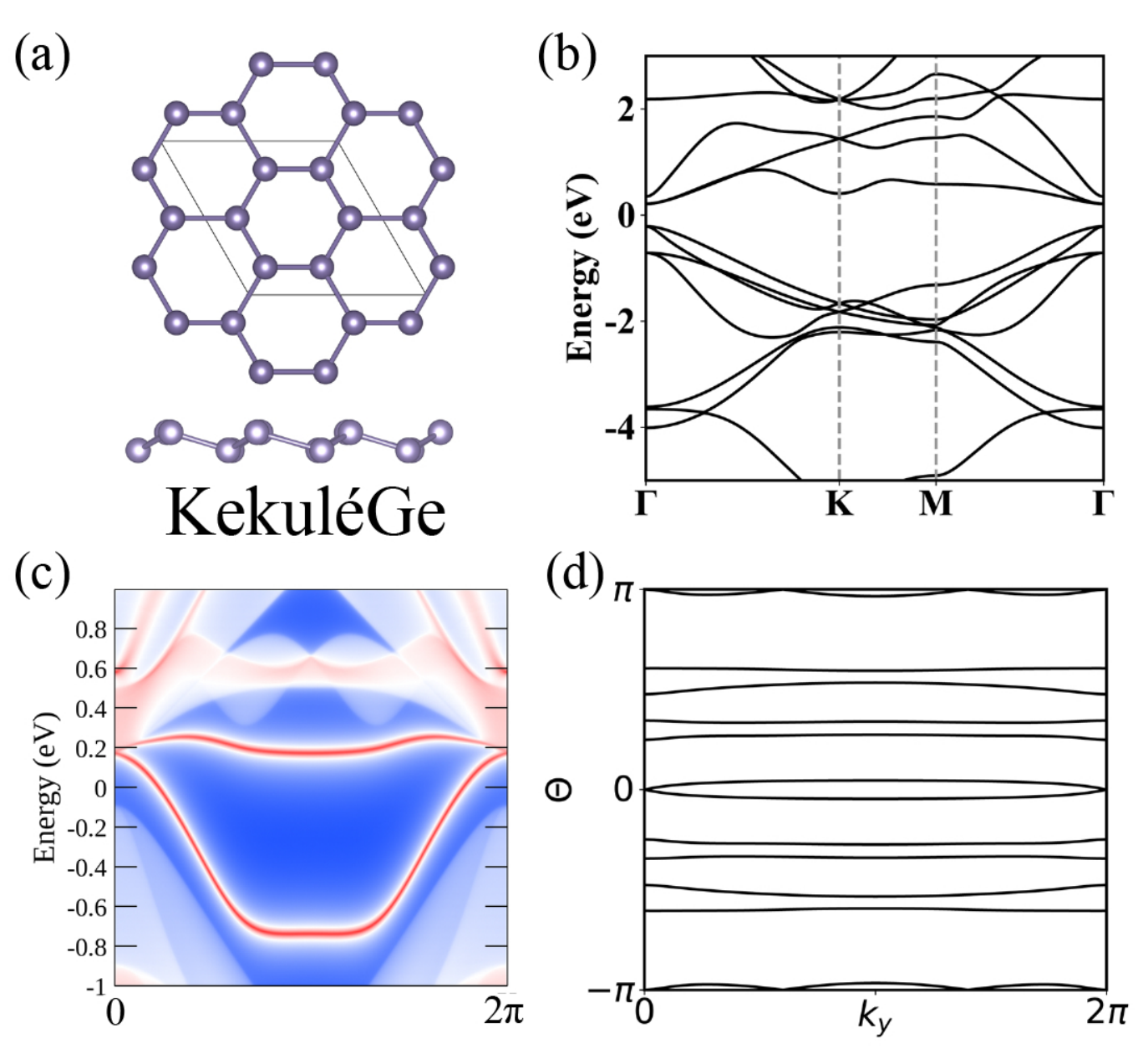}
		\end{center}
		\caption{Top view and side view of geometry structures of Kekul\'e distortion germanene (Kekul\'eGe). The weak (strong) bond length is set as 2.54 (2.30) \AA\. (b) (c) Bulk bands and edge states of Kekul\'eGe from DFT calculation. (d) Wilson loop of Kekul\'eGe. The number of crossing on $\theta=\pi$ is 3 and therefore the second Stiefel-Whitney number $w_2$ is 1. }
	\end{figure}
	
	\begin{figure}[H]
		\begin{center}
			\includegraphics[width=1\linewidth]{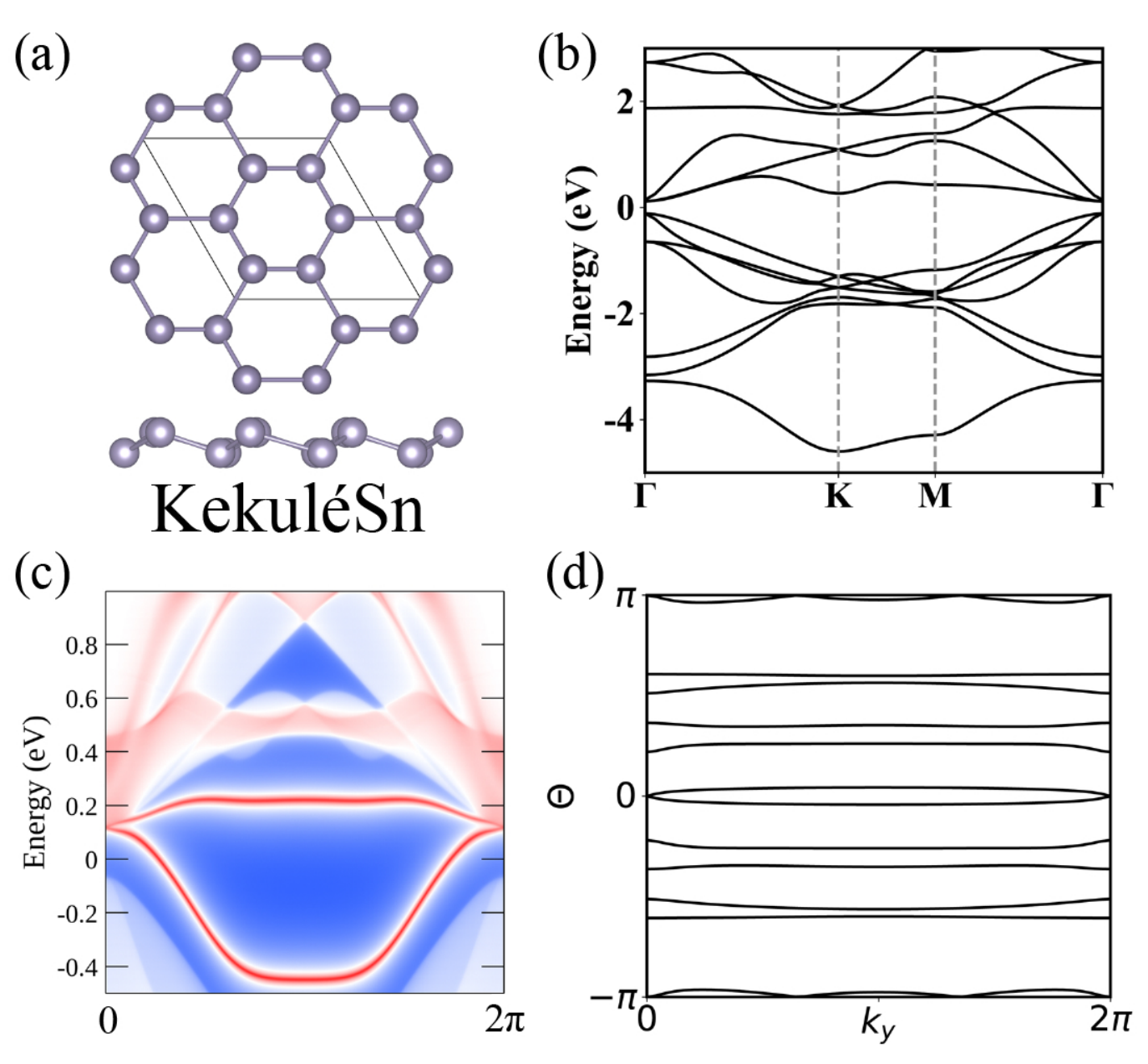}
		\end{center}
		\caption{Top view and side view of geometry structures of Kekul\'e distortion stanene (Kekul\'eSn). The weak (strong) bond length is 3.11 (2.70) \AA. (b) (c) Bulk bands and edge states of Kekul\'eSn from DFT calculation. (d) Wilson loop of Kekul\'eSn. The number of crossing on $\theta=\pi$ is 3 and therefore the second Stiefel-Whitney number $w_2$ is 1.}
	\end{figure}

	\section{$k\cdot p$ Hamiltonian and edge theory}
	We constructed a $k\cdot p$ model to capture the bulk low-energy band structure, which is around $\Gamma$ point and features a band inversion. The $\Gamma$ point of TB model I and model II both have $D_{6h}$ and $T$ symmetry. The generators for $D_{6h}$ are chosen as the three fold rotation $C_{3z}$, inversion $P$ and the vertical mirror $M_y$ perpendicular to $y$ direction. For the buckled hexagonal lattice, the symmetry of $\Gamma$ point changes from $D_{6h}$ to $D_{3d}$, but the generators for $D_{3d}$ can also be chosen as $C_{3z}$, $P$ and $M_y$. 
	We only consider the four lowest energy bands and in the basis of the two doublets $(E_{1g}, E_{2u})^T$ or $(E_{2g}, E_{1u})^T$  [Figs. 1(c, e) in the text], the symmetry operators can be represented by 
	\begin{equation}
		C_{3 z}=\tau_{0}e^{-i(\pi / 3) \sigma_{y}} ,\quad \mathcal{P} (C_{2 z})=\tau_{z}\sigma_{0} , \quad \mathcal{M}_{y}=\tau_{0} \sigma_{z},
	\end{equation}
	where Pauli matrices $\tau$ act on the two doublets and $\sigma$ act on the two degenerate states within each doublet. In addition, $T = K$ in spinless system, where $K$ is the complex conjugation. The form of Hamiltonian is constrained by crystal symmetry and time-reversal symmetry. Under the operation of these symmetry operators, momentum and pseudospin are transformed as follows:
	\begin{equation}
		\begin{aligned}
			&C_{3}: k_{\pm} \rightarrow e^{\pm i 2 \pi / 3} k_{\pm}, \sigma_{\pm} \rightarrow e^{\pm i 2 \pi / 3} \sigma_{\pm}, \sigma_{y} \rightarrow \sigma_{y},\\
			&P (C_2): k_{\pm} \rightarrow -k_{\pm}, \sigma_{\pm} \rightarrow \sigma_{\pm}, \sigma_{y} \rightarrow \sigma_{y},\\
			&M_y: k_{\pm} \rightarrow k_{\mp}, \sigma_{\pm} \rightarrow\sigma_{\mp}, \sigma_{y} \rightarrow -\sigma_{y},\\
		\end{aligned}\label{trans}
	\end{equation}
	where $k_{\pm} = k_x \pm ik_y$ and $\sigma_{\pm} = \sigma_{z} \pm i\sigma_{x}$. The Hamiltonian must be invariant under Eq. \ref{trans}. In addition, the Hamiltonian is is also constrained under time-reversal symmetry
	\begin{equation}
		T H_{eff}(\boldsymbol{k}) T^{-1} = H_{eff}(-\boldsymbol{k}).
	\end{equation}
	Considered by these symmetries, the bulk effective model expanded to $k$-quadratic order is
	\begin{equation}
		\begin{aligned}
			\mathcal{H}_{eff}(\boldsymbol{k})=&E_{0}+ \left(m_{0}-m_{1} k^{2}\right) \tau_{z}+\frac{1}{2}v\left(k_{+} \sigma_{-}+k_{-} \sigma_{+}\right) \tau_{y} \\
			&+\frac{1}{2}\left(k_{+}^2 \sigma_{+}+k_{-}^2 \sigma_{-}\right)\left(\alpha_{1} \tau_{0}+\alpha_{2} \tau_{z}\right),
		\end{aligned}
	\end{equation}
	where $E_{0}=c_{0}+c_{1} k^{2}, k=|\boldsymbol{k}|$; $c_{i}, m_{i}, \alpha_{i}$ and $v$ are real parameters. The $k\cdot p$ Hamiltonian also has the form 
	\begin{equation}
		\begin{aligned}
			\mathcal{H}_{eff}(\boldsymbol{k})=& E_{0}+\left(m_{0}-m_{1} k^{2}\right) \tau_{z}+v\left(k_{x} \sigma_{z}+k_{y} \sigma_{x}\right) \tau_{y} \\
			&+\left[\left(k_{x}^{2}-k_{y}^{2}\right) \sigma_{z}-2 k_{x} k_{y} \sigma_{x}\right]\left(\alpha_{1} \tau_{0}+\alpha_{2} \tau_{z}\right).
		\end{aligned}
	\end{equation}
	The $m_{0,1}<0$ for Mechanism I, while $m_{0,1}>0$ for Mechanism II, since their low-energy bands host opposite parity.  The two mechanisms have an approximate chiral symmetry. The chiral symmetry is represented as $C = \tau_x$, when the first term and the $\alpha_1$ term can be ignored.
	
	To obtain the intuitive understanding of the corner states, we study the edge theory based on the $k\cdot p$ effective model. To simplify the picture, we only consider the first order of the $k\cdot p$ Hamiltonian. We replace $k_{x} \rightarrow-i \partial_{x}$ and the edge states are solved from 
	\begin{equation}
		\tilde{\mathcal{H}} \psi=E \psi, 
	\end{equation}
	where
	\begin{equation}
		\tilde{\mathcal{H}}_{eff}=m(x) \tau_{z}+v\left(-i\partial_{x} \sigma_{z}+ k_y \sigma_{x}\right) \tau_{y}.
	\end{equation}
	with $m(x<0) $ for the sample, $m(x>0) $ for the vacuum side, and $m(x<0)m(x>0)<0$. At $k_y=0$, the states with $\sigma_z= \pm1$ are decoupled with opposite $M_y$ eigenvalues. First, considering the $\sigma_z= +1$ and the equation reduced to a Jackiw-Rebbi \cite{JRmodel} problem with the zero-energy edge mode
	\begin{equation}
		\psi_{+}=\frac{1}{A}\exp \left(-\frac{m(x)x}{v}\right)\left(\begin{array}{l}
			1 \\
			0
		\end{array}\right)_{\sigma} \otimes\left(\begin{array}{c}
			1 \\
			-1
		\end{array}\right)_{\tau},
	\end{equation}
	where A is the normalization factor. The other zero mode with $\sigma_z=-1$ is
	\begin{equation}
		\psi_{-}=\frac{1}{A}\exp \left(-\frac{m(x)x}{v}\right)\left(\begin{array}{l}
			0 \\
			1
		\end{array}\right)_{\sigma} \otimes\left(\begin{array}{l}
			1 \\
			1
		\end{array}\right)_{\tau}.
	\end{equation}
	Then, in the basis of above two states, the edge Hamiltonian is given by
	\begin{equation}
		\mathcal{H}_{\text {edge }}(k)=v k s_{y},
	\end{equation}
	where $k$ is the wave vector along the edge, the Pauli matrices $s$ act on the space of $\left(\psi_{+}, \psi_{-}\right)^{T}$ and $v$ is Fermi velocity. In the basis, the symmetry operators can be represented by $\mathcal{M}_{y}=\mathcal{C}=s_{z}$. In the presence of $M_y$ and $\mathcal{C}$ symmetry, the mass term is forced to be zero at edge Hamiltonian. However, for an edge that does not preserve $M_y$, the edge Hamiltonian will generally be gaped with the mass term $\Delta_{M}=m_{M} s_{x}$. Two edges related by $M_y$ must have opposite Dirac mass. Therefore, the protected 0D corner modes must exit at the intersection between two edges.
	
	\section{Corner charge, second Stiefel-Whitney number, nested Wilson loop and bulk gaps of all the SOTI material candidates}
	The nontrivial topology of these SOTI material candidates is verified by the second Stiefel-Whitney number $w_2$ and fractional corner charge. The number of odd parity of at time-reversal invariant momentum and nested Wilson loop demonstrate the nontrivial value of $w_2$, and the topological invariants of $C_n$ symmetry at high symmetric points determine the fractional corner charge [Table. \ref{table: 2}]. Table \ref{table: 2} also presents the gaps of these SOTI material candidates.
	
	\begin{table*}[htbp]
		\centering
		{\tabcolsep0.078in
			\caption{Topological invariants of $C_n$ symmetry at high symmetric points, the number of occupied bands with odd parity at time-reversal invariant momenta, corner charge $Q_c$, the second Stiefel-Whitney number $w_2$, nested Wilson loop, and gaps of the predicted SOTI material candidates, which include graphane (CH), graphene fluoride (CF), graphene chloride (CCl), silicene hydride (SiH), silicene chloride (SiCl), germanene hydride(GeH), stanene hydride (SnH), blue phosphorene (blue P), blue arsenene (blue As), black phosphorene (black P), black arsenene (black As), anti-Kekul\'e/Kekul\'e distortion graphenes, silicenes, germanenes and stanenes (anti-Kekul\'eG/Kekul\'eG, anti-Kekul\'eSi/Kekul\'eSi, anti-Kekul\'eGe/Kekul\'eGe, anti-Kekul\'eSn/Kekul\'eSn). The $C_n$ symmetry eigenvalues of XY X=(C, Si, Ge, Sn, Y=H, F, Cl) materials are calculated using a $\sqrt{3} \times \sqrt{3}$ supercell. The gap of anti-Kekul\'e distortion graphene is taken as the experimental value of 0.38 eV.} \label{table: 2}
			\begin{tabular}{cccccccccccccc}
				\toprule
				& \multicolumn{2}{c}{} & \multicolumn{3}{c}{Invariants} & \multicolumn{4}{c}{Number of Odd Parity} & $Q_c$    & $w_2$    & det($W_2$) & Gap (eV) \\
				\midrule
				&       &       &  $[K_1^{(3)}]$ & $[M_1^{(2)}]$ &       &  $\#M_2^{(I)}$ &$\#\Gamma_2^{(I)}$  &       &       &       &       &       &  \\
				\cmidrule{2-14}    \multirow{3}[1]{*}{$C_6$} & \multirow{2}[1]{*}{Anti-Kekul\'e} & $sp^2$   & \multicolumn{1}{c}{0} & \multicolumn{1}{c}{2} &       & \multicolumn{1}{c}{5} & \multicolumn{1}{c}{3} &       &       & 1/2   & 1     & $\pi$    & \multirow{2}[1]{*}{0.38} \\
				&       & $p_z$    & \multicolumn{1}{c}{0} & \multicolumn{1}{c}{-2} &       & \multicolumn{1}{c}{1} & \multicolumn{1}{c}{3} &       &       & 1/2   & 1     & $\pi$    &  \\
				& \multicolumn{2}{c}{Kekul\'e} & \multicolumn{1}{c}{0} & \multicolumn{1}{c}{-2} &       &    6   &    4   &       &       & 1/2   & 1     & $\pi$    &  \\
				&       &       & $[K_1^{(3)}]$ &  $[M_1^{(I)}]$      &       &$\#M_2^{(I)}$ & $\#\Gamma_2^{(I)}$ &       &       &       &       &       &  \\
				\cmidrule{2-14}    \multirow{9}[1]{*}{$C_3$} & \multicolumn{2}{c}{CH} & 0     &    -2   &       & 3     & 1     &       &       & 1/2   & 1     & $\pi$    & 3.5 \\
				& \multicolumn{2}{c}{CF} & 0     &   -2    &       & 6     & 4     &       &       & 1/2   & 1     & $\pi$    & 3.1 \\
				& \multicolumn{2}{c}{CCl} & 0     &   -2    &       & 6     & 4     &       &       & 1/2   & 1     & $\pi$    & 1.6 \\
				& \multicolumn{2}{c}{SiH} & 0     &   -2    &       & 3     & 1     &       &       & 1/2   & 1     & $\pi$    & 2.2 \\
				& \multicolumn{2}{c}{SiCl} & 0     &  -2     &       & 6     & 4     &       &       & 1/2   & 1     & $\pi$    & 1.3 \\
				& \multicolumn{2}{c}{GeH} & 0     &    -2   &       & 3     & 1     &       &       & 1/2   & 1     & $\pi$    & 1.0 \\
				& \multicolumn{2}{c}{SnH} & 0     &    -2   &       & 3     & 1     &       &       & 1/2   & 1     & $\pi$    & 0.5 \\
				& \multicolumn{2}{c}{BlueP} & 0     &   -2    &       & 3     & 1     &       &       & 1/2   & 1     & $\pi$    & 1.9 \\
				& \multicolumn{2}{c}{BlueAs} & 0     &   -2    &       & 3     & 1     &       &       & 1/2   & 1     & $\pi$    & 1.6 \\
				& \multicolumn{2}{c}{Anti-KekuleSi} & 0     &   0    &       & 6     & 6     &       &       & 0   & 0     & 0    &   \\
				& \multicolumn{2}{c}{Anti-KekuleGe} & 0     &   0    &       & 6     & 6     &       &       & 0   & 0     & 0    &   \\
				& \multicolumn{2}{c}{Anti-KekuleSn} & 0     &   0    &       & 6     & 6     &       &       & 0   & 0     & 0    &   \\
				& \multicolumn{2}{c}{KekuleSi} & 0     &   -2    &       & 6     & 4     &       &       & 1/2   & 1     & $\pi$    &   \\
				& \multicolumn{2}{c}{KekuleGe} & 0     &   -2    &       & 6     & 4     &       &       & 1/2   & 1     & $\pi$    &   \\
				& \multicolumn{2}{c}{KekuleSn} & 0     &   -2    &       & 6     & 4     &       &       & 1/2   & 1     & $\pi$    &   \\
				&       &       & $[X_1^{(2)}]$ & $[Y_1^{(2)}]$ & $[M_1^{(2)}]$ & $\#M_2^{(I)}$ & $\#\Gamma_2^{(I)}$ &   $\#X_2^{(I)}$ & $\#Y_2^{(I)}$      &       &  \\
				\cmidrule{2-14}    \multirow{2}[2]{*}{$C_2$} & \multicolumn{2}{c}{BlackP} & -2    & -2    & -2    & 4     & 4     & 5     & 5     & 1/2   & 1     & $\pi$    & 0.9 \\
				& \multicolumn{2}{c}{BlackAs} & -2    & -2    & -2    & 4     & 4     & 5     & 5     & 1/2   & 1     & $\pi$    & 0.7 \\
				\bottomrule
			\end{tabular}%
			\label{tab:addlabel}%
		}
	\end{table*}%

	\section{Charge spatial distribution of corner states}
\begin{figure}
		\begin{center}
			\includegraphics[width=1\linewidth]{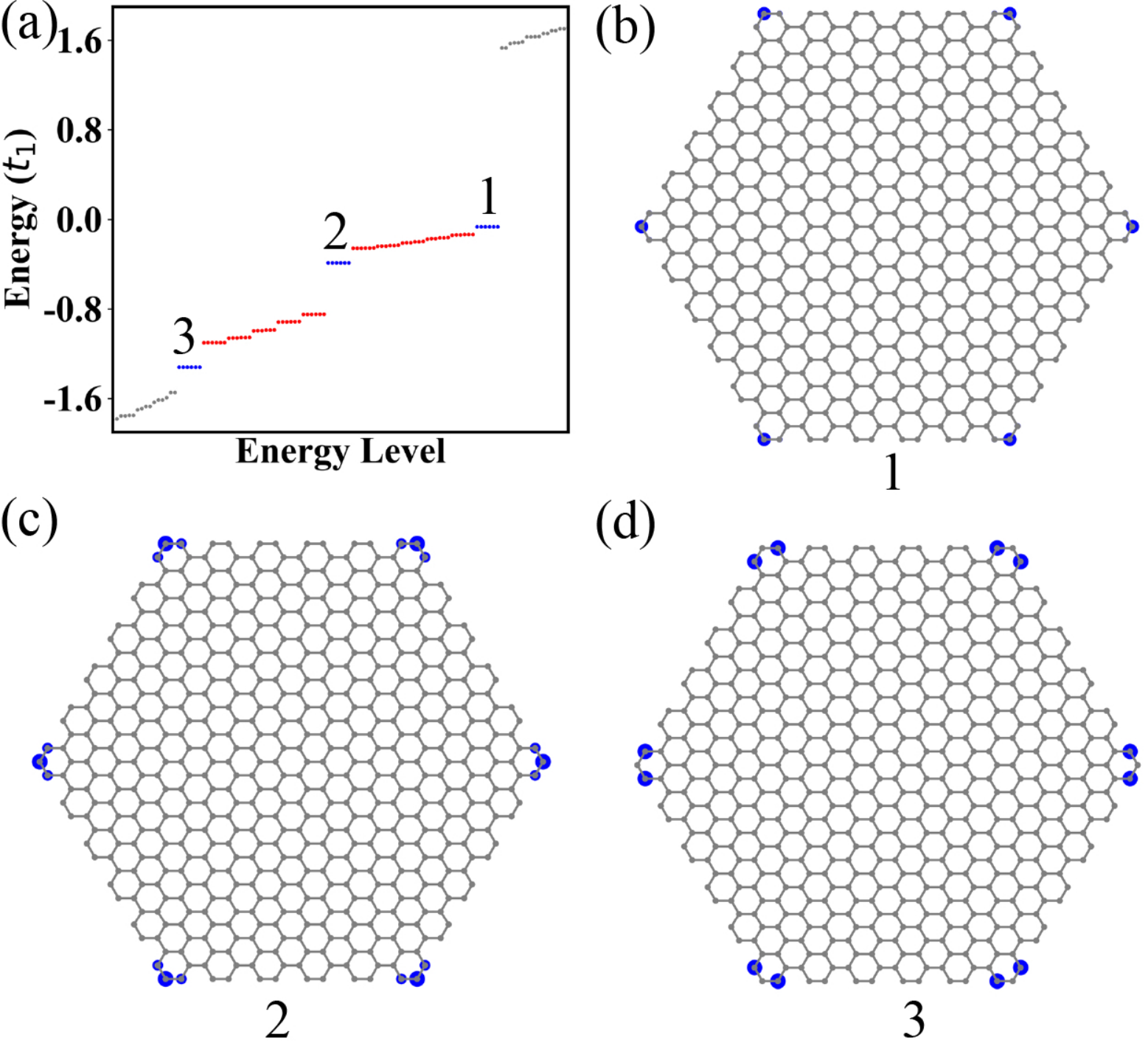}
		\end{center}
		\caption{(a) Energy Spectrum of the finite-size flake of the eight-band TB model. Corner states are located up, middle and below of the edge states, which are labeled as 1, 2, 3 in (a). The charge spatial distribution of the three corner states (b, c, d).} \label{fig: corner}
	\end{figure}

	There are usually more than one corner state at each corner in the two models and SOTI candidate materials. For example, three corner states located at different energy in the eight-band TB model. The three corner states are labeled as 1, 2 and 3 at the Fig. \ref{fig: corner} and the charge distribution pattern of the three corner states are slight different but all are well localized at the corners of the finite-size flake.
	
	\section{Numerical calculation of corner charge}
	\begin{figure}
		\begin{center}
			\includegraphics[width=1\linewidth]{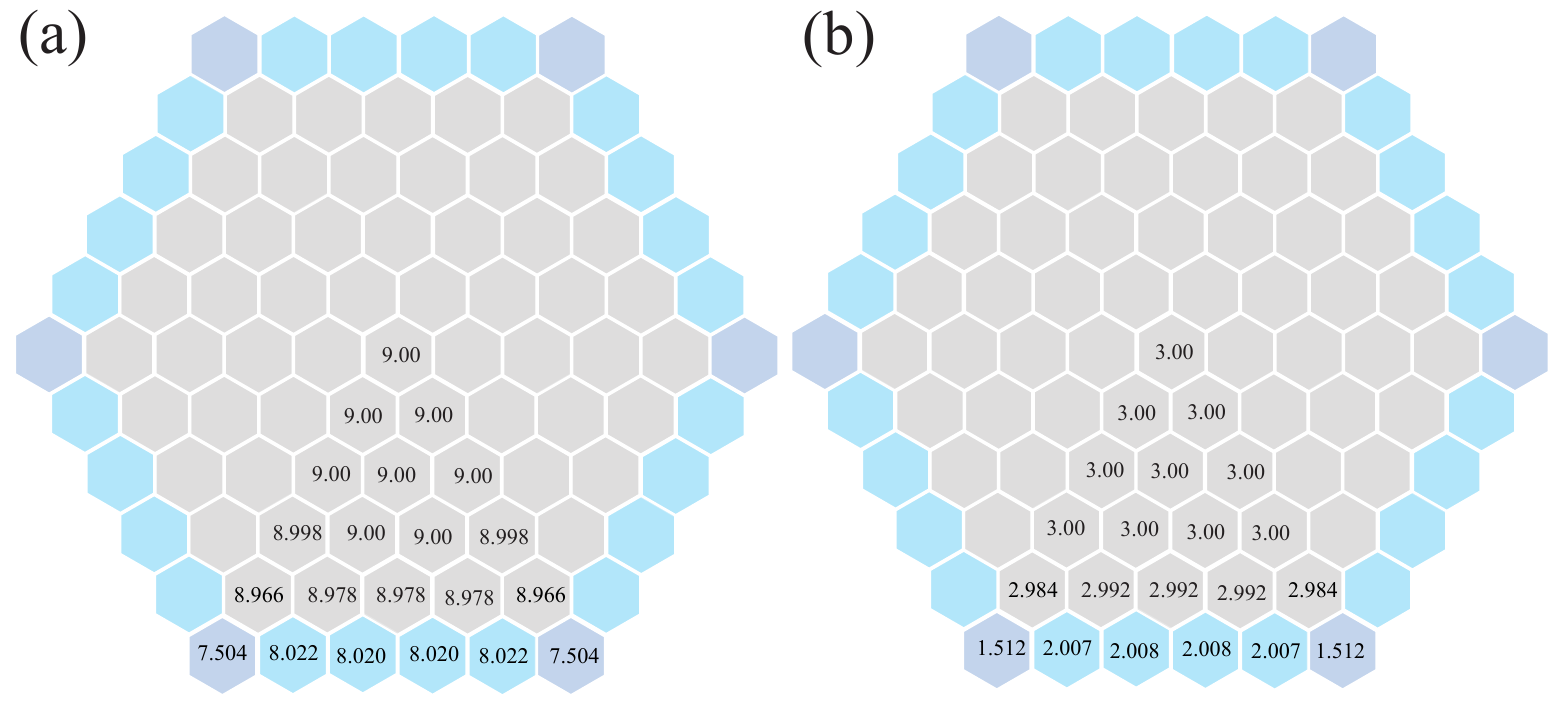}
		\end{center}
		\caption{Charge distribution of model I (a) and model II (b). Light blue and dark blue represent the edge and corner regions. We consider $sp^2$ orbitals in model I.} \label{fig: sfrac}
	\end{figure}
	
	The charge accumulated in a corner region of finite-size flake is 
	\begin{equation}
		Q_{c}=\sum_{i \in \text { occupied }} \int_{Corner} d \boldsymbol{r}\left|\psi_{i}(\boldsymbol{r})\right|^{2} \bmod 1,
	\end{equation}
	where $\psi_{i}(\boldsymbol{r})$ is the wavefunction of the occupied eigenstates and the integration is performed within the corner region. We calculated the charge distribution of every unit cell of finite-size flake of model I and model II (Fig. \ref{fig: sfrac}). The corner charge of model I and model II are 0.504 and 0.512 respectively, which are close to the theoretical value 1/2.

\end{document}